\RequirePackage{snapshot} % Needed by bundledoc
\documentclass[aps,prb,reprint,superscriptaddress,longbibliography
]{revtex4-2}

\usepackage{microtype}
\usepackage[T1]{fontenc}

\usepackage[
	colorlinks=true, citecolor=teal, linkcolor=purple, urlcolor=brown
]{hyperref}% add hypertext capabilities

\usepackage{amsmath}
\usepackage{amssymb}
\usepackage{empheq} % For boxes around multiple equations, among other things
\usepackage{dsfont} % For identity matrix 1

\usepackage[dinkus]{froufrou}
\usepackage{enumitem}

\usepackage{tabularray} % A powerful environment for creating arrays
\UseTblrLibrary{varwidth}

\newcommand{\rb}{\mathrm{b}}

\newcommand{\rd}{\mathrm{d}} % roman d
\newcommand{\re}{\mathrm{e}} % roman e
\newcommand{\rh}{\mathrm{h}}
\newcommand{\ri}{\mathrm{i}} % roman i
\newcommand{\ro}{\mathrm{o}} % roman d
\newcommand{\rp}{\mathrm{p}}
\newcommand{\rs}{\mathrm{s}}
\newcommand{\bB}{\mathbf{B}} % bold E
\newcommand{\bD}{\mathbf{D}} % bold E
\newcommand{\bR}{\mathbf{R}} % bold R
\newcommand{\bS}{\mathbf{S}} % bold S
\newcommand{\ba}{\mathbf{a}} % bold a
\newcommand{\bk}{\mathbf{k}} % bold k
\newcommand{\bp}{\mathbf{p}} % bold p
\newcommand{\bx}{\mathbf{x}} % bold x
\newcommand{\btau}{\boldsymbol{\tau}} % bold tau
\newcommand{\bdelta}{\boldsymbol{\delta}} % bold delta
\newcommand{\tM}{\tilde{M}}
\newcommand{\tQ}{\tilde{Q}}
\newcommand{\tS}{\tilde{S}}
\newcommand{\tV}{\tilde{V}}
\newcommand{\tbB}{\tilde{\bB}}
\newcommand{\tbS}{\tilde{\bS}}
\def\cB {\mathcal{B}}

\def\cP {\mathcal{P}}

\def\barP {\bar{P}}
\def\hn {\hat{n}}
\def\hz {\hat{z}}

\newcommand\set[1]{#1}

\newcommand{\transpose}{\top}

\newcommand\sgn[1]{\text{sgn} \, #1}
\newcommand\abs[1]{\text{abs} \, #1}

\def\idmat {\mathds{1}}

\def\tauzero {\tau^0}
\def\tauone {\tau^1}
\def\tauthree {\tau^3}
\def\taumu {\tau^\mu}

\def\krein {\sharp}
\def\disp {\epsilon}

\newcommand\inlineinner[2]{\langle #1, #2 \rangle}
\newcommand\bd[1]{{#1}_\text{b}}
\newcommand\bod[1]{{#1}_\text{o}}

\newcommand\opnorm[1]{\lVert #1 \rVert}

\newcommand\eq[1]{Eq.~#1}
\newcommand\Eq[1]{Equation~#1}
\newcommand\sect[1]{Sec.~#1}

\newcommand\app[1]{Appendix~#1}
\newcommand\App[1]{Appendix~#1}
\newcommand\reference[1]{Ref.~\onlinecite{#1}}

\newcommand\fig[1]{Fig.~#1}

\newcommand\tab[1]{Table~#1}

\newcommand\mycomment[1]{}

\def\integers  {\mathbb{Z}}

\def\reals     {\mathbb{R}}
\def\complex   {\mathbb{C}}

\newcommand\MI{h_\text{MI}}
\newcommand\EMI{h_\text{EMI}}

\begin{document}

% Hyphenations
\hyphenation{para-unitary non-unitary ortho-normal to-po-lo-gy to-po-lo-gi-cal to-po-lo-gi-cal-ly de-ge-ne-rate de-ge-ne-ra-cy}

%\title{Krein-Hermitian Schrieffer-Wolff transformation and nodal lines in \replaced{space-time-inversion}{PT}-symmetric \replaced{magnon}{bosonic} Hamiltonians}
%\title{Krein-Hermitian Schrieffer-Wolff transformation, space-time-inversion symmetry, and nodal lines in boson Hamiltonians}
%\title{Krein-Hermitian Schrieffer-Wolff transformation and band touchings in bosonic BdG Hamiltonians}
\title{Krein-unitary Schrieffer-Wolff transformation and band touchings in bosonic Bogoliubov--de Gennes and other Krein-Hermitian Hamiltonians}

\author{Geremia Massarelli}
\thanks{These authors contributed equally}
\affiliation{Department of Physics, University of Toronto, Toronto, Ontario M5S 1A7, Canada}

\author{Ilia Khait}
\thanks{These authors contributed equally}
\affiliation{Department of Physics, University of Toronto, Toronto, Ontario M5S 1A7, Canada}

\author{Arun Paramekanti}
\email[Electronic mail address: ]{arunp@physics.utoronto.ca}
\affiliation{Department of Physics, University of Toronto, Toronto, Ontario M5S 1A7, Canada}

\begin{abstract}
	Krein-Hermitian Hamiltonians, i.e., Hamiltonians Hermitian with respect to an indefinite inner product, have emerged as an important class of non-Hermitian Hamiltonians in physics, encompassing both single-particle bosonic Bogoliubov--de Gennes (BdG) Hamiltonians and so-called ``$PT$-symmetric'' non-Hermitian Hamiltonians. 
	In particular, they have attracted considerable scrutiny owing to the recent surge in interest for boson topology.
	Motivated by these developments, we formulate a perturbative Krein-unitary Schrieffer-Wolff transformation for finite-size dynamically stable Krein-Hermitian Hamiltonians, yielding an effective Hamiltonian for a subspace of interest.
	The effective Hamiltonian is Krein Hermitian and, for sufficiently small perturbations, also dynamically stable.
	As an application, we use this transformation to justify codimension-based analyses of band touchings in bosonic BdG Hamiltonians, which complement topological characterization.
	We use this simple approach based on symmetry and codimension to revisit known topological magnon band touchings in several materials of recent interest.
\end{abstract}

\maketitle

%\tableofcontents

\section{Introduction}\label{sec:intro}

Non-Hermitian Hamiltonians have become a central field of study in condensed-matter physics in the last decades~\cite{NonSelfadjoint2015}. 
Important examples of systems governed by non-Hermitian Hamiltonians are quadratic bosonic Bogoliubov--de Gennes (BdG) Hamiltonians~\cite{Colpa1978Diagonalization, BlaizotRipka}, which have been the subject of renewed interest recently in the context of topological systems~\cite{Shivam2017Neutron, McClarty2018Topological, lu2018magnon, Gong2018NonHermitian, Kawabata2019NonHermitian, McClarty2019NonHermitian, Flynn2020Deconstructing, Xu2020Squaring, Kumar2020Dirac, McClarty2021SpinSpaceGroups}, and so-called ``$PT$-symmetric'' non-Hermitian Hamiltonians, which initially attracted interest because they could give rise to real spectra despite not being Hermitian~\cite{bender1999pt, Bender_2007}. 
%Note that in the context of non-Hermitian Hamiltonians, \added{a ``$PT$ symmetry'' has come to refer to any of a certain/particular/specific kind/form of antilinear symmetry---see \reference{Flynn2020Deconstructing}, \sect{3.1} for a discussion---}and not necessarily to space-time inversion symmetry.
Note that in the context of non-Hermitian Hamiltonians, a ``$PT$ symmetry'' has come to refer to, loosely, any antilinear symmetry whose linear part is involutory (see \reference{Flynn2020Deconstructing}, \sect{3.1} for a careful discussion), and not necessarily to space-time inversion symmetry.
%\added{More references on the equivalence of PT symmetry and Krein Hermicity:~\cite{Solombrino2002Weak, Scolarici2003On}.}
Both these examples fall into a class of non-Hermitian Hamiltonians called \emph{Krein Hermitian}~\cite{Flynn2020Deconstructing, Mostafazadeh2002PseudoI, Mostafazadeh2002PseudoII, Mostafazadeh2002PseudoIII, mostafazadeh2006krein, Tanaka2006Krein1, Tanaka2006Krein2, AlbeverioKuzhel2015, elganainy2018nonhermitian}, also referred to in the literature as ``pseudo-Hermitian''~\cite{Mostafazadeh2002PseudoI, Mostafazadeh2002PseudoII, Mostafazadeh2002PseudoIII, mostafazadeh2006krein, Flynn2020Restoring, Flynn2020Deconstructing, Xu2020Squaring} or ``para-Hermitian''~\cite{Lein2019Krein}.

%We mention that Krein-Hermitian Hamiltonians are closely related to $PT$-symmetric non-Hermitian Hamiltonians, which have been studied in recent decades in the context of lossy systems~\cite{elganainy2018nonhermitian} and out of fundamental interest for the axioms of quantum mechanics~\cite{bender1999pt}.

Krein-Hermitian Hamiltonians are Hermitian with respect to an indefinite, nondegenerate inner product, rather than the conventional positive-definite inner product~\cite{Lein2019Krein}.
They have spectral properties that enable them to describe stable physical systems: the phase diagram of a Krein-Hermitian Hamiltonian typically includes \emph{dynamically stable} regions, in which the Hamiltonian is diagonalizable with real eigenvalues and with eigenvectors that are orthonormal with respect to the indefinite inner product.
Lately, the mathematics of Krein-Hermitian Hamiltonians and Krein spaces more generally have seen growing interest and adoption among theoretical physicists as a tool to approach these specific non-Hermitian problems~\cite{AlbeverioKuzhel2015, Lein2019Krein, Flynn2020Deconstructing, Xu2020Squaring}.

% Context - renewed interest in the SW transformation. Non-Hermitian SW transformation.
The Schrieffer-Wolff (SW) transformation~\cite{Bravyi2011Schrieffer}, also known as the L\"owdin partitioning method~\cite{Winkler2003Quasi}, has been the object of renewed interest in recent years. Novel non-perturbative approaches have yielded new results, like the expansion's radius of convergence and closed-form expressions to all orders~\cite{Bravyi2011Schrieffer}.
Moreover, the SW approach has been generalized to obtain non-unitary similarity transformations of (non-Hermitian) Liouvillian operators in order to perturbatively eliminate fast degrees of freedom in dissipative quantum systems~\cite{Kessler2012Generalized, Kessler2012Dissipative}.

Previously, perturbative transformations have been used in bosonic BdG Hamiltonians specifically to eliminate boson pairing terms~\cite{Shindou2013SpinWave, Shindou2014Magnetostatic, McClarty2018Topological, Zhou2020Bosonic, Wan2021Squeezing}.
Other recent works have simply assumed more or less implicitly that, in regards to perturbation theory and effective Hamiltonians, Krein-Hermitian Hamiltonians can be treated the same as Hermitian Hamiltonians.
However, a systematic study highlighting differences and similarities between them in this regard has been lacking.
In \sect{\ref{sec:KreinHermitianSW}}, we formulate a general perturbative SW transformation for dynamically stable, finite-dimensional Krein-Hermitian Hamiltonians, yielding a reduced effective Hamiltonian that is Krein Hermitian in a reduced space and is sure to be Krein-unitarily diagonalizable. 
% Not the full story: if the metric is not diagonal with entries \pm 1, then the Krein-Hermitian Hamiltonian can be transformed into a Hermitian one via a Krein-unitary transformation. An additional, unitary transformation is required to diagonalize this secondary Hermitian matrix.
Unlike previous work on general non-Hermitian SW transformations~\cite{Kessler2012Generalized, Kessler2012Dissipative}, our transformation is explicitly Krein unitary; further, we highlight the implications of Krein stability theory on the effective Hamiltonian and its diagonalizability. 
In particular, it is safe to perturbatively study degenerate and near-degenerate states as long as they have the same (nonzero) Krein signature, since small perturbations cannot drive them to dynamical instability.

Aside from the practical applications of a SW transformation for Krein-Hermitian Hamiltonians, our results allow a justification for the use of reduced Hamiltonians describing a subset of contiguous energy bands in a Krein-Hermitian system, as discussed in \sect{\ref{sec:nodal-lines-magnons}}.
In particular, we clarify why adjacent positive-energy bands of thermodynamically stable systems can simply be described by a Hermitian reduced Hamiltonian, even though the larger Hamiltonian they originate from is Krein Hermitian.

We use this observation to justify codimension-based arguments for band touchings in bosonic BdG Hamiltonians, focusing on magnons for concreteness.
Such arguments are well known for Hermitian Hamiltonians~\cite{Tiwari2020NodalLine}. 
We then revisit known results on topological band touchings and the consequences of space-time inversion symmetry in magnon systems, based solely on codimension and symmetry arguments.
In short, the behavior hinges on the fact that space-time inversion squares to $+1$ for bosons~\cite{Li2017Dirac}.
We focus on Dirac points and gapless lines in the magnon spectrum of Cu\textsubscript{3}TeO\textsubscript{6}~\cite{Li2017Dirac, Yao2018Antiferromagnet, Bao2018Antiferromagnet} and gapless lines in the magnon spectrum of CoTiO\textsubscript{3}~\cite{Yuan2020, Elliot2020Visualization}, both part of a recent drive to identify topological band touchings in bosonic systems.

% Phononic Helical Nodal Lines with PT Protection in MoB2~\cite{Zhang2019Phononic}
% Discovering Topological Surface States of Dirac Points (acoustic crystal)~\cite{Cheng2020SurfaceStates}
% Topological magnon nodal lines and absence of magnon spin Nernst effect in layered collinear antiferromagnets~\cite{Owerre2019MagnonNodalLines}
% Competing phases and topological excitations of spin-1 pyrochlore antiferromagnets~\cite{Li2018PyrochloreAFMs}

\App{\ref{app:SW-expansion}} contains details on the SW transformation for Krein-Hermitian matrices, and \app{\ref{sec:symmetries-implementation}} discusses magnons in linear spin-wave theory as well as the implementation of spin-space-group symmetries on the linear spin-wave Hamiltonian.

To avoid confusion with the aforementioned notion of ``$PT$ symmetry'' in the context of non-Hermitian Hamiltonians, we refer to space-time inversion symmetry as \emph{magnetic inversion} symmetry rather than ``$PT$ symmetry''.
The Krein Hermicity---equivalent to so-called ``$PT$ symmetry''~\cite{Flynn2020Deconstructing}---of bosonic BdG Hamiltonians stems solely from the bosonic commutation relations and holds regardless of the system's physical symmetries.

\section{Bosonic BdG Hamiltonians} \label{sec:bosonic-bdg-intro}
In this section, we review the mathematics of quadratic BdG Hamiltonians of bosons, which will be our main focus in this work, though our results have broader applicability. This keeps the article self-contained, and allows us to introduce necessary notions and definitions.

Consider a bosonic Fock space made up of countably many single-particle states, and a quadratic Hermitian Hamiltonian $\hat{H}$ in that space.
Although the second-quantized Hamiltonian $\hat{H}$ is Hermitian and the bosonic Fock space is necessarily of infinite dimension, the ``single-particle Hamiltonian'' that governs the time evolution and determines the spectrum of the non-interacting bosons is finite dimensional and, in general, as we will see, Krein Hermitian.
For concreteness, we assume the system has (either discrete or continuous) translational invariance, allowing us to write
\begin{equation} \label{eq:boson-bdg-ham}
	\hat{H} = \frac{1}{2} \sum_{\bk} \hat{\Phi}_{\bk}^\dagger H^{\vphantom{\dagger}}_{\bk} \hat{\Phi}^{\vphantom{\dagger}}_{\bk},
\end{equation}
where $\hat{\Phi}_{\bk}^\dagger = [b_{\bk,1}^\dagger, \dots, b_{\bk,n}^\dagger, b_{-\bk,1}^{\vphantom{\dagger}}, \dots, b_{-\bk,n}^{\vphantom{\dagger}}]$ is the Nambu spinor, and $b_{\bk \alpha}$ is the annihilation operator for the state with wave vector $\bk$ and ``flavor'' $\alpha \in \{1, \dots, n\}$; the extension of our results to systems without translational invariance presents no difficulties.
%{state $\alpha$ in the unit cell and with wavevector $\bk$}
In the case of discrete translational invariance, the momentum sum is restricted to a Brillouin zone, while in the case of continuous translational invariance, it is not restricted.
In magnon systems such as those considered in \sect{\ref{sec:KreinHermitianSW}} and \app{\ref{sec:symmetries-implementation}}, $\bk$ is summed over the magnetic Brillouin zone and $\alpha$ labels the different spins within the magnetic unit cell. 
Notice that for clarity, we use hats on Hilbert-space operators like $\hat{H}$, but not on coefficient matrices like $H_{\bk}$.

The redundancy inherent in the Nambu formalism is eliminated by choosing the Hermitian coefficient matrix $H_{\bk}$ to satisfy the particle-hole (PH) constraint, $\tauone H^{\vphantom{\dagger}}_{\bk} \tauone = H_{-\bk}^*$. 
Here and throughout the article, $\taumu$, short for $\taumu \otimes \idmat_{n}$, are the Pauli matrices in Nambu space.

Note that if the system is particle conserving, $H_{\bk}$ is block diagonal in the Nambu space, in which case the spectrum of $\hat{H}$ and the system's time evolution are simply determined by the unitary eigendecomposition of $H_{\bk}$.

\subsection{Diagonalization} \label{sec:bosonic-bdg-intro-diagonalization}

Unlike for fermions, a Bogoliubov transformation for bosons is not simply achieved by the eigendecomposition of the coefficient matrix $H_{\bk}$, as such a transformation does not preserve the bosonic canonical commutation relations~\cite{Colpa1978Diagonalization, BlaizotRipka, Xiao2009Theory, Kawaguchi2012Spinor, Choi2019Nonsymmorphic}.
Writing the transformation to a new bosonic Nambu spinor $\hat{\Gamma}_{\bk}$ as $\hat{\Phi}_{\bk} = T_{\bk} \hat{\Gamma}_{\bk}$, meaning
\begin{equation}
	\hat{H} = \frac{1}{2} \sum_{\bk} \hat{\Gamma}_{\bk}^\dagger \bigl( T_{\bk}^\dagger H^{\vphantom{\dagger}}_{\bk} T^{\vphantom{\dagger}}_{\bk} \bigr) \hat{\Gamma}^{\vphantom{\dagger}}_{\bk},
\end{equation}
the $2n \times 2n$ transformation matrix $T_{\bk}$ must fulfill 
\begin{equation} \label{eq:paraunitary-constraint}
	T_{\bk}^\dagger \tauthree T^{\vphantom{\dagger}}_{\bk} = T^{\vphantom{\dagger}}_{\bk} \tauthree T_{\bk}^\dagger = \tauthree
\end{equation} 
in order to preserve the bosonic commutation relations. 
Hence, in general, $T_{\bk}$ is not unitary, and the transformation on $H_{\bk}$ is not a similarity transformation~\footnote{
	In general, a matrix transformation $P^\dagger A P$, where $P$ is invertible, is known as a congruence transformation. Congruence transformations and similarity transformations overlap when $P$ is unitary.
}.
A matrix satisfying the constraint of \eq{\eqref{eq:paraunitary-constraint}} has been called \emph{paraunitary} in the physics literature~\cite{Colpa1978Diagonalization}. 

In order to give rise to the eigenmodes of $\hat{H}$, the transformation $T_{\bk}$ should be chosen such that $T_{\bk}^\dagger H^{\vphantom{\dagger}}_{\bk} T^{\vphantom{\dagger}}_{\bk}$ is diagonal with real entries.
If this is possible, the Hamiltonian can be written in its eigenbasis as
\begin{equation}
	\hat{H} = \sum_\bk \sum_{\alpha=1}^{n} \disp_{\bk,\alpha} \frac{\gamma_{\bk,\alpha}^\dagger \gamma^{\vphantom{\dagger}}_{\bk,\alpha} + \gamma^{\vphantom{\dagger}}_{\bk,\alpha} \gamma_{\bk,\alpha}^\dagger}{2},
\end{equation}
where we have used the fact that the particle-hole constraint ensures the modes at $\bk$ and $-\bk$ have the same energies and appropriately related eigenvectors.
As we shall see in reviewing Krein theory, that $H_{\bk}$ is positive definite, which we denote $H_{\bk} > 0$, is a sufficient condition for there to exist a paraunitary $T_{\bk}$ such that $T_{\bk}^\dagger H^{\vphantom{\dagger}}_{\bk} T^{\vphantom{\dagger}}_{\bk}$ is diagonal with real, strictly positive entries
% Sylvester's law of inertia?
\footnote{
    It seems this was initially proven in the physics literature using a constructive approach based on the Cholesky decomposition of $H_{\bk}$~\cite{Colpa1978Diagonalization}, rather than using the tools from Krein theory.
}.

The Hamiltonian $\hat{H}$ is \emph{thermodynamically stable}, meaning it is bounded below, if and only if $H_{\bk} \geq 0$ at all $\bk$~\cite{Xu2020Squaring}. 
In a linear spin-wave system, for example, $H_{\bk} \geq 0$ holds as long as the underlying classical ground state is a local minimum of the classical energy.
In many physical systems for which $H_{\bk}$ is positive semidefinite but not positive definite, the zero eigenspace corresponds to the Goldstone mode; in such systems, $H_{\bk} > 0$ for $\bk \neq 0$~\cite{BlaizotRipka, Xu2020Squaring}.

%\added{Further discuss zero modes?}

Finally, we point out that the transformed coefficient matrix $T_{\bk}^\dagger H^{\vphantom{\dagger}}_{\bk} T^{\vphantom{\dagger}}_{\bk}$ can be rewritten using the paraunitarity constraint of \eq{\eqref{eq:paraunitary-constraint}}:
\begin{equation}
	T_{\bk}^\dagger H^{\vphantom{\dagger}}_{\bk} T^{\vphantom{\dagger}}_{\bk} = \tauthree  \Bigl( T_{\bk}^{-1} \bigl(\tauthree H^{\vphantom{\dagger}}_{\bk}\bigr) T^{\vphantom{\dagger}}_{\bk} \Bigr).
\end{equation}
Since $\tauthree$ is diagonal, this form shows that the spectrum and the transformation $T_{\bk}$ can be found from the (nonunitary) eigendecomposition of the non-Hermitian matrix $\tauthree H_{\bk}$. Since it determines the spectrum and governs the system's time evolution~\cite{Flynn2020Deconstructing}, we refer to $\tauthree H_{\bk}$ as the \emph{single-particle Hamiltonian}.
The properties of the single-particle Hamiltonian, and of Krein-Hermitian Hamiltonians more generally, are reviewed in \sect{\ref{subsec:KreinPrimer}}.

\subsection{Unitary transformations}
We also review how a unitary transformation on the Hamiltonian $\hat{H}$ of \eq{\eqref{eq:boson-bdg-ham}} translates to a paraunitary transformation on the single-particle Hamiltonian $\tauthree H_{\bk}$~\cite{BlaizotRipka, McClarty2018Topological, Wan2021Squeezing}. 
To see this, consider the transformed Hamiltonian $\hat{\tilde{H}} = \re^{-\hat{W}} \hat{H} \re^{\hat{W}}$, where $\re^{\hat{W}}$ is a unitary transformation in the Fock space generated by the skew-Hermitian generator $\hat{W}$, and rewrite it using the following expansion:
\begin{equation} \label{eq:expansion}
	\hat{\tilde{H}} = \re^{-\hat{W}} \hat{H} \re^{\hat{W}} = \hat{H} + [\hat{H},\hat{W}] + \frac{1}{2!} [[\hat{H},\hat{W}],\hat{W}] + \dots
\end{equation}
In order for the transformation to keep the Hamiltonian quadratic, its generator $\hat{W}$ must itself be quadratic: $\hat{W} = \frac{1}{2} \sum_{\bk} \hat{\Phi}_{\bk}^\dagger W^{\vphantom{\dagger}}_{\bk} \hat{\Phi}^{\vphantom{\dagger}}_{\bk}$, where $W_{\bk}$ is a skew-Hermitian $2n \times 2n$ matrix. 
Furthermore, imposing the PH constraint on $W_{\bk}$ (i.e., $\tau^1 W^{\vphantom{\dagger}}_{\bk} \tau^1 = W_{-\bk}^\transpose$) will ensure the coefficient matrix for $\hat{\tilde{H}}$ is PH symmetric, in addition to eliminating the redundancy in $W_{\bk}$ intrinsic to the Nambu formalism.

Using the bosonic commutation relations $[\Phi_{\bk}^a, {\Phi_{\bp}^b}^\dagger] = (\tau^3)^{ab} \delta_{\bp,\bk}$, where $a$ and $b \in \{1, \dots, 2n\}$, and the PH constraint $\tau^1 H^{\vphantom{\dagger}}_{\bk} \tau^1 = H_{-\bk}^\transpose$, one finds that the commutator is given by
\begin{equation} \label{eq:W_commutator}
	[\hat{H},\hat{W}] = \frac{1}{2} \sum_{\bk} \hat{\Phi}_{\bk}^\dagger ( H^{\vphantom{\dagger}}_{\bk}, W^{\vphantom{\dagger}}_{\bk} ) \hat{\Phi}^{\vphantom{\dagger}}_{\bk},
\end{equation}
where $ ( H_{\bk}, W_{\bk} ) \coloneqq H_{\bk} \tau^3 W_{\bk} - W_{\bk} \tau^3 H_{\bk}$ is a $\tau^3$-intertwined commutator, which arises from the bosonic commutation relations. 
Since $H_{\bk}$ and $W_{\bk}$ are PH symmetric, so is $(H_{\bk}, W_{\bk})$; hence, the result of \eq{\eqref{eq:W_commutator}} can be iterated to find all orders of $\hat{\tilde{H}}$ and the transformed Hamiltonian of \eq{\eqref{eq:expansion}} takes the form
\begin{equation}\label{eq:transformation-nambuspace}
	\begin{split}
		\hat{\tilde{H}} &= \frac{1}{2} \sum_{\bk} \hat{\Phi}_{\bk}^\dagger
		\Bigl(
		H^{\vphantom{\dagger}}_{\bk} + (H^{\vphantom{\dagger}}_{\bk}, W^{\vphantom{\dagger}}_{\bk}) 
		\\ 
		& \qquad \qquad \qquad + \frac{1}{2!} ((H^{\vphantom{\dagger}}_{\bk}, W^{\vphantom{\dagger}}_{\bk}),W^{\vphantom{\dagger}}_{\bk}) + \dots
		\Bigr) 
		\hat{\Phi}^{\vphantom{\dagger}}_{\bk}
		\\
		&= \frac{1}{2} \sum_{\bk} \hat{\Phi}_{\bk}^\dagger \Bigl( \tau^3 \re^{-\tau^3 W_{\bk}} \tau^3 H^{\vphantom{\dagger}}_{\bk} \re^{\tau^3 W_{\bk}} \Bigr) \hat{\Phi}^{\vphantom{\dagger}}_{\bk}
		\\
		&= \frac{1}{2} \sum_{\bk} \hat{\Phi}_{\bk}^\dagger \ \underbrace{\tau^3 T_{\bk}^{-1} \tau^3}_{T_{\bk}^\dagger} H^{\vphantom{\dagger}}_{\bk} T^{\vphantom{\dagger}}_{\bk} \ \hat{\Phi}^{\vphantom{\dagger}}_{\bk}.
	\end{split}
\end{equation}
In the last line, we have introduced the paraunitary transformation $T_{\bk} \coloneqq \re^{\tau^3 W_{\bk}}$; indeed, any $2n\times2n$ paraunitary matrix can be generated in this way with a skew-Hermitian matrix $W_{\bk}$. 
Furthermore, the transformation $T_{\bk}$ inherits a particle-hole constraint from that of $W_{\bk}$, i.e., $\tau^1 T^{\vphantom{\dagger}}_{\bk} \tau^1 = T_{-\bk}^*$~\footnote{
	The various expressions of PH symmetry all arise in the same way, from the redundancy of the Nambu spinors. On the (Hermitian) Bloch-BdG matrix, it reads as $\tau^1 H_{\bk} \tau^1 = H_{-\bk}^* = H_{-\bk}^\transpose$. On the transformation matrix, it reads as $\tau^1 T_{\bk} \tau^1 = T_{-\bk}^*$, making the transformed Bloch-BdG matrix P-H symmetric. On the (skew-Hermitian) generator $W_{\bk}$, it reads as $\tau^1 W_{\bk} \tau^1 = W_{-\bk}^\transpose = - W_{-\bk}^*$.
}. 

\Eq{\eqref{eq:transformation-nambuspace}} shows that a unitary transformation $\hat{H} \rightarrow \re^{-\hat{W}} \hat{H} \re^{\hat{W}}$ on the second-quantized Hamiltonian has a paraunitary representation $\tauthree H^{\vphantom{\dagger}}_\bk \rightarrow T_\bk^{-1} \tauthree H^{\vphantom{\dagger}}_\bk T^{\vphantom{\dagger}}_\bk$ on the single-particle Hamiltonian.
%Since $H_\bk$ is positive-definite, so is $T_\bk^\dagger H_\bk T_\bk$ as dictated by Sylvester's law of inertia; hence, unitary transformations of a bosonic Hamiltonian preserve its diagonalizability.
Since the single-particle Hamiltonian undergoes a similarity transformation, the invariance of the spectrum is manifest.

Had the ladder operators been fermionic and obeyed anticommutation relations rather than commutation relations, the unitary transformation on the second-quantized Hamiltonian would have had a more familiar unitary representation on the single-particle Hamiltonian.

\section{Krein-Hermitian Hamiltonians and perturbation theory}\label{sec:KreinHermitianSW}

In this section, we show that the SW transformation can be used to formulate a Krein-unitary perturbation theory for finite-size Krein-Hermitian Hamiltonians within a dynamically stable phase.
A Hamiltonian is \emph{dynamically stable}~\footnote{
	Note that dynamical stability and thermodynamic stability are fully independent concepts: Krein-Hermitian Hamiltonians can be both, either, or neither~\cite{Flynn2020Deconstructing}.
}, meaning the time evolution it generates is bounded, iff the Hamiltonian is diagonalizable with real eigenvalues~\cite{Flynn2020Deconstructing}.

Since single-particle BdG Hamiltonians $\tauthree H_\bk$ and paraunitary transformations $T_\bk$ are, respectively, Krein Hermitian and Krein unitary with respect to $\tauthree$, our results allow for a perturbation theory perturbatively yielding bosonic eigenmodes as well as mode energies.

We begin by briefly reviewing necessary aspects of Krein spaces, and then present the perturbation theory.

\subsection{Krein spaces}\label{subsec:KreinPrimer}

Although the single-particle Hamiltonian $\tauthree H_\bk$ is not Hermitian, it is Hermitian with respect to an indefinite inner product with metric $\tauthree$, $\inlineinner{\phi}{\psi}_{\tauthree} \coloneqq \phi^\dagger \tauthree \psi$, in the sense that $\inlineinner{\tauthree H \phi}{\psi}_{\tauthree} = \inlineinner{\phi}{\tauthree H \psi}_{\tauthree}$ for any vectors $\phi$ and $\psi \in \complex^{2n}$. 
%nondegenerate  % IK is it important?
Vector spaces endowed with an indefinite, nondegenerate inner product in addition to a positive-definite inner product are called \emph{Krein spaces}~\cite{Lein2019Krein, Peano2018Topological, SchulzBaldes2017Signatures}.
While $\tauthree H_\bk$ may fail to be diagonalizable or to have real eigenvalues because it is non-Hermitian, these possibilities are narrowly constrained because of its \emph{Krein Hermicity}.

Consider a Krein space of dimension $n_\text{K}$ with indefinite inner product $\inlineinner{\phi}{\psi}_\eta \coloneqq \phi^\dagger \eta \psi$, where the Hermitian metric $\eta$ is indefinite and nondegenerate.
The \emph{Krein adjoint} of a matrix $A$ is given by 
\begin{equation}
	A^\krein \coloneqq \eta^{-1} A^\dagger \eta, 
\end{equation}
since $\inlineinner{A \phi}{\psi}_\eta = \inlineinner{\phi}{A^\krein \psi}_\eta$ for any vectors $\phi$ and $\psi \in \complex^{n_\text{K}}$. 
The concepts of Krein-Hermitian matrices ($K^\krein = K$) and Krein-unitary matrices ($T^\krein T = T\,T^\krein = \idmat$) follow.
Furthermore, a \emph{Krein-antiunitary} operator $Z$ can be written $Z = K T$ in a suitable basis, where $K$ is the operation of complex conjugation and $T$ is a Krein-unitary operator~\cite{Bracci1975Wigner}.
Note that a change of basis (to ``normal coordinates'') can make $\eta$ diagonal with entries $\pm 1$; specifically, diagonalization followed by a scaling transformation
\footnote{
	To see this, write $\eta$ in its eigenbasis, $\eta = U \eta_\rd U^\dagger$, and note that $\eta = (U \sqrt{\abs{\eta_\rd}}) \sgn{\eta_\rd} (\sqrt{\abs{\eta_\rd}} U^\dagger) = V^\dagger \sgn{\eta_\rd} V$, where $V = \sqrt{\abs{\eta_\rd}} U^\dagger$ and the $\text{sgn}$ and $\text{abs}$ functions act elementwise. 
	It is easy to see that $\inlineinner{\phi}{\psi}_\eta = \inlineinner{V \phi}{V \psi}_{\sgn{\eta_\rd}}$, $K$ is Krein Hermitian with respect to $\eta$ iff $VKV^{-1}$ is Krein Hermitian with respect to $\sgn{\eta_\rd}$, etc.
}. From here on, we assume the metric $\eta$ has this form, with $n_+$ entries of $+1$ and $n_- = n_\text{K} - n_+$ entries of $-1$, which considerably simplifies the treatment.

The Krein space that naturally arises in bosonic BdG theory has metric $\eta = \tauthree \otimes \idmat_{n}$, with $n_- = n_+ = n$: 
the single-particle Hamiltonian $K = \tauthree H$ and paraunitary transformations $T$ are, respectively, Krein Hermitian and Krein unitary with respect to $\tauthree$~\footnote{
	When $\eta$ is diagonal with entries $\pm 1$, the columns of a Krein-unitary matrix $T$ are orthonormal with respect to the Krein inner product.
}.
The PH constraint of bosonic systems is not, however, a general feature of Krein-Hermitian Hamiltonians; indeed, the momentum-space single-particle Hamiltonian \emph{at a given $\bk$} does not in general obey such a constraint, unless an inversion-like symmetry relating $\bk$ and $-\bk$ is present.

\paragraph*{Dynamical stability and perturbations}
The eigenvalues of a Krein-Hermitian matrix $K$ are either real or come in complex-conjugate pairs~\cite{BlaizotRipka, Flynn2020Restoring, Flynn2020Deconstructing, Xu2020Squaring}.
An eigenvector $\psi$ corresponding to a non-real eigenvalue has null signature ($\inlineinner{\psi}{\psi}_\eta = 0$), while one corresponding to a real eigenvalue can have positive ($\inlineinner{\psi}{\psi}_\eta > 0$), negative ($\inlineinner{\psi}{\psi}_\eta < 0$), or null signature.
If all the vectors in the eigenspace of an eigenvalue $\lambda$ have the same nonzero signature, the eigenspace is said to have \emph{definite signature}; otherwise, it has \emph{indefinite signature}.
According to Krein stability theory~\cite{Flynn2020Deconstructing, Xu2020Squaring}, if each eigenspace of $K$ has definite signature, a perturbed Hamiltonian $K + \epsilon K'$ for sufficiently small $\epsilon$ is diagonalizable with real eigenvalues; that is, it is dynamically stable~\footnote{
	For Krein-Hermitian matrices, dynamical stability is equivalent to the concept of ``Krein spectrality''~\cite[\sect{III\,A\,1}]{Lein2019Krein}.
}.
Furthermore, its eigenvectors can be chosen to be orthonormal with respect to the Krein inner product.
% IK the following is unclear and seems irrlevant
Conversely, if $K$ has an indefinite eigenspace, either with or without nontrivial Jordan blocks, then $K$ either has non-real eigenvalues or there are arbitrarily small perturbations that make them non-real, meaning $K$ is either dynamically unstable or at a phase boundary of dynamical stability.

If $K = K^\krein$ is positive definite with respect to the Krein inner product, i.e., $\inlineinner{\phi}{K\phi}_\eta > 0$ for any nonzero $\phi$, all its eigenspaces automatically have definite signature, so it is within a dynamically stable phase~\cite{Lein2019Krein, Flynn2020Deconstructing, Xu2020Squaring}. 
Furthermore, it has $n_+$ eigenstates with positive signature and positive energy, and $n_- = n_\text{K} - n_+$ eigenstates with negative signature and negative energy.
Taking $K = \tauthree H$, $K$ is clearly positive definite with respect to $\tauthree$ iff $H > 0$. 
Once the eigenstates $\psi$ of $K$ are chosen to be orthonormal with respect to the Krein inner product, the claim made in \sect{\ref{sec:bosonic-bdg-intro-diagonalization}}---that $H>0$ implies there is a paraunitary $T$ such that $T^\dagger H T$ is diagonal with real, positive entries---follows. (Note, however, that dynamically stable phases can also exist if $K$ is not positive definite with respect to the Krein inner product, as exemplified in \reference{Flynn2020Deconstructing}.)

\subsection{Schrieffer-Wolff transformation for Krein-Hermitian Hamiltonians} \label{subsec:SW-Krein}

\begin{figure*}
    \centering
    \includegraphics[scale=1]{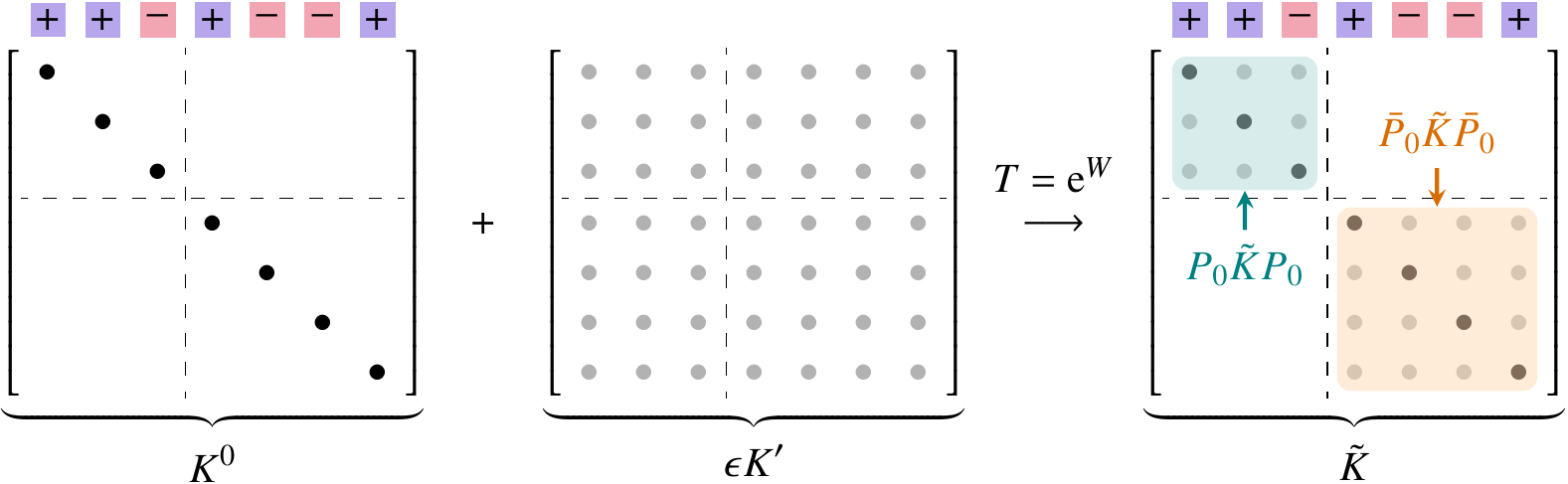}
    \caption{
    	Schematic depiction of the SW transformation for Krein-Hermitian matrices, shown in the eigenbasis of the unperturbed Hamiltonian $K^0$.
    	The Krein-unitary transformation $T$ block-diagonalizes the perturbed Hamiltonian with respect to the subspace $\cP_0$, and the effective Hamiltonian in the subspace $\cP_0$ is $P_0 \tilde{K} P_0$.
    	The signs above the columns of $K^0$ show the Krein signatures of its eigenvectors; in the case shown here, $m_+ = 2$ and $m_-=1$.
    	This implies that the eigenstates of $P_0 \tilde{K} P_0$ will have the same signatures, as illustrated schematically by the signs above the columns of $\tilde{K}$.
    }
    \label{fig:schematicSW}
\end{figure*}

% Our novel findings
We now demonstrate that within a dynamically stable phase, the perturbative SW transformation can be generally formulated in terms of Krein-unitary transformations. Applied to single-particle boson Hamiltonians, the transformation yields a reduced single-particle Hamiltonian with proper bosonic eigenmodes.

% Overview - what is the SW transformation
The SW transformation is used to study the effect of a perturbation on a subset of contiguous energy states of a (traditionally) Hermitian Hamiltonian~\cite{Bravyi2011Schrieffer, Winkler2003Quasi}. 
This \emph{subspace of interest} is often, but not necessarily, made up of the low-energy states of the unperturbed Hamiltonian. 
The transformation is defined as a unitary rotation that block-diagonalizes the perturbed Hamiltonian; the block corresponding to the subspace of interest then yields an effective Hamiltonian for those states.
The transformation as well as the effective Hamiltonian can be expressed perturbatively, order by order in terms of the perturbation. 
One of the considerable strengths of this method is its ability to deal on an equal footing with perturbations of degenerate as well as nondegenerate states, avoiding the complications that arise in Rayleigh-Schr\"odinger perturbation theory, for example~\cite{Winkler2003Quasi, sakurai}.
%\cite{Lowdin1951Perturbation, LuttingerKohn1955, Winkler2003Quasi}.

The canonical transformations for Krein-Hermitian Hamiltonians are Krein-unitary transformations: they preserve Krein Hermicity, and leave eigenspace signatures unchanged.
We adapt the SW procedure to Krein-Hermitian Hamiltonians within a dynamically stable regime, showing that it naturally yields a Krein-unitary transformation that block-diagonalizes the Krein-Hermitian Hamiltonian order by order in the perturbation, producing an effective Hamiltonian that is also Krein Hermitian.
Krein stability theory ensures that for sufficiently small perturbations, the effective Hamiltonian at any order is also dynamically stable.
Applied to single-particle boson Hamiltonians, the transformation will yield canonical boson modes. 

Consider a perturbed $\eta$-Krein-Hermitian $n_\text{K} \times n_\text{K}$ Hamiltonian
\begin{equation}
	K = K^0 + \epsilon K',
\end{equation}
where $K^0$ is within a dynamically stable regime and has a known Krein-orthonormal eigendecomposition, $K^0 t^{\vphantom{\dagger}}_i = E^0_i t^{\vphantom{\dagger}}_i$ with $t_i^\dagger \tauthree t^{\vphantom{\dagger}}_j = \pm \delta_{ij}$, and $\epsilon \in \reals$ controls the smallness of the perturbation.
Let $S_0$ denote a subset of energy-contiguous eigenstates of $K^0$, comprised of $m_+$ positive-signature states and $m_-$ negative-signature states, and gapped from the rest of the spectrum. 
Let $\cP_0$ denote the subspace spanned by the eigenstates in $S_0$.
We wish to block-diagonalize the perturbed Hamiltonian $K$ in the subspace $\cP_0$ using a Krein-unitary transformation (see the schematic depiction in \fig{\ref{fig:schematicSW}}).
Associated with the subspace $\cP_0$ is a unique Krein-orthogonal projector $P_0 = P_0^2 = P_0^\krein$~\footnote{
	In general, a non-orthonogal projection can be written $P = \sum_{i \in \set{S}} e_i^{\vphantom{\dagger}} \, \omega_i^\dagger$, where $\{e_i\}$ is a basis of the vector space and $\set{S}$ is a subset of the basis vectors. Furthermore, $\{\omega_i\}$ is the basis of the dual space, chosen to satisfy the biorthogonality condition $\omega_i^\dagger e_j^{\vphantom{\dagger}} = \delta_{ij}$. For a Krein-orthogonal basis $\{t_i\}$, the dual basis vector $\omega_i$ depends only on the basis vector $t_i$: $\omega_i = \eta \, t_i^{\vphantom{\dagger}} / (t_i^\dagger \eta t_i^{\vphantom{\dagger}})$.
}, where
\begin{equation}
	P_0 = \sum_{i \in \set{S}_0} \frac{t^{\vphantom{\dagger}}_i t_i^\dagger \eta}{t_i^\dagger \eta t^{\vphantom{\dagger}}_i}.
\end{equation}
Schematically, in the eigenbasis of $K^0$, any matrix $A$ can be written
\begin{equation}
\renewcommand*{\arraystretch}{1.3}
	\begin{split}
		A &= P_0 A P_0 + P_0 A \bar{P}_0 + \bar{P}_0 A P_0 + \bar{P}_0 A \bar{P}_0
		\\
		&=\left[
		\begin{array}{c|c}
			P_0 A P_0 & P_0 A \bar{P}_0 \\
			\hline
			\bar{P}_0 A P_0 & \bar{P}_0 A \bar{P}_0
		\end{array}
		\right],
	\end{split}
\end{equation}
where $\bar{P}_0 \coloneqq \idmat - P_0$.
We use the notation $\bd{A} \coloneqq P_0 A P_0 + \barP_0 A \barP_0$ and $\bod{A} \coloneqq P_0 A \barP_0 + \barP_0 A P_0$ for the parts of $A$ that are, respectively, block diagonal and block off-diagonal in the subspace $\cP_0$~\footnote{
	Please bear in mind the difference between the digit ``$0$'' (zero) and the letter ``o''.
}.

We seek a paraunitary transformation $T$, generated by a skew-Krein-Hermitian matrix $W = -W^\krein$,
\begin{equation}\label{eq:KU-transf}
	\tilde{K}
	=
	T^\krein K T
	=
	\re^{-W} K \re^{W},
\end{equation}
to block-diagonalize the Hamiltonian in the subspace $\cP_0$, meaning $\tilde{K}_\ro=0$.
This can be achieved by choosing the generator $W$ to be purely block off-diagonal, $W_\rb = 0$~\footnote{
	Clearly, this condition does not uniquely specify $W$ or $T$. In \app{\ref{app:SW-expansion}}, we perturbatively derive the ``canonical'' choice~\cite{Kessler2012Generalized} of $W$, with which $W$ is block off-diagonal.
}.
The parameter $\epsilon$ is assumed small enough to keep $K$ dynamically stable, and to preserve an energy gap between the states in $\cP_0$ and those not in $\cP_0$; hence, there is an unambiguous correspondence between the unperturbed states in $\cP_0$ and the smoothly connected eigenstates of the effective Hamiltonian.

The $\cP_0$ block of the transformed Hamiltonian, $P_0 \tilde{K} P_0$, is Krein Hermitian with metric $P_0 \eta P_0$, meaning it has $m_+$ positive-signature eigenstates and $m_-$ negative-signature eigenstates (see \fig{\ref{fig:schematicSW}});
these states are smoothly connected to the unperturbed states in $\set{S}_0$.
We note that if the subset of interest is comprised only of positive-signature states, meaning $m_-=0$, the effective Hamiltonian for the $\cP_0$ subspace is Hermitian, since $P_0 \eta P_0 = \idmat_{m_+}$.

\paragraph{Perturbative expansion} \label{subsubsec:pert-expansion}
We adapt the procedure of \reference{Winkler2003Quasi} to Krein-Hermitian and Krein-unitary matrices.
Employing the expansion 
\begin{equation} \label{eq:transf-expansion}
	\begin{split}
		\tilde{K} = T^\krein K T &= \re^{-W} K \re^{W} 
		\\
		&= K + [K,W] + \frac{1}{2!} [[K,W],W] + \dots,
	\end{split}
\end{equation}
using the assumption $W_\rb = 0$, and demanding that $\tilde{K}_\ro=0$, we find the following equations defining $W = \epsilon W^{(1)} + \epsilon^2 W^{(2)} + \dots$ order by order in the perturbation $K'$:
\begin{subequations} \label{eq:pert-constraints-W}
	\begin{align}
		[K^0, W^{(1)}] &= -K'_\ro,
		\\
		[K^0, W^{(2)}] &= -[K'_\rb, W^{(1)}],
		\\
		[K^0, W^{(3)}] &= -[K'_\rb, W^{(2)}] - \frac{1}{3} [[K'_\ro, W^{(1)}], W^{(1)}],
		\\
		&\ \, \vdots \nonumber
	\end{align}
\end{subequations}
These equations are compatible with ${W^{(j)}}^\krein = - W^{(j)}$, i.e., with the assumption that $T$ is Krein unitary at all orders.
Indeed, the first few orders of $W = \sum_{l=1}^{+\infty} \epsilon^l W^{(l)}$ and of the transformed Hamiltonian $\tilde{K} \eqqcolon \sum_{l=0}^{+\infty} \epsilon^l \tilde{K}^{(l)}$ are explicitly solved for in \app{\ref{app:SW-expansion}}, showing that the generator $W$ is indeed skew-Krein-Hermitian, and the effective Hamiltonian $P_0 \tilde{K} P_0$ is Krein Hermitian with respect to $P_0 \eta P_0$.

We note that if the Hamiltonian $K$ is purely real, the transformation yields a purely real effective Hamiltonian $\tilde{K}$ 
to all orders in the perturbation. This is verified explicitly for the first few orders in \app{\ref{app:SW-expansion}}, and holds to all orders.

Our findings agree with prior results on the first-order contribution $\tilde{K}^{(1)}$ found by applying degenerate perturbation theory to two \emph{degenerate} states of a para-Hermitian Hamiltonian~\cite{Sindou2013Magnonic, Shivam2017Neutron, Kondo2019Magnon}. We stress once again, however, that the SW approach does not have certain limitations of a Rayleigh-Schr\"odinger perturbation theory and treats degenerate, quasidegenerate, and nondegenerate states on the same footing~\cite{Winkler2003Quasi, sakurai}.

\paragraph{Range of validity}

For Hermitian matrices, the perturbative SW transformation is well controlled as long as the subspace of interest is gapped from the rest of the spectrum~\cite{Winkler2003Quasi, Bravyi2011Schrieffer}. For Krein-Hermitian matrices, we additionally require that the perturbed Hamiltonian remain within a dynamically stable region in order for eigenvalues to remain real, and for eigenvectors to be orthonormal with respect to the Krein inner product. Furthermore, at points of non-diagonalizability, eigenvectors have a non-analytic dependence on perturbations~\cite{TrefethenEmbreePseudospectra, stewart1990matrix}, and since dynamical-stability phase boundaries are mostly composed of non-diagonalizable points (called \emph{exceptional points})~\cite{Flynn2020Deconstructing}, they can constitute an obstruction to the convergence of a perturbation series.

In previous work, ranges of validity for the SW transformation were determined in terms of bounds on the norm $\opnorm{H'}$ of a perturbation $H'$ ensuring that the gaps between the subspace of interest and the rest of the spectrum remain sufficiently large~\cite{Bravyi2011Schrieffer}.
This simple picture was possible because the eigenvalues of a Hermitian matrix $H$ (and of normal matrices more generally) shift at most by $\opnorm{H'}$ under the effect of a perturbation $H'$, regardless of the details of $H$~\cite{TrefethenEmbreePseudospectra, stewart1990matrix, Bravyi2011Schrieffer}.
Here, $\opnorm{\cdot}$ is the operator norm, equal to the largest singular value, but equal in general to the largest eigenvalue (in absolute value) \emph{only} for a normal matrix.

In contrast, the eigenvalues of a non-normal matrix $A$ can shift by more than $\opnorm{A'}$ under the effect of a perturbation $A'$ (contra \reference{Kessler2012Generalized}), and the set of points at which the eigenvalues can end up depends on the details of $A$~\cite{TrefethenEmbreePseudospectra, stewart1990matrix}.
This makes the task of formulating sufficient conditions for the non-closing of relevant gaps more complex.
In mathematics, the notion of \emph{pseudospectra} explores this striking difference between normal and non-normal matrices in regards to perturbations and eigenvalues~\cite{TrefethenEmbreePseudospectra, stewart1990matrix}.

For a Krein-Hermitian matrix $K^0$ satisfying $\eta K^0 > 0$ \footnote{
	It is easy to convince oneself that $\eta K^0$ is positive definite in the ordinary sense if and only if $K^0$ is positive definite with respect to the Krein inner product, with the consequences laid out at the end of \sect{\ref{subsec:KreinPrimer}}.
}, there is a simple condition under which a perturbed Hamiltonian $K^0 + K'$ remains dynamically stable and Krein-unitarily diagonalizable~\cite[\sect{IV\,C}]{Xu2020Squaring}: assuming $\opnorm{K'} < \opnorm{(K^0)^{-1}}^{-1}$, no eigenvalues of $K^0 + K'$ can reach zero, implying the perturbed Hamiltonian remains within the dynamically stable phase. 
Note that $\opnorm{(K^0)^{-1}}^{-1}$---the smallest singular value of $K^0$~\cite[\sect{II}]{TrefethenEmbreePseudospectra}---constitutes a lower bound for the smallest eigenvalue (in absolute value) of $K^0$, but since $K^0$ is non-normal, it is not in general equal to its smallest eigenvalue (contra \reference{Xu2020Squaring}).

The formulation of more general bounds on allowable perturbations $K'$, and the derivation of a radius of convergence such as that which exists for perturbative SW transformations on Hermitian Hamiltonians~\cite{Bravyi2011Schrieffer}, are left for future work.

%More generally, if $z$ is not in the spectrum of $K^0$, it cannot be in the spectrum of $K^0 + K'$ as long as $\opnorm{K'} < \opnorm{(K^0 - z \idmat)^{-1}}^{-1}$.

\section{Band touchings and codimension in magnon systems} \label{sec:nodal-lines-magnons}
%\section{Nodal lines in magnetic-inversion-symmetric magnon systems}

In Hermitian systems, the standard SW transformation allows one to obtain, at least locally in momentum space, an effective Hamiltonian describing a contiguous subset of bands~\cite{Kormanyos2013Monolayer, Kormanyos2014SpinOrbit, Korm_nyos_2015, *Korm_nyos_2015_corr, Beiranvand2018TwoBand}. 
The transformation, which can in principle be carried out to any order, also justifies the existence of an effective reduced Hamiltonian that exactly describes contiguous bands if sufficiently large gaps separate them from the rest of the spectrum, again, locally in momentum space.

In linear spin-wave theory, magnons are generally described by a bosonic BdG Hamiltonian of the form of \eq{\eqref{eq:boson-bdg-ham}}, with a Krein-Hermitian single-particle Hamiltonian (see \app{\ref{sec:symmetries-implementation}} for an overview of linear spin-wave theory). 
The Krein-Hermitian SW transformation we formulate explains why such effective reduced descriptions can also be used for Krein-Hermitian systems, and clarifies the properties of the effective reduced Hamiltonian in such cases.
With these facts in hand, we revisit results on the codimensions of generic band touchings (i.e., those not at high-symmetry momenta) of noninteracting magnon systems with and without magnetic inversion symmetry---a term which, as explained in \sect{\ref{sec:intro}}, we use to refer to space-time inversion symmetry.
We comment on high-symmetry momenta at the end of the section.
% Footnote on ``PT symmetry'' was here.

Quadratic magnon Hamiltonians in three dimensions (3D) with magnetic inversion symmetry are known to exhibit either topologically protected Dirac points or topologically protected gapless lines, depending on the classical spin order and on other symmetries of the system~\cite{Li2017Dirac}. 
This behavior makes them in part analogous to quadratic Hamiltonians of spinless electrons in 3D, which also generically exhibit gapless lines in the presence of space-time-inversion symmetry~\cite{Fang2015Topological, Ahn2018Band, Rui2018NodalLine, Li2018CuTeO3, Tiwari2020NodalLine}.
The source of this commonality is that both are integer-spin particles, for which space-time inversion squares to $+1$~\cite{Ludwig_2015, Zhao2017PT};
however, the magnon case can exhibit richer behavior because of the larger set of possible symmetries. 
The symmetries of magnetically ordered systems and their implementation on magnon Hamiltonians are discussed in \app{\ref{sec:symmetries-implementation}}.

Since our findings justify an effective two-band model for bosonic BdG Hamiltonians, we can use simple codimension counting approaches familiar from fermionic systems~\cite{Tiwari2020NodalLine, Shuichi_Murakami_2007, RMP2016Classification} to understand the emergence of generic gapless features in magnon spectra, giving a useful complementary understanding of the topologically protection.

Given a Bloch coefficient matrix function $H_\bk$, we first maximally block-diagonalize it using all the ($\bk$-conserving) unitary symmetries of the Hamiltonian; for example, the spin-rotation symmetry in the example from \reference{Li2017Dirac} mentioned above, further elaborated on below. This gives rise to ``symmetryless'' blocks $\{H_\bk^a\}$ whose dependence on $\bk$ should be generic~\cite{Ludwig_2015}. 
Then, we consider the action of magnetic inversion symmetries---which are antiunitary---on the symmetryless blocks to understand the generic band touchings that occur within them.

\subsection{Two-band effective Hamiltonian for magnon systems}

Consider a symmetryless block as described above, at least of size $2 \times 2$.
At a wave vector $\bk_0$ at which $H_{\bk_0} > 0$, the magnon spectrum at $\bk_0$ and neighboring wave vectors is given by the set of positive eigenvalues of the $\tauthree$-Krein-Hermitian matrix $\tauthree H_{\bk}$. 
Assume two positive eigenvalues of $\tauthree H_{\bk}$, spanning a subspace $\cP_0$, are separated from the rest of the spectrum by nonzero gaps.
In the neighborhood of $\bk_0$, we can write
\begin{equation}
	\tau^3 H_{\bk} = \tau^3 H_{\bk_0} + \tau^3\left( H_{\bk} - H_{\bk_0}\right),
\end{equation}
with the second term on the right-hand side acting as a ``small'' perturbation of the first term.

For every $\bk$ in a sufficiently small neighborhood of $\bk_0$, the SW transformation for Krein-Hermitian Hamiltonians presented in the previous section gives an effective Hamiltonian $P_0 \bigl( \widetilde{\tauthree H}_{\bk} \bigr) P_0$ (as defined in and after \eq{\eqref{eq:KU-transf}}) for the subspace $\cP_0$, which faithfully reproduces the band structure.
We reiterate that the SW method is especially well suited for the study of band touchings and their neighborhoods.
%\highlight{For matrices of the form $\tau^3 H = \tau^3 H^0 + \tau^3 H'$, with $H$ and $H^0$ Hermitian and positive definite, it specifies a paraunitary transformation that decouples the subspace $\cP_0$.}
Since $\cP_0$ contains only positive-signature bands, the effective Hamiltonian is Hermitian. 
In the eigenbasis of $\tauthree H_{\bk_0}$, it is given by
\begin{equation}\label{eq:two-band-Ham}
	P_0 \bigl( \widetilde{\tauthree H}_{\bk} \bigr) P_0 = d^0_{\bk} \sigma^0 + \vec{d}^{\vphantom{0}}_{\bk} \cdot \vec{\sigma},
\end{equation}
where $(\sigma^0, \vec{\sigma})$ are the Pauli matrices in the two-dimensional subspace $\cP_0$ and where $d^\mu_{\bk}, \mu\in\{0,1,2,3\}$ are real.

Importantly, the expressions for the effective Krein-Hermitian Hamiltonian (\app{\ref{app:SW-expansion}}) reveal that if $H_\bk$ is real, so is the effective Hamiltonian $P_0 \bigl( \widetilde{\tauthree H}_{\bk} \bigr) P_0$, forcing the term proportional to $\sigma^2$ to vanish, and meaning that the two bands touch iff the functions $d_\bk^1$ and $d_\bk^3$ vanish simultaneously. 
In that case, except at high-symmetry momenta, band touchings that are stable against symmetry-preserving perturbations will generically have a codimension of two, with twofold degeneracy.
Otherwise, they would have a codimension of three, with twofold degeneracy.

\subsection{Constraint of magnetic inversion symmetry and examples}

\begin{figure}
	\centering
	\includegraphics[width=\columnwidth]{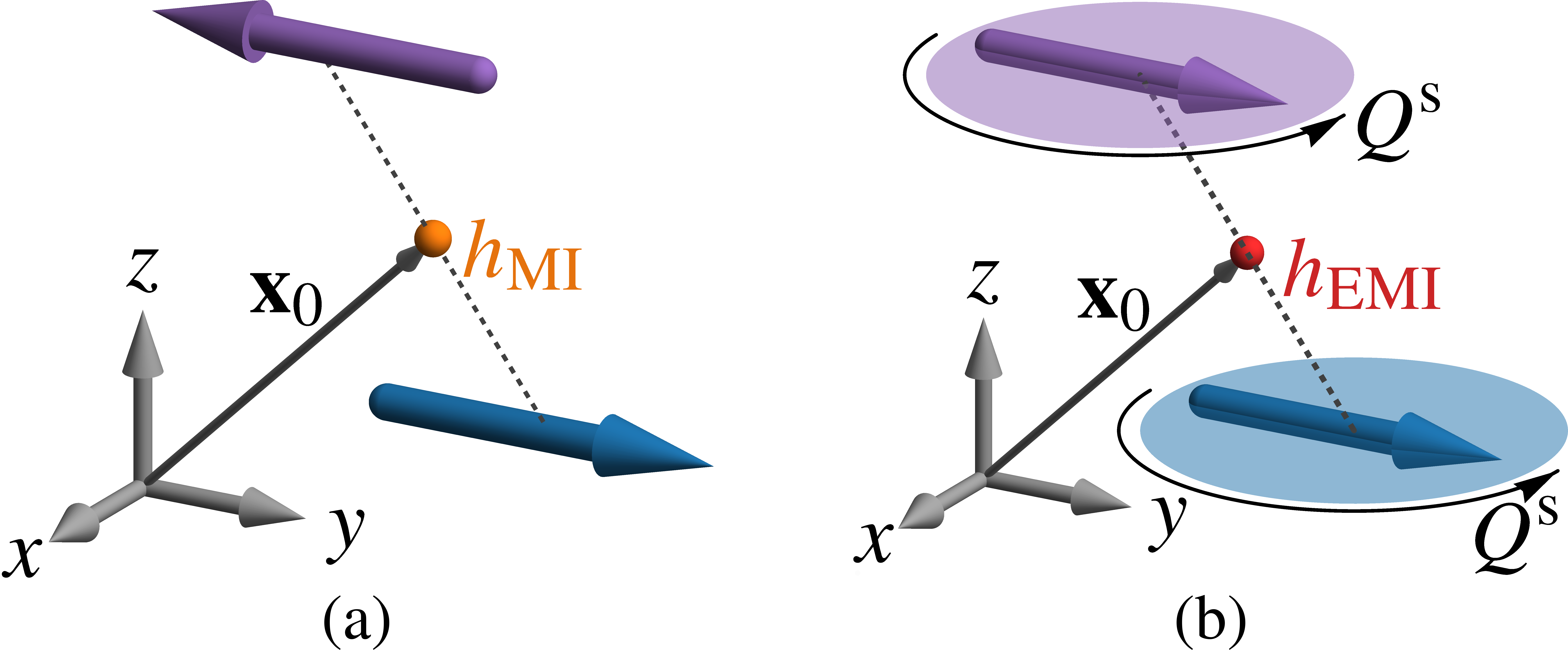}
	\caption{
		(a) Schematic depiction of spins left invariant by a magnetic inversion symmetry $\MI = \{ P | 2 \bx_0 \}T$ whose inversion center, shown in orange, is located at $\bx_0$.
		(b) Spins left invariant by an effective magnetic inversion symmetry $\EMI = \{ P; Q^\rs | 2\bx_0 \} T$ whose inversion center, shown in red, is located at $\bx_0$. 
		The spin rotation $Q^\rs$, which is only allowed if the spin Hamiltonian lacks certain SOCs, effectively ``undoes'' the effect of time reversal and returns the spins to their initial configuration, as detailed in \sect{\ref{subsubsec:effective_magnetic_inversion}}.
	}
	\label{fig:magnetic_inversion}
\end{figure}

If a spin Hamiltonian and its classical ground state are invariant under a symmetry transformation, then the Bloch coefficient matrix $H_\bk$ of the associated magnon Hamiltonian admits a representation of that symmetry (see \app{\ref{sec:symmetries-implementation}} for a general discussion). 
Magnetic inversion symmetry plays a prominent role in the present discussion, and \fig{\ref{fig:magnetic_inversion}(a)} depicts a pair of spins that are left invariant by magnetic inversion symmetry: spatial inversion interchanges the spins (while leaving the spin direction unaffected), and time reversal reverses the spin directions, returning the system to its initial configuration. 
Hence, provided the spin Hamiltonian has only terms even in spin operators, magnetic inversion symmetry will have an implementation on $H_\bk$ in such a scenario. 
In Seitz notation, a magnetic inversion symmetry with inversion center $\bx_0$ (see \fig{\ref{fig:magnetic_inversion}(a)}) takes the form $\MI = \{P | 2\bx_0 \} T$~\cite{BradleyCracknell}, where $P$ denotes inversion with respect to the origin and $T$ denotes time reversal. 
The translational part $2\bx_0$ depends on the choice of origin and will mostly be unimportant in the present discussion.

As shown in \app{\ref{sec:symmetries-implementation}}, the constraint due to magnetic inversion in a magnon system is generally $H_\bk = V^\dagger H_\bk^* V$, where $V$ is a paraunitary, unitary, particle-hole symmetric, and also symmetric matrix.
This implies that $V V^* = +\idmat$, i.e., that magnetic inversion squares to $+1$~\cite{Ludwig_2015}; indeed, this must be the case for integer-spin excitations~\cite{sakurai, Tiwari2020NodalLine}, like the magnon. 
Furthermore, because $V V^* = +\idmat$, there must exist a basis in which the constraint takes the simple form $H^{\vphantom{*}}_\bk = H_\bk^*$; see \app{\ref{sec:symmetries-implementation}}.

This basis, however, may not coincide with that in which $H_\bk$ is maximally block-diagonalized into symmetryless blocks. 
If, under magnetic inversion, a symmetryless block $H_\bk^a$ is mapped to itself ($H_\bk^a = \tV^\dagger {H_\bk^a}^* \tV$ for some $\tV$), then magnetic inversion admits a representation within that block, and its generic band touchings are twofold degenerate and have codimension two instead of three. 
If two blocks $H_\bk^a$ and $H_\bk^{a'}$ are mapped to each other ($H_\bk^a = \tV^\dagger {H_\bk^{a'}}^* \tV$ for some $\tV$), then they have the same energy eigenvalues, so their generic band touchings are fourfold degenerate with codimension three.

We now discuss some examples of magnon band touchings from the literature and illustrate how the codimension approach can be useful for quickly understanding the generic features.

\subsubsection{Magnetic inversion and spin-rotation symmetries} \label{subsubsec:AFMexample}
The scenario considered in \reference{Li2017Dirac} is that of a 3D collinear antiferromagnetic spin system with magnetic inversion symmetry (see \fig{1} of \reference{Li2017Dirac}), with and without spin-rotation symmetry.
If the system has a spin-rotation symmetry about an axis, say $\hat{z}$, the excitations conserve the quantum number $S_z$. This is a $\bk$-independent symmetry that allows a block-diagonalization of $H_\bk$,
\begin{equation} \label{eq:spin-conserving-blocks}
\renewcommand*{\arraystretch}{1.5}
	H_\bk = \left[
	\begin{array}{c|c}
		H_\bk^+ & \\
		\hline
		 & H_\bk^-
	\end{array}
	\right],
\end{equation}
into sectors corresponding to excitations with $S_z = \pm 1$~\cite{Li2017Dirac} (see \app{\ref{sec:symmetries-implementation}} for details). 
In this case, magnetic inversion interchanges $+\hz$ spins and $-\hz$ spins, hence mapping the two blocks to each other and making the spectrum everywhere doubly degenerate.
Then, assuming the blocks are symmetryless, the generically occurring band touchings are fourfold-degenerate isolated points; as explained in \reference{Li2017Dirac}, they are made up of Weyl points with opposite monopole charges, making them Dirac points.

If the spin-rotation symmetry is broken by certain spin-orbit-coupling (SOC) terms in the Hamiltonian or by canting of the spins, but the magnetic inversion symmetry is preserved, then $H_\bk$ can no longer be block-diagonalized into definite-spin sectors. 
Assuming $H_\bk$ is a symmetryless block, its generic gapless manifolds are doubly-degenerate lines.

Of course, an analysis based simply on the codimension of the generic gapless manifolds does not tell the whole story~\cite{Tiwari2020NodalLine}. 
For instance, it does not capture the $\integers_2$ monopole charge carried by the the Dirac nodes and gapless lines in this example~\cite{Li2017Dirac}, which explains why the two can evolve into each other, but cannot---contrary to other gapless lines---gap out without encountering an opposite charge~\cite{Fang2015Topological, Ahn2018Band, Tiwari2020NodalLine}.
Nonetheless, the simplicity of this analysis makes it complementary to other approaches.

In \reference{Li2017Dirac}, the authors propose Cu\textsubscript{3}TeO\textsubscript{6} as a material nearly at the cusp between these two scenarios. Indeed, Heisenberg interactions (which have spin-rotation symmetry) appear to be dominant in this compound; thus, in the absence of subdominant interactions, we expect Dirac points in the band structure. However, weak SOC likely breaks the spin-rotation symmetry and deforms the Dirac points to small gapless lines. Indeed, Dirac points have recently been observed in experiment, though they may be gapless lines that are too small to be resolved with current resolutions~\cite{Yao2018Antiferromagnet, Bao2018Antiferromagnet}.

\subsubsection{Effective magnetic inversion and Heisenberg ferromagnets} \label{subsubsec:effective_magnetic_inversion}
We can contrast the scenario of \sect{\ref{subsubsec:AFMexample}} with one in which the magnetic inversion symmetry is replaced with an \emph{effective} magnetic inversion symmetry that leaves the spin direction unchanged, an example of which is considered in \reference{Mook2017Ferromagnets}. 
Such a symmetry is possible in the absence of certain SOCs, and necessarily acts differently in position space and spin space. 
Using a modified Seitz notation (see \app{\ref{sec:symmetries-implementation}}), we denote the effective magnetic inversion as $\EMI = \{ P; Q^\rs | 2\bx_0 \} T$; here, $Q^\rs$ is a \emph{spin} rotation by $\pi$ about some suitable axis that serves to return the spin to its original direction and undo the effect of time reversal on the classical order, as depicted schematically in \fig{\ref{fig:magnetic_inversion}(b)}. 

If spin-rotation symmetry about an axis is present, the Bloch Hamiltonian can again be block-diagonalized into spin-conserving sectors, as in \eq{\eqref{eq:spin-conserving-blocks}}. 
However, rather than interchanging the sectors, $h$ now maps each sector to itself.
Therefore, the intra-sector band touchings will generically have codimension two and twofold degeneracy (assuming each sector is symmetryless). 
Of course, inter-sector band touchings can also occur; since they involve bands with different spin quantum numbers, their touchings will generically be of codimension one with twofold degeneracy. 

On the other hand, if the spin-rotation symmetry is broken, leaving the Bloch Hamiltonian symmetryless, any codimension-one band touchings either gap out or are reduced to codimension two.

The situation considered in \reference{Mook2017Ferromagnets}, a spin model with Heisenberg interactions and ferromagnetic order, falls within this category.
The effective magnetic inversion $h = \{P; R(\pi\hn) | \btau\} T$, where $R(\pi\hn)$ is a spin rotation of $\pi$ about an axis $\hn$ perpendicular to the spin direction and $\btau$ is some origin-dependent translation, protects the nodal lines. 
As the authors mention, certain SOCs, like the Dzyaloshinskii-Moriya (DM) interaction, would gap out the nodal lines completely or downgrade them to Weyl points. 
In our present language, it is those interactions that are incompatible with the symmetry $h$ that will generically gap out nodal lines.

%\added{The symmetry $h$ can also be thought of as the composition of ordinary inversion, $\{P | \btau\}$, with an effective time reversal, $\{\idmat; R(\pi\hn) | 0 \} T$.}

\subsubsection{Nodal lines in a stacked honeycomb quantum magnet} \label{subsubsec:cobalt_titanate}

\begin{figure}
	\centering
	\includegraphics[width=\columnwidth]{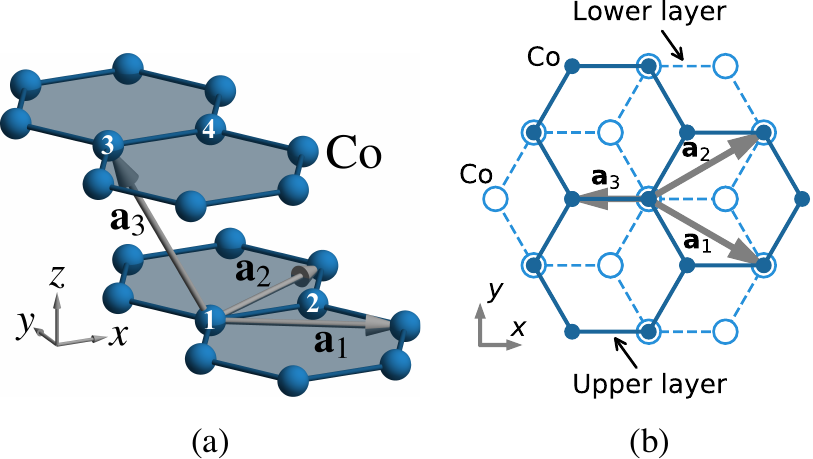}
	\caption{
		Crystal structure of the magnetic Co\textsuperscript{2+} ions in CoTiO\textsubscript{3}. %model used in \sect{\ref{subsubsec:XXZ}}. 
		%The structure is the same as that of CoTiO\textsubscript{3}, but the parameter values are not faithful to the material. 
		Two layers are shown, both as a 3D illustration in panel (a) and as a projection along the stacking direction $\hat{z}$ in panel (b). 
		The crystal's primitive lattice vectors $\ba_1$, $\ba_2$, and $\ba_3$ are shown; the magnetic order (depicted in \fig{\ref{fig:magnetic_order}}) doubles the third one to $2 \ba_3$.
		Sublattice indices 1, 2, 3, and 4 of the magnetic unit cell are shown in panel (a).
	}
	\label{fig:cryst_struct}
\end{figure}

We next discuss CoTiO\textsubscript{3}, an example of a stacked honeycomb quantum magnet that has been the object of recent interest.
CoTiO\textsubscript{3} has an ilmenite structure and space group $\mathit{R\bar{3}}$: its magnetic Co\textsuperscript{2+} ions form stacked honeycomb layers with ABC-type stacking~\cite{Yuan2020, Elliot2020Visualization, Newnham1964}, depicted in \fig{\ref{fig:cryst_struct}}.
It becomes magnetically ordered below $T_\text{N} \approx 38\,\mathrm{K}$, with ferromagnetic order within each layer and antiferromagnetic stacking of the layers. 
The ordering direction is in the plane of the layers~\cite{Yuan2020, Elliot2020Visualization, Newnham1964}, as exemplified in \fig{\ref{fig:magnetic_order}(a)}.
The onset of magnetic order considerably lowers the symmetry: the corresponding magnetic space group is $\mathit{P_S\bar{1}}$, which contains only inversions, magnetic inversions, and magnetic translations~\cite{Elliot2020Visualization, Yuan2020, Newnham1964, BradleyCracknell}.

The dominant exchange interactions are not rotation invariant about the ordering direction, since they have an easy-plane anisotropy~\cite{Yuan2020, Elliot2020Visualization}. 
Hence, there is no approximate spin-rotation symmetry, and we expect gapless lines to be generically stable because of the magnetic inversion symmetry $\MI$ present in $\mathit{P_S\bar{1}}$.
Indeed, seemingly gapless lines have been reported in recent experiments~\cite{Yuan2020, Elliot2020Visualization}.

\subsubsection{XXZ model of a stacked honeycomb quantum magnet} \label{subsubsec:XXZ}

Let us now consider a simple spin model for CoTiO\textsubscript{3} whose linear spin-wave theory has been found to capture all but the fine details of the magnon spectrum, including the observed gapless lines~\cite{Yuan2020, Elliot2020Visualization}.
It consists of effective spin-one-half moments on the Co\textsuperscript{2+} ions, coupled by easy-plane XXZ interactions; see \app{\ref{app:XXZ}} for details.

The effect of including additional symmetry-allowed SOC terms, which are not rotationally symmetric about the $\hat{z}$ axis, is to eliminate certain extraneous symmetries of the XXZ interactions. 
The strength of the various spin interactions in CoTiO\textsubscript{3} and other honeycomb magnets, including Kitaev, Gamma, and DM interactions, is still a subject of active research~\cite{Liu2020, Yuan2020, Elliot2020Visualization, Das2021Cobaltates}.
Therefore, rather than predicting the aspect of the gap closures in the presence of all such subdominant SOC terms, we merely focus on the general effects of lowering the symmetry. 
For this purpose, we use the DM interaction to illustrate the general effects of lowering the XXZ model symmetries, while recognizing that this is not the dominant SOC term relevant to CoTiO\textsubscript{3}.
The four magnetic Co\textsuperscript{2+} ions per magnetic unit cell (identified in \fig{\ref{fig:cryst_struct}(a)})~\cite{Elliot2020Visualization, Newnham1964} give rise to four magnon bands, which we label 1, 2, 3, and 4 in order of increasing energy.

\begin{figure}
	\centering
	\includegraphics[width=\columnwidth]{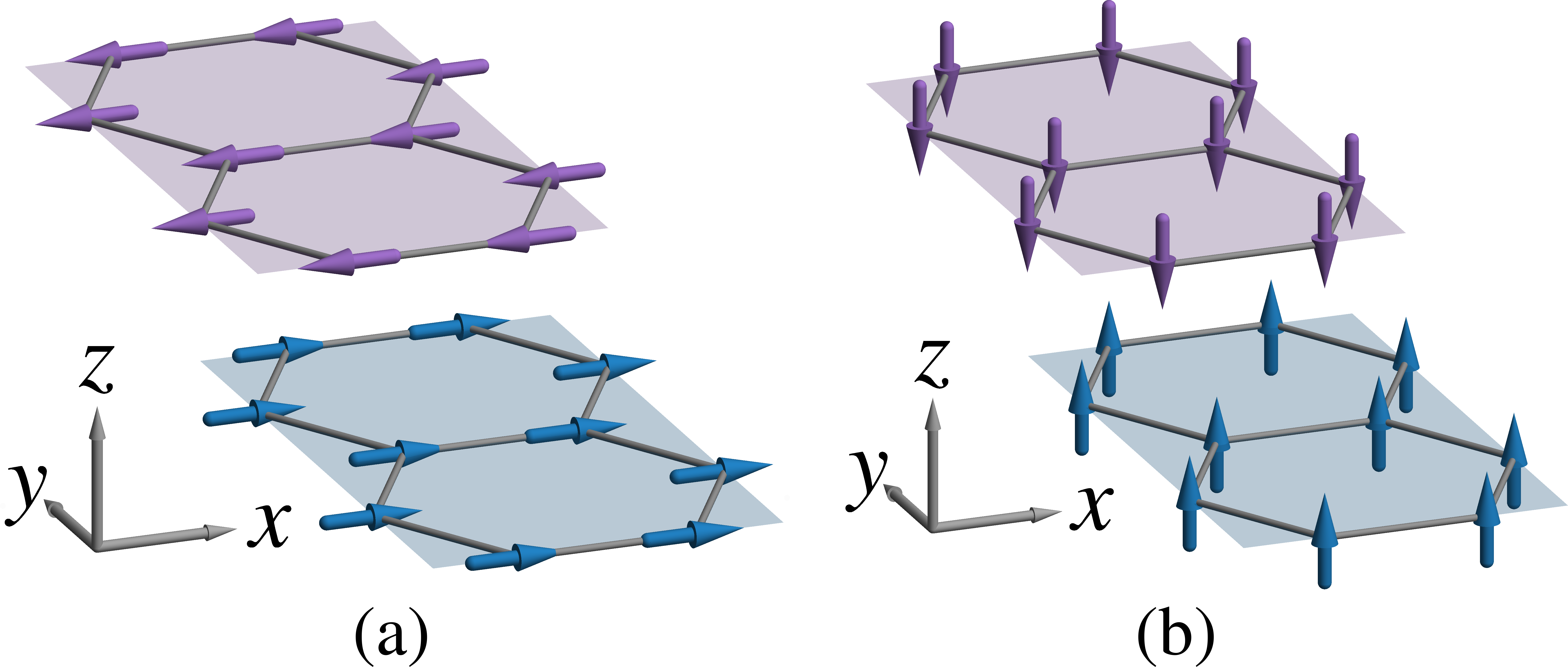}
	\caption{
		Magnetic orders in the stacked-honeycomb systems of Secs.~\ref{subsubsec:cobalt_titanate} and \ref{subsubsec:XXZ}. 
		(a) When easy-plane interactions dominate, the order is in plane; the specific case of $x$-direction order, used in our numerical calculations, is shown.
		(b) When easy-axis interactions dominate, the order is parallel to the stacking direction $\hat{z}$.
	}
	\label{fig:magnetic_order}
\end{figure}

\paragraph{Pure XXZ model} \label{par:XXZ_easyplane}
With no additional interactions present, the XXZ model has more symmetry than required by the magnetic space group $\mathit{P_S\bar{1}}$; in particular, the ordering direction can be freely rotated about the stacking axis $\hat{z}$ at no energy cost~\cite{Elliot2020Visualization}.
Consequently, for a fixed in-plane order (like that shown in \fig{\ref{fig:magnetic_order}(a)}), the model's spin space group (see \app{\ref{sec:symmetries-implementation}}) includes the unitary symmetry $g = \{\idmat; R(\pi\hat{z}) | \btau_g\}$, where the translation $\btau_g$ connects like sublattices of the honeycomb in adjacent layers. 
The coefficient matrix $H_\bk$ can hence be block-diagonalized into two sectors corresponding to the eigenvalues of $g$, and the band touchings from different sectors are generically surfaces (codimension one)~\footnote{
Equivalently, the symmetry $g$ allows a Fourier transform in a unit cell smaller than the magnetic unit cell, essentially ``unfolding'' the band structure~\cite{Elliot2020Visualization, Yuan2020} (see also \app{\ref{sec:symmetries-implementation}}).
}.
Indeed, with the parameters of \app{\ref{app:XXZ}}, we find that gapless surfaces occur between bands 1 and 2 as well as between bands 3 and 4.

Another extraneous symmetry of the XXZ model is the effective magnetic inversion symmetry $\EMI = \{P; R(\pi \hz) | \btau_h \} T$, distinct from the true magnetic inversion $\MI$ present in $\mathit{P_S\bar{1}}$.
Indeed, $\EMI$ interchanges sites with the same ordering directions, while $\MI$ interchanges sites with opposite ordering directions. 
Both magnetic inversions work \emph{within} the sectors that arise from the symmetry $g$, meaning either can stabilize magnon gapless lines, and both would have to be broken in order to destroy the gapless lines in the model, which occur between bands 2 and 3 (these are the gapless lines described in Refs.~\onlinecite{Yuan2020, Elliot2020Visualization}).
We emphasize, however, that $\MI$, which is part of $\mathit{P_S\bar{1}}$, is model independent, presuming the magnetic space group has been accurately identified~\cite{Elliot2020Visualization, Newnham1964}.

\paragraph{Adding SOC terms} \label{par:XXZ_DM}

\begin{figure*}
	\centering
	\includegraphics[width=\textwidth]{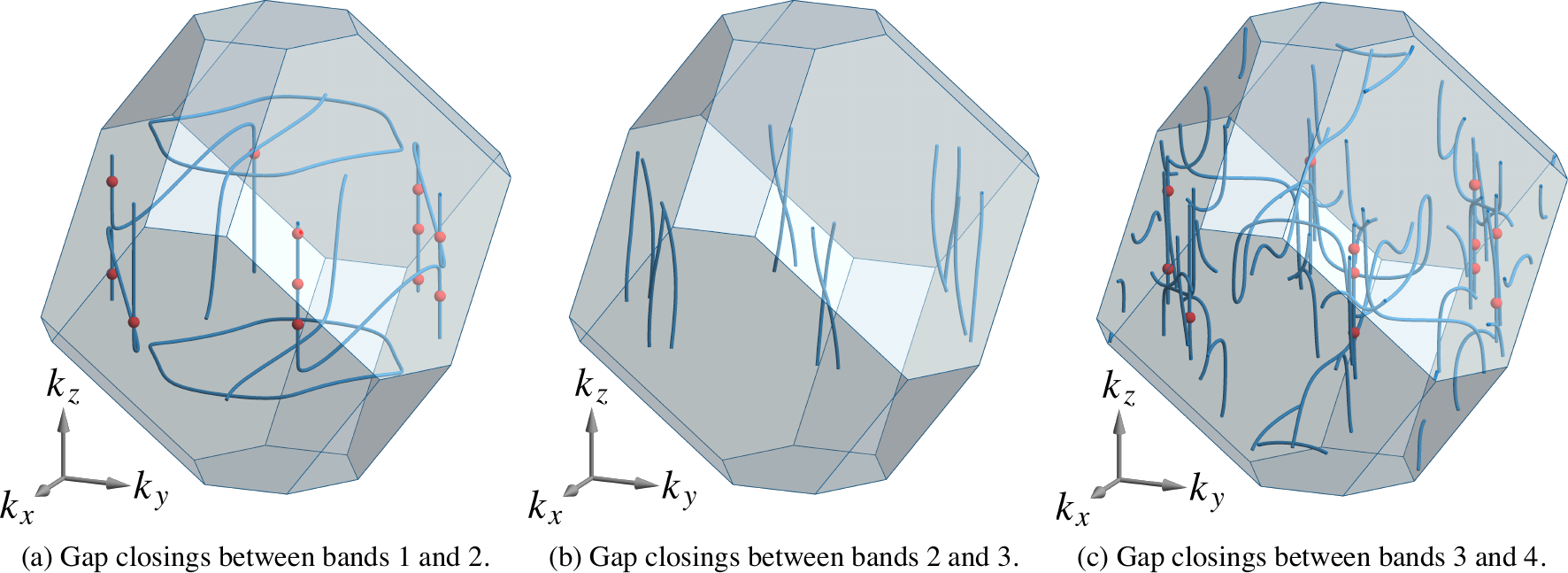}
	\caption{
		Magnonic gap closings in a stacked honeycomb magnet modeled with an XXZ interaction together with a weak DM interaction, with parameter values given in \app{\ref{app:XXZ}}.
		As argued in the text, the gap closings in the presence of magnetic inversion should generically be gapless lines, and indeed a plethora of gapless lines (shown in blue) are found.
		If magnetic inversion symmetry is broken (see main text and \app{\ref{app:XXZ}}), the gap closings are Weyl points and are shown in red.
		Note that some of the gapless lines in (a) and (c) and all the lines in (b) reach the first-Brillouin-zone boundary and re-emerge at the opposite face.
	}
	\label{fig:gapless_pts}
\end{figure*}

The addition of the DM interaction breaks the symmetries $g$ and $\EMI$ of the pure XXZ model (see \app{\ref{app:XXZ}} for further details).
The remaining magnetic inversion symmetry $\MI$ stabilizes gapless lines, which are shown in blue in \fig{\ref{fig:gapless_pts}} for the parameter set given in \app{\ref{app:XXZ}} and with the specific $x$-direction order of \fig{\ref{fig:magnetic_order}(a)}.
%\Fig{\ref{fig:gapless_pts}} shows the gapless lines between pairs of adjacent magnon bands, drawn in blue.
The gapless spirals already present in the pure XXZ model~\cite{Yuan2020, Elliot2020Visualization} are qualitatively unchanged, and appear between bands 2 and 3 (\fig{\ref{fig:gapless_pts}(b)}). 
This is expected, since here the DM interaction is a relatively small perturbation to the XXZ interaction, and the nodal lines are stable against small symmetry-preserving perturbations.
On the other hand, the gapless lines between bands 1 and 2 (\fig{\ref{fig:gapless_pts}(a)}, blue lines) and 3 and 4 (\fig{\ref{fig:gapless_pts}(c)}, blue lines) are new and are the remnants of the gapless surfaces of the pure XXZ model, which have been reduced to lines by the DM interaction.
We expect that including other symmetry-allowed subdominant interactions would significantly alter these gapless lines, though they would still appear as the remnants of the gapless surfaces as long as the subdominant interactions are small.

If the magnetic inversion symmetry $\MI$ were somehow broken, gapless lines would no longer be stable, and since no spin-rotation symmetry is present, the generically occurring band touching would become Weyl points.
To demonstrate this, we introduce a small Zeeman field on one of the four atoms in the magnetic unit cell (see \app{\ref{app:XXZ}}), hence explicitly breaking $\MI$. 
(Incidentally, this perturbation also breaks inversion symmetry and the magnetic translation symmetry.) 
The remaining band touchings, all of which are Weyl points, are shown in red in \fig{\ref{fig:gapless_pts}}: some of the gapless lines have been reduced to Weyl points, while others have fully disappeared.

\paragraph{Easy-axis XXZ model}\label{par:XXZ_easyaxis}
It is informative to contrast the easy-plane XXZ model discussed above with its easy-axis counterpart, in which the ordering direction is parallel to $\hat{z}$, with the order otherwise unchanged (see \fig{\ref{fig:magnetic_order}(b)}). 
%even though not related to CoTiO3
The magnetic space group associated with this order is $\mathit{R_I\bar{3}}$~\cite{Elliot2020Visualization}, which also contains a model-independent magnetic inversion symmetry $\MI$.

As in the easy-plane case, the easy-axis XXZ model gives rise to an additional, effective magnetic inversion symmetry $\EMI$. 
Crucially, whereas the easy-plane case has no spin-rotation symmetry, the present model has spin-rotation symmetry about the $\hat{z}$ axis, meaning $H_\bk$ can be block-diagonalized as in \eq{\eqref{eq:spin-conserving-blocks}}.
Since $\MI$ interchanges the blocks $H_\bk^+$ and $H_\bk^-$, it causes them to have the same eigenvalues, making the bands everywhere at least doubly degenerate. 
Since $\EMI$ maps each block to itself, its effect is to generically give rise to gapless lines rather than gapless points. 

The symmetries $\MI$ and $\EMI$ can be broken independently using different Zeeman fields on the four sublattices of the magnetic unit cell. 
The effects of breaking either or both of these symmetries are presented in \tab{\ref{tab:XXZ_easyaxis}}.
The explicit form of the Hamiltonian and the parameters used in the numerics are presented in \app{\ref{app:XXZ}}.

\begin{table*}[t]
\SetTblrInner{rowsep=4pt} % Gives more stretch to compensate for stretch=-1
\begin{tblr}{width=\textwidth, colspec = {l || X[c] | X[c]}, rowspec={Q[t] || Q[t] | Q[t]}, 
			 measure = vbox, % Allows enums to work (together with \UseTblrLibrary{varwidth}): https://github.com/lvjr/tabularray/issues/36
			 stretch=-1 % For top-aligning with enums: https://github.com/lvjr/tabularray/issues/99
			 }
    & $\MI$ unbroken & $\MI$ broken 
  \\
    $\EMI$ unbroken
    &
    \begin{itemize}[leftmargin=*, itemsep=3pt]
      \item All bands are at least doubly degenerate because of $\MI$.
      \item Band touchings within each sector are generically \emph{lines} because of $\EMI$.
      \item In numerics, gapless lines are observed between the two pairs of degenerate bands.
    \end{itemize}
    & 
    \begin{itemize}[leftmargin=*, itemsep=3pt]
      \item The two sectors $H_\bk^+$ and $H_\bk^-$ are independent. Global double degeneracy is gone, and bands from different sectors generically cross on surfaces (codimension one).
      \item Band touchings within each sector are generically \emph{lines} because of $\EMI$.
      \item In numerics, gapless surfaces are observed between bands 2 and 3; gapless lines are observed between bands 1 and 2 and between bands 3 and 4.
    \end{itemize}
  \\
    $\EMI$ broken
    &
    \begin{itemize}[leftmargin=*, itemsep=3pt]
      \item All bands are at least doubly degenerate because of $\MI$.
      \item Band touchings within each sector are generically \emph{points}.
      \item In numerics, \emph{no} gapless lines are observed between the two pairs of degenerate bands.
    \end{itemize}
    & 
    \begin{itemize}[leftmargin=*, itemsep=3pt]
      \item The two sectors $H_\bk^+$ and $H_\bk^-$ are independent. Global double degeneracy is gone, and bands from different sectors generically cross on surfaces (codimension one).
      \item Band touchings within each sector are generically \emph{points}.
      \item In numerics, gapless surfaces were found between each pair of adjacent bands.
    \end{itemize}
  \\
\end{tblr}
\caption{The effects of the symmetries $\MI$ and $\EMI$ on the band touchings in the easy-axis stacked honeycomb XXZ model, which has spin-rotation symmetry about the $\hat{z}$ axis. See \sect{\ref{par:XXZ_easyaxis}} for details and \app{\ref{app:XXZ}} for parameter values.}
\label{tab:XXZ_easyaxis}
\end{table*}

\froufrou*

The previous examples all pertain to band touchings at generic momenta, i.e., not at high-symmetry points of the Brillouin zone.
Of course, as is commonly done in fermion systems~\cite{Kane2015DiracSemimetals, Tiwari2020NodalLine}, these methods can also be applied specifically to high-symmetry lines or planes, which are left invariant by additional unitary symmetries~\cite{Choi2019Nonsymmorphic}.
Within these manifolds, the additional symmetries can give rise to a block structure different from that at generic momenta.

We close the present section with the following remark: in \app{\ref{sec:symmetries-implementation}}, we show generally that unitary and antiunitary symmetries of the spin Hamiltonian that are compatible with the ground state have, respectively, Krein-unitary and Krein-antiunitary representations on the Krein-Hermitian single-particle magnon Hamiltonian.
It is a deep result in quantum mechanics, known as Wigner's theorem, that the representations of physical symmetries in an ordinary Hilbert space can be either unitary or antiunitary~\cite{Wigner1959Group}.
The natural analogues of these operators in Krein spaces are Krein-unitary and Krein-antiunitary operators, respectively.
However, the extension of Wigner's theorem to indefinite metric spaces~\cite{Bracci1975Wigner} turns out to be richer than its Hilbert-space counterpart: in addition to Krein-unitary and Krein-antiunitary representations, two other possible types of representations arise, called Krein-pseudounitary and Krein-antipseudounitary, which have no analogue in ordinary Hilbert spaces.
Whether such exotic symmetry representations arise in magnon or other bosonic BdG Hamiltonians and what physical consequences they may have are interesting questions left for future work.

\section{Conclusion}

In summary, after reviewing Krein-Hermitian Hamiltonians, particularly those arising as single-particle bosonic BdG Hamiltonians, we formulated a Krein-unitary SW transformation, paying special attention to the consequences of Krein stability theory and to the form of the effective reduced Hamiltonian. 
For simplicity, we assumed the Krein metric $\eta$ is diagonal with entries $\pm1$. 
We found that if the unperturbed Hamiltonian $K^0$ is \emph{within} a dynamically stable phase, then for sufficiently small perturbations, the effective Hamiltonian for a subspace of interest $\cP_0$, given by $P_0 \tilde{K} P_0$, is dynamically stable and Krein Hermitian with respect to the metric $P_0 \eta P_0$. 
If the subspace of interest is composed only of positive-signature states, then the effective Hamiltonian is Hermitian.
Also, if the (perturbed) Hamiltonian is purely real, then so is the effective Hamiltonian.

We argued that, in a translation-invariant Krein-Hermitian system, the SW transformation we formulated justifies the description of adjacent bands by reduced effective Hamiltonians, at least locally in momentum space. 
For a single-particle bosonic Hamiltonian $\tauthree H_\bk$ with $H_{\bk_0} > 0$, a subset of contiguous positive-energy bands that is gapped from the rest of the spectrum near $\bk_0$ can locally be described by a Hermitian reduced Hamiltonian.
In particular, by considering two adjacent positive-energy bands, the generic codimension of band touchings is easily predicted, much as for Hermitian Hamiltonians.
We then reviewed this line of reasoning in the context of magnon systems using examples from the literature, and emphasizing differences with electron systems.

In the linear spin-wave approximation, which we have adopted here, interaction between the Holstein-Primakoff bosons are neglected and the eigenstates of the quadratic Hamiltonian are taken to be long-lived quasiparticles. 
This simplification works well for many magnetic materials~\cite{Chernyshev2013Colloquium}, though it can fail even at zero temperature for systems with strong quantum fluctuations~\cite{Chernyshev2013Colloquium}, \cite{Mook2021Interaction}.
Systems in which magnon interactions play an essential role are at the forefront of the field of topological magnons~\cite{Mook2021Interaction}.
%{Magnon decay, for example, can give rise to topological magnon systems without any fermion analogues, because of number conservation.}
On the other hand, whereas electrons can be spinless only in an approximate sense, 
for magnons and other bosonic partcles, space-time inversion truly squares to $+1$ regardless of additional material details~\cite{Zhang2019Phononic}.

%\added[comment={resolve}]{Applicability: in real materials, the spin magnitude $S$ is often not large, calling into question the omission of magnon interactions in the HP transformation (see appendix)~\cite{Yao2018Antiferromagnet}, \cite{Mook2021Interaction}. Quantum fluctuations, etc.,  can invalidate the harmonic picture\dots}

\section*{Acknowledgements}
This research was funded by the Natural Sciences and Engineering Research Council of Canada and the Canadian Institute for Advanced Research. G.~M.~thanks the \emph{Fonds de recherche du Qu\'ebec -- Nature et technologies} for its support. 
Computations were partly performed on the Niagara supercomputer at the SciNet HPC Consortium~\cite{Ponce2019Deploying, Loken2010SciNet}, part of Compute Ontario and the Digital Research Alliance of Canada. 
SciNet is funded by the Canada Foundation for Innovation; the Government of Ontario; Ontario Research Fund -- Research Excellence; and the University of Toronto.

\appendix

\section{Details of perturbative SW transformation} \label{app:SW-expansion}

This section provides more details on the perturbatively defined transformation presented in \sect{\ref{subsec:SW-Krein}}. The presentation follows that of \reference{Winkler2003Quasi}, adapting it to Krein-Hermitian matrices.

We seek an effective Hamiltonian $\tilde{K}$ that is block diagonal in the subspace $\cP_0$, meaning $\bod{\tilde{K}} = 0$, and that reproduces the spectrum of $K$ order by order in the perturbation $\epsilon K'$ (see \eq{\eqref{eq:transf-expansion}}). 
We assume that there exists a skew-Krein-unitary $W = -W^\krein$ that block-diagonalizes the perturbed Hamiltonian $K$ order by order in the perturbation $\epsilon K'$, and later confirm our assumption by establishing that explicit expressions for such a $W$ can be found to all orders.

We further assume that $W$ is block off-diagonal; i.e., $\bd{W} = 0$.
Because of this, $[ \bd{A}, W ]$ is purely block off-diagonal and $[ \bod{A}, W ]$ is purely block diagonal for any $A$. 
Hence, the transformed Hamiltonian can be split into block-diagonal and block-off-diagonal parts:
\begin{subequations}
	\begin{align}
		\tilde{K}_\rb &= \sum_{l=0}^{+\infty} \frac{1}{(2l)!} [K_\rb, W]^{(2l)} + \sum_{l=0}^{+\infty} \frac{1}{(2l\!+\!1)!} [K_\ro, W]^{(2l+1)},
		\\
		\tilde{K}_\ro &= \sum_{l=0}^{+\infty} \frac{1}{(2l)!} [K_\ro, W]^{(2l)} + \sum_{l=0}^{+\infty} \frac{1}{(2l\!+\!1)!} [K_\rb, W]^{(2l+1)},
	\end{align}
\end{subequations}
where
\begin{equation}
	[A,B]^{(l)} \coloneqq [\dots [[A \underbrace{,B],B]\dots,B]}_{l\text{ repetitions}}.
\end{equation}

We then further constrain the generating matrix $W$ by demanding that $\tilde{K}_\ro=0$ order by order in the perturbation $\epsilon K'$, making the transformed Hamiltonian $\tilde{K} = \tilde{K}_\rb$ purely block diagonal. Letting $W = \epsilon W^{(1)} + \epsilon^2 W^{(2)} + \dots$, this requirement leads to the constraints of \eq{\eqref{eq:pert-constraints-W}}. As noted in the main text, these constraints are compatible with the assumption that $W$ is skew-Krein-Hermitian to all orders.

Working in the eigenbasis of $K^0$, $K^0 t_i = E^0_i t_i$,
\eq{\eqref{eq:pert-constraints-W}} can be solved order by order. For instance, 
\begin{subequations}
\begin{equation}
	\begin{split}
		[K^0, W^{(1)}] &= -K'_\ro
		\\
		\Leftrightarrow \sum_{c} \Bigl( K^0_{ac} W^{(1)}_{cb} - W^{(1)}_{ac} K^0_{cb} \Bigr) &= - (K'_\ro)_{ab}
		\\
		\Leftrightarrow \left( E^0_{a} - E^0_{b} \right) W^{(1)}_{ab} &= - (K'_\ro)_{ab}.
	\end{split}
\end{equation}
Letting $m$ index states in $\cP_0$ and $l$ index states not in $\cP_0$ (or vice versa), we may then write
\begin{align}
	W^{(1)}_{ml} &= - \frac{K'_{ml}}{E^0_{m} - E^0_{l}},
	&
	W^{(1)}_{mm'} = W^{(1)}_{ll'} = 0.
\end{align}
\end{subequations}
By the same token, one can find the expressions to all orders:
\begin{equation}
	\begin{split}
		W^{(2)}_{ml} &= \frac{1}{E^0_{m} - E^0_{l}} \Biggl( \sum_{m'} \frac{K'_{mm'} K'_{m'l}}{E^0_{m'} - E^0_{l}} - \sum_{l'} \frac{K'_{ml'} K'_{l'l}}{E^0_{m} - E^0_{l'}} \Biggr)
		\\
		& \ \, \vdots
	\end{split}
\end{equation}

Using these expressions and that for $\tilde{K} = \bd{\tilde{K}}$, we find the effective Hamiltonian $P_0 \tilde{K} P_0 = P_0 \big(\sum_{l=0}^{+\infty} \epsilon^l \tilde{K}^{(l)} \big) P_0$, whose matrix elements are $\tilde{K}_{mm'}$, order by order:
\begin{subequations}
	\begin{align}
		\tilde{K}^{(0)}_{mm'} &= K^0_{mm'},
		\\
		\tilde{K}^{(1)}_{mm'} &= K'_{mm'},
		\\
		\tilde{K}^{(2)}_{mm'} &= \frac{1}{2} \! \sum_{l} \! K'_{ml} K'_{lm'} \! \biggl( \! \frac{1}{E^0_{m} - E^0_{l}} \!+\! \frac{1}{E^0_{m'} - E^0_{l}} \! \biggr),
		\\
		\vdots \nonumber
	\end{align}
\end{subequations}
The zeroth- and first-order expressions are manifestly Krein Hermitian with respect to $P_0 \eta P_0$; the higher orders, which involve mixing with states outside $\cP_0$, are too, as can be seen using $K'_{ab} = \eta_{aa} {K'_{ba}}\!^* \; \eta_{bb}$ and the fact that $\eta$ is diagonal with entries $\pm 1$.

The formulae for $W^{(j)}$ and $\tilde{K}^{(j)}$ at all orders have the same form as those for a Hermitian Hamiltonian, more of which are given in Refs.~\onlinecite{Winkler2003Quasi} and \onlinecite{Bravyi2011Schrieffer}.
The expansion is controlled as long as the perturbation is small compared to energy separations between states in $\cP_0$ and states outside $\cP_0$.

\section{Details and parameter values for XXZ model} \label{app:XXZ}

This appendix provides more details on the XXZ spin model and parameter values used in \sect{\ref{subsubsec:XXZ}}. 
As explained in the main text, it is based on models of CoTiO\textsubscript{3} wherein effective spin-one-half moments at each Co\textsuperscript{2+} ion (seen in \fig{\ref{fig:cryst_struct}}) are coupled by XXZ interactions~\cite{Yuan2020, Elliot2020Visualization}.
In general, the Hamiltonian has the form
\begin{equation} \label{eq:general_XXZ}
H_\text{XXZ} = \sum_{i<j} \Big( J_{ij}\left(S_i^x S_j^x + S_i^y S_j^y\right) + J^\perp_{ij} \, S_i^z S_j^z \Big),
\end{equation}
where $i$ and $j$ label spins and ``\,$i<j$\,'' indicates that pairs of spins are not double counted.
In the stacked honeycomb geometry under study, the $\hat{z}$ direction is taken to be the stacking direction (see \fig{\ref{fig:cryst_struct}}).
Hence, an XXZ interaction is of easy plane if $|J|>|J^\perp|$, and of easy axis if $|J|<|J^\perp|$; the magnetic ground-state order depends on the relative strength of the nonzero interactions and their coordination numbers.

\paragraph{Easy plane}

Our choice of which spins to couple with the interactions of \eq{\eqref{eq:general_XXZ}} is loosely based on Refs.~\onlinecite{Yuan2020, Elliot2020Visualization}, which model the magnon dispersion of CoTiO\textsubscript{3}. 
An additional, weak coupling, $J_9$, is included to avoid accidental degeneracies along six lines in the Brillouin zone~\footnote{
	See Supplementary Figure~4 of \reference{Elliot2020Visualization} for a depiction of these lines, which, as explained in the reference, project onto the corner K points of the 2D hexagonal Brillouin zone.
}. 
The couplings are as follows:
\begin{itemize}
	\item If $i$ and $j$ are NN in-plane spins, 
	$J_{ij} \eqqcolon J_1 = -1$ and $J^\perp_{ij} \eqqcolon J^\perp_1 = 0.25\, J_1$ (see \reference{Yuan2020} for schematic representation);
	
	\item If $i$ and $j$ are next-nearest-neighbor (NNN) sites in adjacent layers, 
	$J_{ij} \eqqcolon J_2 = 0.5$ and $J^\perp_{ij} \eqqcolon J^\perp_2 = 0.25\, J_2$ (see \reference{Yuan2020});
	
	\item If $i$ and $j$ are NN sites in adjacent layers, 
	$J_{ij} \eqqcolon J_3 = 0.5$ and $J^\perp_{ij} \eqqcolon J^\perp_3 = 0.25\, J_3$ (see \reference{Yuan2020});
	
	\item If $i$ and $j$ are NN sites three layers apart, 
	$J_{ij} \eqqcolon J_9 = 0.02$ and $J^\perp_{ij} \eqqcolon J^\perp_9 = J_9$.
\end{itemize}
With these parameter values, the magnetic order is manifestly in plane, with ferromagnetic order within each layer and antiferromagnetically stacked layers. 
The ordering direction is chosen to be $\hat{x}$ (see \fig{\ref{fig:magnetic_order}(a)}), and the Holstein-Primakoff transformation is carried out as shown in \app{\ref{sec:symmetries-implementation}}.

\paragraph{Easy-plane XXZ with DM}

As explained in \sect{\ref{subsubsec:XXZ}}, a weak DM interaction,
\begin{equation}
H_\text{DM} = \sum_{i<j} \bD_{ij} \cdot \left( \bS_i \times \bS_j \right),
\end{equation}
is added to the easy-plane XXZ model to break certain extraneous symmetries of the pure XXZ model. 

In the space group $\mathit{R\bar{3}}$ of the stacked honeycomb CoTiO\textsubscript{3}, the geometry of which we are using here, inversion centers forbid DM interactions on the intra-layer NN bonds. 
Hence, the simplest choice is to include it on the intra-layer NNN bonds; this leaves the magnetic order unchanged. 
The vector $\bD_{ij}$ on one such bond is not constrained by symmetry; we choose it to be in the $\hat{x}$-$\hat{y}$ plane and perpendicular to the bond direction, pointing toward the center of each hexagonal plaquette, and with magnitude $0.25$.
The interactions on different bonds are constrained by the threefold rotation, inversion, and translation symmetries of $\mathit{R\bar{3}}$.
The gapless lines of the associated linear spin-wave dispersion are represented in \fig{\ref{fig:gapless_pts}}.

%Since DM interactions between intra-layer NNs are forbidden because of mid-bond inversion centers, the simplest choice is to include it on the NNN intra-layer bonds. 

In order to break the magnetic inversion symmetry $\MI$ and destabilize the gapless lines, a Zeeman field $\bB_1 = B_1 \hat{x}$, with $B_1 = 0.0008$, is introduced on sublattice 1 in the magnetic unit cell (see \fig{\ref{fig:cryst_struct}}), in the direction of the ground-state magnetic moment. 
This lifts the line degeneracies, leaving behind gapless points, as shown in \fig{\ref{fig:gapless_pts}}.

%Remark that the XXZ \& DM model still has higher symmetry than $\mathit{P_S\bar{1}}$, though not independent spin transformations.

\paragraph{Easy-axis XXZ}
In \sect{\ref{par:XXZ_easyaxis}}, we consider an easy-axis version of the model of \eq{\eqref{eq:general_XXZ}}, using the following parameters:
\begin{itemize}
	\item If $i$ and $j$ are NN in-plane spins, 
	$J_{ij} \eqqcolon J_1 = -1$ and $J^\perp_{ij} \eqqcolon J^\perp_1 = 1.5\, J_1$;
	
	\item If $i$ and $j$ are NNN sites in adjacent layers, 
	$J_{ij} \eqqcolon J_2 = 0.5$ and $J^\perp_{ij} \eqqcolon J^\perp_2 = 1.5\, J_2$;
	
	\item If $i$ and $j$ are NN sites in adjacent layers, 
	$J_{ij} \eqqcolon J_3 = 0.5$ and $J^\perp_{ij} \eqqcolon J^\perp_3 = 1.5\, J_3$;
	
	\item If $i$ and $j$ are NN sites three layers apart, 
	$J_{ij} \eqqcolon J_9 = 0.02$ and $J^\perp_{ij} \eqqcolon J^\perp_9 = J_9$.
\end{itemize}
With these values, the magnetic order is clearly parallel to the stacking direction $\hat{z}$, with ferromagnetic order within each layer and antiferromagnetically stacked layers (\fig{\ref{fig:magnetic_order}}(b)). The Holstein-Primakoff transformation is carried out as shown in \app{\ref{sec:symmetries-implementation}}.

In the results presented in the main text, Zeeman fields pointing along the moments are used to selectively break the symmetries $\MI$ and $\EMI$. 
Labelling the four sublattices in the magnetic unit cell as shown in \fig{\ref{fig:cryst_struct}}, the Zeeman fields are $\bB_1 = B_1 \hat{z}$, $\bB_2 = B_2 \hat{z}$, $\bB_3 = -B_3 \hat{z}$, and $\bB_4 = -B_4 \hat{z}$.
Since $\MI$ constrains $B_1 = B_4$ and $B_2 = B_3$, and $\EMI$ constrains $B_1 = B_2$ and $B_3 = B_4$, the symmetries can be selectively broken in the following way:
\begin{itemize}
	\item $B_1 = B_2 = 0.01$ and $B_3 = B_4 = 0$ to break $\MI$;
	\item $B_1 = B_3 = 0.01$ and $B_2 = B_4 = 0$ to break $\EMI$;
	\item $B_1 = B_3 = 0.01$ and $B_2 = B_4 = 0.03$ to break both $\MI$ and $\EMI$.
\end{itemize}

\section{Implementation of symmetries in magnon Hamiltonians} \label{sec:symmetries-implementation}

In this appendix, we provide an overview of how symmetries of a spin Hamiltonian translate to symmetries of the magnon Hamiltonian in a compatible ordered state.
The spin Hamiltonians we consider describe localized moments in a crystal lattice and, accordingly, have the translational symmetry of the lattice.

We highlight the constraints that unitary and antiunitary symmetries place on the single-particle Bloch Hamiltonian (Eqs.~\eqref{eq:Bloch-constrain-unitary} and \eqref{eq:Bloch-constrain-antiunitary}).
%\added{Unitary symmetries that leave momentum invariant can be used to block-diagonalize the Bloch Hamiltonian---see \sect{\ref{subsubsec:translation-spinrot}} for an example.}
We show that, in the right basis, magnetic inversion symmetry constrains the single-particle magnon Hamiltonian to be purely real.

\subsection{Overview}

The \emph{spin space group} of a spin Hamiltonian together with a compatible classical ground state is the set of symmetries of the Hamiltonian which also leave the classical order invariant~\cite{BradleyCracknell, brinkman1966space, brinkman1966theory, brinkman1967magnetic}.
Such symmetries have a well-defined implementation in the magnon Hamiltonian, which describes excitations above a specific classical ground state.
Symmetries of the spin space group come in two types: those that include the operation of time reversal and those that do not~\cite{BradleyCracknell,lu2018magnon}, which are represented, respectively, by Krein-antiunitary and Krein-unitary transformations of the single-particle magnon Hamiltonian~\cite{Bracci1975Wigner}, as shown below. 
Time reversal alone is never part of the spin space group.

Spin space groups are closely related to, though distinct from, \emph{magnetic space groups}: a magnetic space group is the symmetry group of a crystal structure endowed with a magnetic order~\cite{BradleyCracknell}.
For example, since an effective spin Hamiltonian can have more symmetry than the physical system it models, the spin space group may contain symmetries not present in the magnetic space group. 
In particular, unlike the magnetic space group, the spin space group can have symmetries that act differently on spin and position space.
We will denote such operations using a modified Seitz notation~\cite{BradleyCracknell}: $g = \left\{Q_g; Q_g^\rs\ | \btau_g\right\}$, where the first entry denotes the O(3) operation acting in position space, the second entry denotes the O(3) operation acting on spin, and the third denotes the translation in position space. 
Given a spin labeled $i$, we refer to the site to which $i$ is mapped by (the spatial part of) $g$ as $g (i)$: $\bx_{g(i)} = \left\{Q_g| \btau_g\right\} \bx_i = Q_g \bx_i + \btau_g$.

%We find the action of the symmetry operations on the magnon operators by first finding their action in the local basis (in which the local $\hat{\tilde{z}}$ axis is aligned with the classical spin direction). 
%The key is that the transformation leaves the ``classical'' spin component unchanged, so that the rotation $R_i$ at a site $i$ is the same before and after the transformation.

We will find that in the local frame, transformations in the spin space group are of the form 
\begin{align}
	\tbS_i \rightarrow \tbS_i' = \left[\begin{array}{cc|c}
		*&*&0
		\\
		*&*&0
		\\ \hline
		0&0&1
	\end{array}\right]
	\begin{bmatrix}
		\tS^x_j \\ \tS^y_j \\ \tS^z_j
	\end{bmatrix}.
\end{align}
Hence, these symmetry transformations mix together only the components transverse to the local classical order.

\subsection{Spin Hamiltonian}
We begin with a spin Hamiltonian with a corresponding classical ground state. 
The order of the ground state may lower the symmetry of the system, giving rise to a \emph{magnetic lattice} and a corresponding \emph{magnetic unit cell}, which is potentially larger than the initial unit cell~\cite{BradleyCracknell}.
Spin interactions beyond bilinear order can also be included, and would not change our conclusions at the level of noninteracting bosons.

\begin{equation}
\hat{H} = \frac{1}{2} \sum_{ij} \bS_i^\transpose M_{ij} \bS_j - \sum_{i} \bB_i^\transpose \bS_i.
\end{equation}
We may choose $M_{ij} = M_{ji}^\transpose$. 
Furthermore, the Hermicity of $\hat{H}$ and the assumption of time-reversal symmetry (absent the Zeeman term) makes $M_{ij} = M_{ji}^\transpose$ real for all $i$ and $j$. 

Consider a classical ground state compatible with the spin Hamiltonian.
To carry out the Holstein-Primakoff transformation, each spin is expressed in a local basis whose $\hat{\tilde{z}}$ direction (and quantization axis) is aligned with the classical ordering direction of that spin.
Hence, we let
\begin{align}
\bS_i &= R_i \tbS_i, 
&
\tbS_i = \begin{bmatrix}
\tS^x_i & \tS^y_i & \tS^z_i
\end{bmatrix}^\transpose,
\end{align}
where $R_i$ is an appropriate SO(3) rotation matrix. 
There is, of course, freedom in the choice of $R_i$. 
The Hamiltonian becomes
\begin{subequations}
	\begin{align}
		\hat{H} &= \frac{1}{2} \sum_{ij} \tbS_i^\transpose \tM_{ij} \tbS_j - \sum_i \tbB_i^\transpose \tbS_i,
		\\
		&\begin{cases}
			\tM_{ij} \coloneqq R_i^\transpose M_{ij} R_j = \tM_{ji}^\transpose,
			\\
			\tbB_i \coloneqq R_i^\transpose \bB_i.
		\end{cases}
	\end{align}
\end{subequations}
The stability of the classical order implies that linear contributions (in $\tS^x_i$ and $\tS^y_i$) to the Hamiltonian vanish.

We now proceed to rewrite the spin operators in terms of Holstein-Primakoff bosons. Letting $\phi_i = \begin{bmatrix} b_i & b_i^\dagger \end{bmatrix}^\transpose$, in the linearized Holstein-Primakoff transformation, the spin operators become
\begin{equation}
\begin{cases}
\tS_i^+ &= \sqrt{2 S_i} b_i + \dots,
\\
\tS_i^- &= \sqrt{2 S_i} b_i^\dagger + \dots,
\\
\tS_i^z &= S_i - b_i^\dagger b_i = S_i + \frac{1}{2} - \frac{1}{2} \phi_i^\dagger \phi_i,
\end{cases}
\end{equation}
where $\tS_i^\pm \coloneqq \tS_i^x \pm \ri \tS_i^y$ are the raising and lowering operators in the local bases, and where $S_i$ is the spin magnitude at site $i$, formally considered a large parameter controlling the expansion.
This way of writing $\tS^z_i$ is key in ensuring it is invariant under symmetries.
Discarding terms beyond quadratic in boson operators, the linearized Hamiltonian takes the form
\begin{align}
\hat{H} &= E_\text{cl} + \frac{1}{2} \sum_{ij} \phi_i^\dagger h_{ij} \phi_j + \mathcal{O}(\phi^3),
%\\ E_\text{cl} &= \frac{1}{2} \sum_{ij} \left(S_i + \frac{1}{2}\right) \tM^{xx}_{ij} \left(S_j + \frac{1}{2}\right) - \sum_{i} \tB^x_i \left(S_i + \frac{1}{2}\right).
%\\ h_{IJ} &= \sqrt{S_IS_J} m_{IJ} + \delta_{IJ}\left( \tB^x_I - \sum_{I'} \left(S_{I'} + \frac{1}{2}\right) \frac{\tM^{xx}_{II'}+\tM^{xx}_{I'I}}{2}\right) \tau^0
\end{align}
where $E_\text{cl}$ is the classical energy and the $h_{ij}$ are $2 \times 2$ coefficient matrices.
Note that $h_{ji} = h_{ij}^\dagger$ as a consequence of $\tM_{ji} = \tM_{ij}^\transpose$.
Furthermore, the particle-hole symmetry of $h_{ij}$, i.e., $h_{ij} = \tauone h_{ij}^* \tauone$, stems from the reality of $\tM_{ij}$ and the symmetric way of writing $\tS_i^z$.
This constraint leaves no redundancy in $h_{ij}$.
%It constrains $h_{ij} = c^0_{ij} \tau^0 + c^1_{ij} \tau^1 + c^2_{ij} \tau^2 + \ri c^3_{ij} \tau^3$, with $c^\mu_{ij} \in \mathbb{R}$.

\subsubsection*{Quadratic boson Hamiltonian in Fourier space}
The Fourier transformation is a paraunitary (and unitary) transformation that partially diagonalizes the Hamiltonian. Letting $i = (\bR, \alpha)$, where $\bR$ denotes the magnetic unit cell (of which there are $N$) and $\alpha$ denotes the magnetic sublattice (of which there are $n$),
\begin{subequations}
	\begin{align}
		\phi_{\bk\alpha} &\coloneqq \begin{bmatrix} b_{\bk\alpha} \\ b_{-\bk\alpha}^\dagger \end{bmatrix} = \frac{1}{\sqrt{N}} \sum_\bR \re^{-\ri \bk \cdot \bx_{\bR\alpha}} \phi_{\bR\alpha}, 
		\\
		\phi_{\bR \alpha} &= \frac{1}{\sqrt{N}} \sum_{\bk} \re^{\ri \bk \cdot \bx_{\bR \alpha}} \phi_{\bk\alpha}.
	\end{align}
\end{subequations}
With this notation, the Hamiltonian becomes
\begin{equation}
	\begin{split}
		\hat{H} &= E_\text{cl} + \frac{1}{2} \sum_{ij} \phi_i^\dagger h_{ij} \phi_j
		\\
		&= E_\text{cl} + \frac{1}{2} \sum_{\bR \bR'} \sum_{\alpha\beta} \phi_{\bR \alpha}^\dagger h_{\bR \bR';\alpha\beta} \phi_{\bR' \beta}.
		%\\&= E_\text{cl} + \frac{1}{2} \sum_{\bk\bp} \sum_{\alpha\beta} \phi_{\bk\alpha}^\dagger \biggl( \frac{1}{N} \sum_{\bR \bR'} \re^{-\ri\bk\cdot\bx_{\bR\alpha}} h_{\bR \bR';\alpha\beta} \re^{\ri\bp\cdot\bx_{\bR'\beta}} \biggr) \phi_{\bp\beta}.
	\end{split}
\end{equation}
Using the translational invariance of the magnetic lattice (and assuming the SO(3) matrices $R_i \eqqcolon R_\alpha$ were chosen with the same translational invariance), we define $h_{\bR-\bR';\alpha\beta} \coloneqq h_{\bR \bR';\alpha\beta}$, as well as its Fourier transform 
\begin{equation}
h_{\bk;\alpha\beta} \coloneqq \re^{-\ri\bk\cdot\left(\bdelta_\alpha-\bdelta_\beta\right)} \sum_{\bR} \re^{-\ri\bk\cdot\bR} h_{\bR;\alpha\beta}.
\end{equation}
Note that $h_{ij} = h_{ji}^\dagger$ implies that $h_{\bk;\alpha\beta}^\dagger = h_{\bk;\beta\alpha}$, and that the particle-hole symmetry $h_{ij} = \tauone h_{ij}^* \tauone$ manifests itself in momentum space as $\tau^1 h_{\bk;\alpha\beta}^* \tau^1 = h_{-\bk;\alpha\beta}$.

Finally, the magnon Hamiltonian becomes
\begin{subequations}
\begin{align}
\hat{H} &= E_\text{cl} + \frac{1}{2} \sum_{\bk} \sum_{\alpha\beta} \phi_{\bk\alpha}^\dagger h_{\bk;\alpha\beta} \phi_{\bk\beta} \\
&= E_\text{cl} + \frac{1}{2} \sum_{\bk} \Phi_{\bk}^\dagger H_{\bk} \Phi_{\bk}, \quad \Phi_{\bk} \coloneqq \begin{bmatrix}
\phi_{\bk A} \\ \phi_{\bk B} \\ \vdots
\end{bmatrix},
\end{align}
\end{subequations}
where we have introduced the spinor $\Phi_{\bk}$ and the $2n\times 2n$ coefficient matrix $H_{\bk}$, and where $\bk$ is summed over momenta in the magnetic Brillouin zone~\cite{BradleyCracknell}. The matrix $H_{\bk}$ is Hermitian and (assuming the classical order is stable) positive semidefinite. Also, the PH symmetry now takes the form $\tau^1 H_{\bk}^* \tau^1 = H_{-\bk}$.

\subsection{Unitary symmetries}

Let $g$ be a unitary symmetry of the spin space group, with unitary Hilbert-space implementation $U_g$. 
These are the operations of the spin space group that do not contain time reversal. 
In modified Seitz notation, we let $g = \left\{Q_g; Q_g^\rs| \btau_g\right\}$.
Letting $\bS_i'$ denote the transformed spin at site $i$,
\begin{equation}
		\bS_i' = U_g \bS_i U_g^\dagger
		=  (\det Q_g^\rs) Q_g^\rs \bS_{g^{-1}i}.
\end{equation}
Because spins are axial vectors, they transform with $(\det Q_g^\rs)Q_g^\rs$ , which is always an SO(3) rotation matrix.

By definition of spin-space-group symmetries, the transformed spin $\bS_i'$ has the same ``classical'' direction as $\bS_i$; therefore, $\bS_i' = R_i \tbS_i'$, like $\bS_i = R_i \tbS_i$.
This way, in the transformation in the local basis from $\tbS_i$ to $\tbS_i'$,
\begin{subequations}
	\begin{gather}
		\tbS_i' = U_g \tbS_i U_g^\dagger
		= \tQ_g^i \tbS_{g^{-1}i},
		\\
		\tQ_g^i \coloneqq R_i^\transpose (\det Q_g^\rs) Q_g^\rs R_{g^{-1}i},
	\end{gather}
\end{subequations}
$\tQ_g^i$ is just a rotation about the local $\hat{\tilde{z}}$ axis.

Denoting the rotation angle at site $i$ about the local $\hat{\tilde{z}}$ axis as $\theta_g^i$, the ladder operators transform as $U_g \tS_i^\pm U_g^\dagger = \re^{\pm \ri \theta_g^i} \tS_{g^{-1}i}^\pm$, while $U_g \tS_i^z U_g^\dagger = \tS_{g^{-1}i}^z$.
The rotation angle $\theta_g^i$ is determined by $g$ and the rotation matrices $\{R_i\}$.
The bosons transform as
\begin{subequations}
	\begin{gather}
		U_g \phi_i U_g^\dagger = u_g^i \phi_{g^{-1}i},
		\\
		u_g^i = \begin{bmatrix}
			\re^{\ri \theta_g^i} & \\ & \re^{-\ri \theta_g^i}
		\end{bmatrix}
		= \tauzero \cos\theta_g^i + \ri \tauthree \sin\theta_g^i.
	\end{gather}
\end{subequations}

Note that in addition to being unitary, $u_g^i$ is paraunitary (a unitary matrix that commutes with $\tau^3$ is also paraunitary), so it gives valid boson operators.
Furthermore, $u_g^i$ is particle-hole symmetric, $\tau^1 \left(u_g^i\right)^*\tau^1 = u_g^i$, so the transformed Hamiltonian will be particle-hole symmetric, as it must.

\subsubsection*{In Fourier space}
Next, we consider the transformation in Fourier space. Although $\btau_g$, the translational part of $g$, appears in the transformation for $\phi_{\bk\alpha}$, it will not appear in the constraint on the Hamiltonian at the quadratic level. For notational convenience, let $(\bR'\alpha') = g^{-1} (\bR\alpha)$, meaning $\bx_{\bR'\alpha'} = \{Q_g | \btau_g\}^{-1} \bx_{\bR\alpha}$.

We will use the fact that the rotation matrix $R_i \eqqcolon R_\alpha$ (and hence $u_g^i \eqqcolon u_g^\alpha$) does not depend on $\bR$, but only on $\alpha$, and also that the transformed sublattice index $\alpha'$ depends only on $\alpha$ (and not on $\bR$). 
For this reason, we can think of the action of $g$ (or $g^{-1}$) on the $n$ sublattice indices $\alpha$ as a permutation.
\begin{equation}
	\begin{split}
		U_g \phi_{\bk\alpha} U_g^\dagger &= \frac{1}{\sqrt{N}} \sum_{\bR} \re^{-\ri \bk \cdot \bx_{\bR\alpha}} \underbrace{U_g \phi_{\bR\alpha} U_g^\dagger}_{u_g^\alpha \phi_{\bR'\alpha'}}
		\\
		&= \frac{u_g^\alpha}{\sqrt{N}} \sum_{\bR} \re^{-\ri \bk \cdot \{Q_g | \btau_g\} \bx_{\bR'\alpha'}} \phi_{\bR'\alpha'}
		\\
		&= \frac{u_g^\alpha}{\sqrt{N}} \sum_{\bR} \re^{-\ri \bk \cdot 
			\left(Q_g \bx_{\bR'\alpha'} + \btau_g\right) } \phi_{\bR'\alpha'}.
	\end{split}
\end{equation}
We can sum over $\bR'$ instead of $\bR$ because the relation between $\bR$ and $\bR'$ is bijective:
\begin{equation}
	\begin{split}
		U_g \phi_{\bk\alpha} U_g^\dagger &= \re^{-\ri \bk \cdot \btau_g } \frac{u_g^\alpha}{\sqrt{N}} \sum_{\bR'} \re^{-\ri \left(Q_g^\transpose \bk\right) \cdot \bx_{\bR'\alpha'} } \phi_{\bR'\alpha'}
		\\
		&= \re^{-\ri \bk \cdot \btau_g } \ u_g^\alpha \ \phi_{Q_g^\transpose \bk,\alpha'}.
	\end{split}
\end{equation}
In terms of the $2n$-component spinor $\Phi_{\bk}$,
\begin{equation}
\boxed{U_g \Phi_{\bk} U_g^\dagger = \re^{-\ri \bk \cdot \btau_g }  V_g \Phi_{Q_g^\transpose\bk},}
\end{equation}
where $V_g$ is a $2n \times 2n$ unitary, paraunitary, and particle-hole-symmetric ($V_g$, $\tau^1 \left(V_g\right)^* \tau^1 = V_g$) matrix:
\begin{equation}
	V_g = C_g P_{g^{-1}} \otimes \tauzero - \ri S_g P_{g^{-1}} \otimes \tauthree.
\end{equation}
In the above, $P_{g^{-1}}$ is the $n \times n$ permutation matrix for the permutation of sublattices $g^{-1}(\alpha)$, with components $P_{g^{-1}}^{\alpha\beta} = \delta^{\beta,g^{-1}(\alpha)}$, and $C_g$ and $S_g$ are $n \times n$ diagonal matrices with entries $\{\cos \theta_g^\alpha\}$ and $\{\sin \theta_g^\alpha\}$, respectively.

%In this form, it is manifest that $[V_g, \tau^3] = 0$, meaning that $V_g$ is not only unitary but also paraunitary, as must be the case for a transformation on bosonic Nambu spinors.

Finally, we find the transformed magnon Hamiltonian:
\begin{equation}
	\begin{split}
		U_g \hat{H} U_g^\dagger &= E_\text{cl} + \frac{1}{2} \sum_{\bk} \left(U_g \Phi_{\bk}^\dagger U_g^\dagger\right) H_{\bk} \left(U_g \Phi_{\bk} U_g^\dagger\right)
		\\
		&= E_\text{cl} + \frac{1}{2} \sum_{\bk} \left( \Phi_{Q_g^\transpose\bk}^\dagger V_g^\dagger \right) H_{\bk} \left( V_g \Phi_{Q_g^\transpose\bk} \right)
		\\
		&= E_\text{cl} + \frac{1}{2} \sum_{\bk} \Phi_{\bk}^\dagger \ V_g^\dagger \ H_{Q_g \bk} \ V_g \ \Phi_{\bk},
	\end{split}
\end{equation}
so the constraint on $H_{\bk}$ due to $g$ is 
\begin{empheq}[box=\fbox]{equation} \label{eq:Bloch-constrain-unitary}
H_{\bk} = V_g^\dagger \ H_{Q_g \bk} \ V_g,
\end{empheq}
or in terms of the single-particle magnon Hamiltonian $\tauthree H_{\bk}$,
\begin{empheq}[box=\fbox]{equation}
	\tauthree H_{\bk} = V_g^\krein \ \tauthree H_{Q_g \bk} \ V_g,
\end{empheq}
making this a Krein-unitary transformation on the single-particle Hamiltonian.

For a unitary symmetry with $Q_g = \mathds{1}$, the Nambu-Bloch Hamiltonian can be block-diagonalized at each $\bk$ into sectors corresponding to the eigenvalues of $V_g$.

\def \aui {\Theta} % Symbol for the anti-unitary Hilbert space implementation

\subsection{Antiunitary symmetries}
Let $h$ be an antiunitary symmetry in the spin space group. Such a transformation is made up of a spatial transformation (with unitary implementation $U_h$) along with time reversal. Let $\aui_h = U_h T$ be the antiunitary operator that implements it in the Hilbert space. In modified Seitz notation, $h = \left\{Q_h; Q_h^\rs | \btau_h\right\} T$. Since $\aui_h^{-1} = T^{-1} U_h^\dagger$ and spin is odd under time reversal, we have
\begin{equation}
	\begin{split}
		\bS_i' &= \aui_h \bS_i \aui_h^{-1}
		\\
		&= U_h \underbrace{T \bS_i T^{-1}}_{-\bS_i} U_h^\dagger
		\\
		&= -(\det Q_h^\rs) Q_h^\rs \bS_{h^{-1}i}.
	\end{split}
\end{equation}
%It is understood that site $h^{-1}i$ is the site located at $\bx_{h^{-1}i} = \left\{Q_h | \btau_h\right\}^{-1} \bx_i$.

Once again, $\bS_i'$ has the same classical direction as $\bS_i$, so $\bS_i' = R_i \tbS_i'$, like $\bS_i = R_i \tbS_i$; this is the key point that allows a well-defined implementation of $h$ on the magnon Hamiltonian and enables us to write the transformation for $\tbS_i$:
\begin{subequations}
	\begin{gather}
		\tbS_i' = \aui_h \tbS_i \aui_h^{-1} = \tQ_h^i \tbS_{h^{-1}i},
		\\
		\tQ_h^i \coloneqq -R_i^\transpose (\det Q_h^\rs) Q_h^\rs R_{h^{-1}i}.
	\end{gather}
\end{subequations}
In this case, the transformation $\tQ_h^i$ is an improper rotation about the local $\hat{\tilde{z}}$ axis; i.e., an $\mathcal{O}(3)$ matrix with determinant $-1$ that leaves the $\hat{\tilde{z}}$ direction invariant.

Nevertheless, the ladder operators transform as $\aui_h \tS_i^\pm \aui_h^{-1} = \re^{\pm \ri \theta_h^i} \tS_{h^{-1}i}^\pm$, while $\aui_h \tS_i^z \aui_h^{-1} = \tS_{h^{-1}i}^z$, where the parameter $\theta_h^i$ is determined by $h$ and by the SO(3) matrices $\{R_i\}$.
The bosons transform as
\begin{subequations}
	\begin{gather}
		\begin{cases}
			\aui_h \phi_i \aui_h^{-1} &= u_h^i \phi_{h^{-1}i},
			\\
			\aui_h \phi_i^\dagger \aui_h^{-1} &= \phi_{h^{-1}i}^\dagger {u_h^i}^\dagger,
		\end{cases}
		\\
		u_h^i = \begin{bmatrix}
			\re^{\ri \theta_h^i} & \\ & \re^{-\ri \theta_h^i}
		\end{bmatrix}
		= \tauzero \cos\theta_h^i + \ri \tauthree \sin\theta_h^i.
	\end{gather}
\end{subequations}
Like $u_g^i$, $u_h^i$ is paraunitary, unitary, and particle-hole symmetric.

\subsubsection*{In Fourier space}
Next, we consider the effect in Fourier space. 
%Though $\btau_h$, the translational part of $h$, appears in the transformation for $\phi_{\bk \alpha}$, it will not appear in the constraint on the Hamiltonian \added{at the quadratic level}. 
Again, for notational convenience, we let $(\bR'\alpha') = h^{-1} (\bR\alpha)$, meaning $\bx_{\bR'\alpha'} = \{Q_h | \btau_h\}^{-1} \bx_{\bR\alpha}$.
\begin{equation}
	\begin{split}
		\aui_h \phi_{\bk \alpha} \aui_h^{-1} &= \frac{1}{\sqrt{N}} \sum_{\bR} \re^{\ri \bk \cdot \bx_{\bR\alpha}} \underbrace{\theta_h \phi_{\bR\alpha} \theta_h^{-1}}_{u_h^\alpha \phi_{\bR'\alpha'}}
		\\
		&= \frac{u_h^\alpha}{\sqrt{N}} \sum_{\bR} \re^{\ri \bk \cdot \{Q_h | \btau_h\} \bx_{\bR'\alpha'}} \phi_{\bR'\alpha'}
		\\
		&= \frac{u_h^\alpha}{\sqrt{N}} \sum_{\bR} \re^{\ri \bk \cdot \left(Q_h \bx_{\bR'\alpha'} + \btau_h\right)} \phi_{\bR'\alpha'}
		\\
		&= \re^{\ri \bk \cdot \btau_h} \frac{u_h^\alpha}{\sqrt{N}} \sum_{\bR'} \re^{-\ri \left(-Q_h^\transpose\bk\right) \cdot \bx_{\bR'\alpha'}} \phi_{\bR'\alpha'},
	\end{split}
\end{equation}
where in the last line, we have changed summation variable from $\bR$ to $\bR'$.
Hence, we find
\begin{subequations}
\begin{align}
\aui_h \phi_{\bk \alpha} \aui_h^{-1} &= \re^{\ri \bk\cdot\btau_h} \ u_h^\alpha \ \phi_{-Q_h^\transpose\bk,\alpha'}, \\
\aui_h \phi_{\bk \alpha}^\dagger \aui_h^{-1} &= \re^{-\ri \bk\cdot\btau_h} \  \phi_{-Q_h^\transpose\bk,\alpha'}^\dagger \ {u_h^\alpha}^\dagger.
\end{align}
\end{subequations}

In terms of the $2n$-component spinor $\Phi_{\bk}$,
\begin{subequations}
\begin{empheq}[box=\fbox]{align}
\aui_h \Phi_{\bk} \aui_h^{-1} &= \re^{\ri \bk\cdot\btau_h} \ V_h \ \Phi_{-Q_h^\transpose\bk}, \\
\aui_h \Phi_{\bk}^\dagger \aui_h^{-1} &= \re^{-\ri \bk\cdot\btau_h} \  \Phi_{-Q_h^\transpose\bk}^\dagger \ V_h^\dagger,
\end{empheq}
\end{subequations}
where $V_h$ is a $2n \times 2n$ unitary, paraunitary, and particle-hole-symmetric matrix:
\begin{equation}
	V_h = C_h P_{h^{-1}} \otimes \tauzero - \ri S_h P_{h^{-1}} \otimes \tauthree,
\end{equation}
where $P_{h^{-1}}$, $C_h$, and $S_h$ are defined in analogy with the unitary case.

Finally, the transformed magnon Hamiltonian is
\begin{equation}
	\begin{split}
		\aui_h \hat{H} \aui_h^{-1} &= E_\text{cl} + \frac{1}{2} \sum_{\bk} \left(\aui_h \Phi_{\bk}^\dagger \aui_h^{-1}\right) H_{\bk}^* \left(\aui_h \Phi_{\bk} \aui_h^{-1}\right)
		\\
		&= E_\text{cl} + \frac{1}{2} \sum_{\bk} \left( \Phi_{-Q_h^\transpose\bk}^\dagger V_h^\dagger \right) H_{\bk}^* \left( V_h \Phi_{-Q_g^\transpose\bk} \right)
		\\
		&= E_\text{cl} + \frac{1}{2} \sum_{\bk} \Phi_{\bk}^\dagger \ V_h^\dagger \ H_{-Q_h \bk}^* \ V_h \ \Phi_{\bk},
	\end{split}
\end{equation}
so the constraint on $H_{\bk}$ due to $h$ is 
\begin{empheq}[box=\fbox]{equation} \label{eq:Bloch-constrain-antiunitary}
H_{\bk} = V_h^\dagger \ H_{-Q_h \bk}^* \ V_h,
\end{empheq}
or in terms of the single-particle magnon Hamiltonian $\tauthree H_\bk$,
\begin{empheq}[box=\fbox]{equation}
	\tauthree H_{\bk} = V_h^\krein \ \big(\tauthree H_{-Q_h \bk}\big)^* \ V_h,
\end{empheq}
making this a \emph{Krein-antiunitary} transformation on the single-particle Hamiltonian~\cite{Bracci1975Wigner}.

\subsubsection{When is $V_h$ symmetric?} \label{sec:symmetric-conditions}

Here, we identify necessary and sufficient conditions for $V_h = V_h^\transpose$ (or, equivalently, $V_h V_h^* = + 1$).

First, since $V_h$ is block diagonal in Nambu space, we can write it as a direct sum of its particle and hole sectors, $V_h = V_h^\rp \oplus V_h^\rh$; here, $(V_h^\rp)^{\alpha\beta} = \re^{-\ri\theta_h^\alpha} \ \delta^{\beta,h^{-1}(\alpha)}$ and $V_h^\rh = (V_h^\rp)^*$.

\paragraph{Necessary condition}
First, assume $V_h$ is symmetric. Then,
\begin{equation}
	\begin{split}
		(V_h^\rp)^{\alpha\beta} &\overset{!}{=} (V_h^\rp)^{\beta\alpha} \quad \forall \ \alpha,\beta
		\\
		\Leftrightarrow \quad \re^{-\ri\theta_h^\alpha} \ \delta^{\beta,h^{-1}(\alpha)} &\overset{!}{=} \re^{-\ri\theta_h^\beta} \ \delta^{\alpha,h^{-1}(\beta)} \quad \forall \ \alpha,\beta.
	\end{split}
\end{equation}
Since the equality of two numbers implies the equality of their absolute values, we have
\begin{equation}
	\begin{split}
		\Rightarrow \quad \Big| \re^{-\ri\theta_h^\alpha} \ \delta^{\beta,h^{-1}(\alpha)} \Big| &\overset{!}{=} \Big| \re^{-\ri\theta_h^\beta} \ \delta^{\alpha,h^{-1}(\beta)} \Big| \quad \forall \ \alpha,\beta
		\\
		\Leftrightarrow \quad \delta^{\beta,h^{-1}(\alpha)} &\overset{!}{=} \delta^{\alpha,h^{-1}(\beta)} \quad \forall \ \alpha,\beta
		\\
		\Leftrightarrow \quad \delta^{\beta,h^{-1}(\alpha)} &\overset{!}{=} \delta^{\beta,h(\alpha)} \quad \forall \ \alpha,\beta
		\\
		\Leftrightarrow \quad h^{-1}(\alpha) &\overset{!}{=} h(\alpha)\quad \forall \ \alpha
		\\
		\Leftrightarrow \quad h^2(\alpha) &\overset{!}{=} \alpha \quad \forall \ \alpha.
	\end{split}
\end{equation}
In other words, the permutation engendered on the sublattice indices by $h$ must square to identity (called a \emph{transposition}), i.e., it must be a permutation with only 1-cycles and 2-cycles. 
%Physically, this corresponds to a transformation that only exchanges sublattices within pairs (or not at all); $PT$ for example.

Hence, the condition becomes
\begin{equation}
\left(\re^{-\ri\theta_h^\alpha} - \re^{-\ri\theta_h^\beta}\right) \delta^{\beta,h^{-1}(\alpha)} \overset{!}{=} 0 \qquad \forall \ \alpha,\beta.
\end{equation}
\begin{itemize}
\item For $h^{-1}(\alpha) = \alpha$ (a 1-cycle), there is no constraint on $\re^{-\ri\theta_h^\alpha}$.
\item For $h^{-1}(\alpha) = \beta$ (and $h^{-1}(\beta) = \alpha$) (a 2-cycle), $\re^{-\ri\theta_h^\alpha} = \re^{-\ri\theta_h^\beta}$.
\end{itemize}
Hence, if $V_h$ is symmetric, the permutation is a transposition (i.e., made up of 1-cycles and 2-cycles) and the phases $\re^{-\ri\theta_h^\alpha}$ are the same within each cycle of the permutation.

\paragraph{Sufficient condition}
Assuming the permutation is a transposition and the phases $\re^{-\ri\theta_h^\alpha}$ are the same within each cycle of the permutation, it is straightforward to see that $V_h$ is symmetric.

\subsection{Examples}

\subsubsection{Magnetic inversion symmetry}
We consider magnetic inversion symmetry, i.e., $h = \{P | \btau \} T$. In this case, $(\det R_h^\rs) R_h^\rs = (\det P) P = \idmat$.

Consider a pair of sublattices $\alpha$ and $\alpha'$ related by $h$. We can always choose
\begin{align}
	&\begin{cases}
		R_\alpha &= R\big(\eta_\alpha \hat{m}_\alpha\big),
		\\
		R_{\alpha'} &= R\big((\eta_\alpha + \pi) \hat{m}_\alpha\big),
	\end{cases}
	&
	&\hat{m}_\alpha \perp \hat{z},
\end{align}
where $\eta_\alpha$ and $\hat{m}_\alpha$ are an appropriate angle and unit vector, respectively. 
In the local frames, the transformation on the spins are
\begin{subequations}
\begin{align}
\tQ_h^\alpha &= -R_\alpha^\transpose R_{\alpha'} =  -R\left(\pi \hat{m}_\alpha\right),
\\
\tQ_h^{\alpha'} &= -R_{\alpha'}^\transpose R_{\alpha} =  -R\left(-\pi \hat{m}_\alpha\right) = -R\left(\pi \hat{m}_\alpha\right),
\end{align}
\end{subequations}
where the last equality holds because SO(3) rotation matrices have period $2\pi$.

The fact that $\tQ_h^\alpha = \tQ_h^{\alpha'}$ implies that $\theta_h^{\alpha} = \theta_h^{\alpha'}$. Hence, because the sublattice permutation is made up of 2-cycles and the phases $\theta_h^{\alpha}$ are the same within each cycle, the matrix $V_h$ is symmetric based on \app{\ref{sec:symmetric-conditions}}.

Hence, the constraint on the Hamiltonian is
\begin{equation}
	H_{\bk} = V_h^\dagger H_{\bk}^* V_h, \qquad V_h = V_h^\transpose.
\end{equation}

\subsubsection{Time reversal with spin rotation} \label{sec:TR+spinrot}

For coplanar spins, time reversal with spin rotation yields a time-reversal-like constraint on the magnon Hamiltonian. Specifically, if we consider spins in the $y$-$z$ plane, $h = \{\idmat ;R(\pi\hat{x}) | 0 \} T$ and the rotation matrices can be chosen to be
\begin{equation}
R_\alpha = R(\eta_\alpha\hat{x}) = \begin{bmatrix}
	1&&
	\\
	&\cos \eta_\alpha & -\sin\eta_\alpha 
	\\ 
	&\sin\eta_\alpha & \cos\eta_\alpha
\end{bmatrix}.
\end{equation}
With this choice, the local transformation on the spins is
\begin{subequations}
	\begin{align}
		\tQ_h^\alpha &= - R(\eta_\alpha\hat{x})^\transpose R(\pi\hat{x}) R(\eta_\alpha\hat{x})
		\\
		&= \left[\begin{array}{c|c}
			-\tau^3&
			\\ \hline
			&1
		\end{array}\right].
	\end{align}
\end{subequations}
After transforming to the basis of ladder operators, with this specific choice of $R_\alpha$, we find $\theta_h^\alpha = \pi$ for all $\alpha$ and
\begin{equation}
H_{\bk} = H_{-\bk}^*.
\end{equation}
A different choice of $\{R_\alpha\}$ would have yielded different phases $\theta_h^\alpha$, and a more complicated matrix $V_h$.

\subsubsection{Effective magnetic inversion symmetry}
For coplanar spins with separate inversion and time-reversal+spin-rotation symmetries, one can have an effective magnetic inversion symmetry: $h = \{P;R_h^\rs | \btau\} T$, with any appropriate $R_h^\rs$.

For example, for spins in the $y$-$z$ plane, $h = \{P; R(\pi\hat{x}) | \btau\} T$. Let $\alpha$ and $\alpha'$ denote a pair of sublattices related by $P$. These two sublattices have the same classical direction, so we can choose
\begin{equation}
R_\alpha = R_{\alpha'} = \begin{bmatrix}
	1&&
	\\
	&\cos \eta_\alpha & -\sin\eta_\alpha 
	\\ 
	&\sin\eta_\alpha & \cos\eta_\alpha
\end{bmatrix}.
\end{equation}
This choice gives us the exact same result as in \app{\ref{sec:TR+spinrot}}: $\theta_h^\alpha = \theta_h^{\alpha'} = \pi$; more generally, we will find $\theta_h^\alpha = \theta_h^{\alpha'}$. Therefore,
\begin{equation}
H_{\bk} = V_h^\dagger H_{\bk}^* V_h, \qquad V_h = V_h^\transpose
\end{equation}
because the permutations are all 2-cycles with the same phases within each cycle.

\subsubsection{Translation with spin rotation} \label{subsubsec:translation-spinrot}
Suppose that translation followed by spin rotation is an element of the spin space group; for example, $g = \{\idmat; R(\pi\hat{z}) | \btau\}$. Since this transformation has no spatial O(3) part, it acts locally in momentum space, and we get
\begin{equation}
H_{\bk} = V_g^\dagger H_{\bk} V_g, \qquad V_g \text{ unitary}.
\end{equation}
But this is a unitary symmetry of the $H_{\bk}$ \emph{at a fixed momentum}. This means that $H_{\bk}$ can be block-diagonalized, with the different blocks corresponding to different eigenvalues of $V_g$.

What does this additional structure correspond to, and why in this case does Fourier transforming leave remaining local unitary symmetries? Suppose $g^2$ is a primitive translation of the magnetic lattice. Then, one way of understanding this is to note that one can define modified translation operators (containing a global spin rotation by $\pi$ about $\hat{z}$) whose eigenvalues span a Brillouin zone twice as large as the magnetic Brillouin zone. This is simply an alternate way of labeling the states, as it halves the bands and doubles the number of momenta.

\subsubsection{Axial spin-rotation symmetry} \label{subsubsec:spinrot}
Consider a global spin-rotation symmetry $g$ about an axis; we can take it to be the $\hat{z}$ axis, in which case $g = \{\idmat; R(\theta \hat{z}) | 0\}$ for any $\theta$. This implies the classical spin order is collinear, with spins pointing either in the $\hat{z}$ or $-\hat{z}$ directions.

It is easy to see that $\tQ_g^\alpha = R(\theta\hat{z})$ and $\theta_g^\alpha = \theta$ if the spin at sublattice $\alpha$ points in $+\hat{z}$ direction, whereas $\tQ_g^\alpha = R(-\theta\hat{z})$ and $\theta_g^\alpha = -\theta$ if the spin at sublattice $\alpha$ points in $-\hat{z}$ direction. 

Hence, the associated matrix $V_g$ is diagonal with eigenvalues of $\re^{\ri\theta}$ and $\re^{-\ri\theta}$. This implies that the coefficient matrix $H_\bk$ is block diagonal, with the blocks corresponding to $S_z$-conserving sectors: one block contains annihilation operators for $+\hat{z}$ spins and creation operators for $-\hat{z}$ spins, and vice versa for the other block.

\subsection{Magnetic inversion and reality condition}

We show that in the appropriate basis, the constraint imposed by magnetic inversion symmetry on the Bloch Hamiltonian $H_{\bk}$ is that it is purely real for each $\bk$. 
This is expected for integer-spin excitations like magnons~\cite{sakurai}; we nonetheless discuss how this arises for completeness.

Consider any (effective) magnetic inversion symmetry $h = \{P; Q^\rs | \btau\} T$, where $Q^\rs$ is some O(3) operation and $\btau$ is some spatial translation. 
We have seen that the constraint on $H_{\bk}$ is local in momentum space, $H_{\bk} = V_h^\dagger H_{\bk}^* V_h$, with $V_h$ symmetric.
%\deleted{This property of $V$ is specific to magnetic inversion symmetries---see Appendix~\ref{app:magnon_symmetries}.}
Since $V_h = V_h^\rp \oplus V_h^\rh$, this implies $V_h^\rp$ and $V_h^\rh = (V_h^\rp)^*$ are symmetric. 
The symmetric unitary matrix $V_h^\rp$ can be factorized as $V_h^\rp = \ell^\transpose \ell$ using Autonne-Takagi decomposition, where $\ell$ is unitary. 
We then find $V_h = L^\transpose L$, where $L \coloneqq \ell \oplus \ell^*$ is unitary, paraunitary, and particle-hole symmetric.

In the new basis defined by $\Phi_{\bk} = L^\dagger \Phi_{\bk}'$ the transformation $\aui_h \Phi_{\bk} \aui_h^{-1} = \re^{\ri \bk \cdot \btau} V_h \Phi_{\bk}$ becomes
\begin{equation}
	\begin{split}
		\aui_h L^\dagger \Phi_{\bk}' \aui_h^{-1} &= \re^{\ri \bk \cdot \btau} V_h L^\dagger \Phi_{\bk}'
		\\
		\Leftrightarrow \quad \aui_h \Phi_{\bk}' \aui_h^{-1} &= \re^{\ri \bk \cdot \btau} L^* V_h L^\dagger \Phi_{\bk}'.
	\end{split}
\end{equation}
Therefore,
\begin{equation}
	V_h' = L^* V_h L^\dagger = L^* L^\transpose L L^\dagger = \mathds{1},
\end{equation}
and the constraint due to magnetic inversion symmetry reads as $H_{\bk}' = \left(H_{\bk}'\right)^*$. 

Hence, we have shown that there exists a ($\bk$-independent) change of basis such that in that basis, the constraint due to magnetic inversion is simply that $H_{\bk}'$ is purely real.

\bibliography{references}

%apsrev4-2.bst 2019-01-14 (MD) hand-edited version of apsrev4-1.bst
%Control: key (0)
%Control: author (8) initials jnrlst
%Control: editor formatted (1) identically to author
%Control: production of article title (0) allowed
%Control: page (0) single
%Control: year (1) truncated
%Control: production of eprint (0) enabled
\begin{thebibliography}{91}%
\makeatletter
\providecommand \@ifxundefined [1]{%
 \@ifx{#1\undefined}
}%
\providecommand \@ifnum [1]{%
 \ifnum #1\expandafter \@firstoftwo
 \else \expandafter \@secondoftwo
 \fi
}%
\providecommand \@ifx [1]{%
 \ifx #1\expandafter \@firstoftwo
 \else \expandafter \@secondoftwo
 \fi
}%
\providecommand \natexlab [1]{#1}%
\providecommand \enquote  [1]{``#1''}%
\providecommand \bibnamefont  [1]{#1}%
\providecommand \bibfnamefont [1]{#1}%
\providecommand \citenamefont [1]{#1}%
\providecommand \href@noop [0]{\@secondoftwo}%
\providecommand \href [0]{\begingroup \@sanitize@url \@href}%
\providecommand \@href[1]{\@@startlink{#1}\@@href}%
\providecommand \@@href[1]{\endgroup#1\@@endlink}%
\providecommand \@sanitize@url [0]{\catcode `\\12\catcode `\$12\catcode
  `\&12\catcode `\#12\catcode `\^12\catcode `\_12\catcode `\%12\relax}%
\providecommand \@@startlink[1]{}%
\providecommand \@@endlink[0]{}%
\providecommand \url  [0]{\begingroup\@sanitize@url \@url }%
\providecommand \@url [1]{\endgroup\@href {#1}{\urlprefix }}%
\providecommand \urlprefix  [0]{URL }%
\providecommand \Eprint [0]{\href }%
\providecommand \doibase [0]{https://doi.org/}%
\providecommand \selectlanguage [0]{\@gobble}%
\providecommand \bibinfo  [0]{\@secondoftwo}%
\providecommand \bibfield  [0]{\@secondoftwo}%
\providecommand \translation [1]{[#1]}%
\providecommand \BibitemOpen [0]{}%
\providecommand \bibitemStop [0]{}%
\providecommand \bibitemNoStop [0]{.\EOS\space}%
\providecommand \EOS [0]{\spacefactor3000\relax}%
\providecommand \BibitemShut  [1]{\csname bibitem#1\endcsname}%
\let\auto@bib@innerbib\@empty
%</preamble>
\bibitem [{\citenamefont {Bagarello}\ \emph {et~al.}(2015)\citenamefont
  {Bagarello}, \citenamefont {Gazeau}, \citenamefont {Szafraniec},\ and\
  \citenamefont {Znojil}}]{NonSelfadjoint2015}%
  \BibitemOpen
  \bibinfo {editor} {\bibfnamefont {F.}~\bibnamefont {Bagarello}}, \bibinfo
  {editor} {\bibfnamefont {J.-P.}\ \bibnamefont {Gazeau}}, \bibinfo {editor}
  {\bibfnamefont {F.~H.}\ \bibnamefont {Szafraniec}},\ and\ \bibinfo {editor}
  {\bibfnamefont {M.}~\bibnamefont {Znojil}},\ eds.,\ \href@noop {} {\emph
  {\bibinfo {title} {Non‐Selfadjoint Operators in Quantum Physics}}}\
  (\bibinfo  {publisher} {John Wiley \& Sons, Ltd},\ \bibinfo {year}
  {2015})\BibitemShut {NoStop}%
\bibitem [{\citenamefont {Colpa}(1978)}]{Colpa1978Diagonalization}%
  \BibitemOpen
  \bibfield  {author} {\bibinfo {author} {\bibfnamefont {J.}~\bibnamefont
  {Colpa}},\ }\bibfield  {title} {\bibinfo {title} {Diagonalization of the
  quadratic boson {Hamiltonian}},\ }\href
  {https://doi.org/10.1016/0378-4371(78)90160-7} {\bibfield  {journal}
  {\bibinfo  {journal} {Physica A: Statistical Mechanics and its Applications}\
  }\textbf {\bibinfo {volume} {93}},\ \bibinfo {pages} {327 } (\bibinfo {year}
  {1978})}\BibitemShut {NoStop}%
\bibitem [{\citenamefont {Blaizot}\ and\ \citenamefont
  {Ripka}(1986)}]{BlaizotRipka}%
  \BibitemOpen
  \bibfield  {author} {\bibinfo {author} {\bibfnamefont {J.-P.}\ \bibnamefont
  {Blaizot}}\ and\ \bibinfo {author} {\bibfnamefont {G.}~\bibnamefont
  {Ripka}},\ }\href@noop {} {\emph {\bibinfo {title} {Quantum theory of finite
  systems}}}\ (\bibinfo  {publisher} {The MIT Press},\ \bibinfo {address}
  {Cambridge, MA},\ \bibinfo {year} {1986})\BibitemShut {NoStop}%
\bibitem [{\citenamefont {{Shivam}}\ \emph {et~al.}(2017)\citenamefont
  {{Shivam}}, \citenamefont {{Coldea}}, \citenamefont {{Moessner}},\ and\
  \citenamefont {{McClarty}}}]{Shivam2017Neutron}%
  \BibitemOpen
  \bibfield  {author} {\bibinfo {author} {\bibfnamefont {S.}~\bibnamefont
  {{Shivam}}}, \bibinfo {author} {\bibfnamefont {R.}~\bibnamefont {{Coldea}}},
  \bibinfo {author} {\bibfnamefont {R.}~\bibnamefont {{Moessner}}},\ and\
  \bibinfo {author} {\bibfnamefont {P.}~\bibnamefont {{McClarty}}},\ }\bibfield
   {title} {\bibinfo {title} {{Neutron Scattering Signatures of Magnon {Weyl}
  Points}},\ }\href@noop {} {\bibfield  {journal} {\bibinfo  {journal} {arXiv
  e-prints}\ ,\ \bibinfo {eid} {arXiv:1712.08535}} (\bibinfo {year} {2017})},\
  \Eprint {https://arxiv.org/abs/1712.08535} {arXiv:1712.08535
  [cond-mat.str-el]} \BibitemShut {NoStop}%
\bibitem [{\citenamefont {McClarty}\ \emph {et~al.}(2018)\citenamefont
  {McClarty}, \citenamefont {Dong}, \citenamefont {Gohlke}, \citenamefont
  {Rau}, \citenamefont {Pollmann}, \citenamefont {Moessner},\ and\
  \citenamefont {Penc}}]{McClarty2018Topological}%
  \BibitemOpen
  \bibfield  {author} {\bibinfo {author} {\bibfnamefont {P.~A.}\ \bibnamefont
  {McClarty}}, \bibinfo {author} {\bibfnamefont {X.-Y.}\ \bibnamefont {Dong}},
  \bibinfo {author} {\bibfnamefont {M.}~\bibnamefont {Gohlke}}, \bibinfo
  {author} {\bibfnamefont {J.~G.}\ \bibnamefont {Rau}}, \bibinfo {author}
  {\bibfnamefont {F.}~\bibnamefont {Pollmann}}, \bibinfo {author}
  {\bibfnamefont {R.}~\bibnamefont {Moessner}},\ and\ \bibinfo {author}
  {\bibfnamefont {K.}~\bibnamefont {Penc}},\ }\bibfield  {title} {\bibinfo
  {title} {Topological magnons in {Kitaev} magnets at high fields},\ }\href
  {https://doi.org/10.1103/PhysRevB.98.060404} {\bibfield  {journal} {\bibinfo
  {journal} {Phys. Rev. B}\ }\textbf {\bibinfo {volume} {98}},\ \bibinfo
  {pages} {060404(R)} (\bibinfo {year} {2018})}\BibitemShut {NoStop}%
\bibitem [{\citenamefont {{Lu}}\ and\ \citenamefont
  {{Lu}}(2018)}]{lu2018magnon}%
  \BibitemOpen
  \bibfield  {author} {\bibinfo {author} {\bibfnamefont {F.}~\bibnamefont
  {{Lu}}}\ and\ \bibinfo {author} {\bibfnamefont {Y.-M.}\ \bibnamefont
  {{Lu}}},\ }\bibfield  {title} {\bibinfo {title} {{Magnon band topology in
  spin-orbital coupled magnets: classification and application to
  $\alpha$-RuCl$_3$}},\ }\href@noop {} {\bibfield  {journal} {\bibinfo
  {journal} {arXiv e-prints}\ ,\ \bibinfo {eid} {arXiv:1807.05232}} (\bibinfo
  {year} {2018})},\ \Eprint {https://arxiv.org/abs/1807.05232}
  {arXiv:1807.05232 [cond-mat.str-el]} \BibitemShut {NoStop}%
\bibitem [{\citenamefont {Gong}\ \emph {et~al.}(2018)\citenamefont {Gong},
  \citenamefont {Ashida}, \citenamefont {Kawabata}, \citenamefont {Takasan},
  \citenamefont {Higashikawa},\ and\ \citenamefont
  {Ueda}}]{Gong2018NonHermitian}%
  \BibitemOpen
  \bibfield  {author} {\bibinfo {author} {\bibfnamefont {Z.}~\bibnamefont
  {Gong}}, \bibinfo {author} {\bibfnamefont {Y.}~\bibnamefont {Ashida}},
  \bibinfo {author} {\bibfnamefont {K.}~\bibnamefont {Kawabata}}, \bibinfo
  {author} {\bibfnamefont {K.}~\bibnamefont {Takasan}}, \bibinfo {author}
  {\bibfnamefont {S.}~\bibnamefont {Higashikawa}},\ and\ \bibinfo {author}
  {\bibfnamefont {M.}~\bibnamefont {Ueda}},\ }\bibfield  {title} {\bibinfo
  {title} {Topological phases of {Non-Hermitian} systems},\ }\href
  {https://doi.org/10.1103/PhysRevX.8.031079} {\bibfield  {journal} {\bibinfo
  {journal} {Phys. Rev. X}\ }\textbf {\bibinfo {volume} {8}},\ \bibinfo {pages}
  {031079} (\bibinfo {year} {2018})}\BibitemShut {NoStop}%
\bibitem [{\citenamefont {Kawabata}\ \emph {et~al.}(2019)\citenamefont
  {Kawabata}, \citenamefont {Shiozaki}, \citenamefont {Ueda},\ and\
  \citenamefont {Sato}}]{Kawabata2019NonHermitian}%
  \BibitemOpen
  \bibfield  {author} {\bibinfo {author} {\bibfnamefont {K.}~\bibnamefont
  {Kawabata}}, \bibinfo {author} {\bibfnamefont {K.}~\bibnamefont {Shiozaki}},
  \bibinfo {author} {\bibfnamefont {M.}~\bibnamefont {Ueda}},\ and\ \bibinfo
  {author} {\bibfnamefont {M.}~\bibnamefont {Sato}},\ }\bibfield  {title}
  {\bibinfo {title} {Symmetry and topology in non-{Hermitian} physics},\ }\href
  {https://doi.org/10.1103/PhysRevX.9.041015} {\bibfield  {journal} {\bibinfo
  {journal} {Phys. Rev. X}\ }\textbf {\bibinfo {volume} {9}},\ \bibinfo {pages}
  {041015} (\bibinfo {year} {2019})}\BibitemShut {NoStop}%
\bibitem [{\citenamefont {McClarty}\ and\ \citenamefont
  {Rau}(2019)}]{McClarty2019NonHermitian}%
  \BibitemOpen
  \bibfield  {author} {\bibinfo {author} {\bibfnamefont {P.~A.}\ \bibnamefont
  {McClarty}}\ and\ \bibinfo {author} {\bibfnamefont {J.~G.}\ \bibnamefont
  {Rau}},\ }\bibfield  {title} {\bibinfo {title} {{Non-Hermitian} topology of
  spontaneous magnon decay},\ }\href
  {https://doi.org/10.1103/PhysRevB.100.100405} {\bibfield  {journal} {\bibinfo
   {journal} {Phys. Rev. B}\ }\textbf {\bibinfo {volume} {100}},\ \bibinfo
  {pages} {100405(R)} (\bibinfo {year} {2019})}\BibitemShut {NoStop}%
\bibitem [{\citenamefont {Flynn}\ \emph
  {et~al.}(2020{\natexlab{a}})\citenamefont {Flynn}, \citenamefont {Cobanera},\
  and\ \citenamefont {Viola}}]{Flynn2020Deconstructing}%
  \BibitemOpen
  \bibfield  {author} {\bibinfo {author} {\bibfnamefont {V.~P.}\ \bibnamefont
  {Flynn}}, \bibinfo {author} {\bibfnamefont {E.}~\bibnamefont {Cobanera}},\
  and\ \bibinfo {author} {\bibfnamefont {L.}~\bibnamefont {Viola}},\ }\bibfield
   {title} {\bibinfo {title} {Deconstructing effective non-{Hermitian} dynamics
  in quadratic bosonic {Hamiltonians}},\ }\href
  {https://doi.org/10.1088/1367-2630/ab9e87} {\bibfield  {journal} {\bibinfo
  {journal} {New Journal of Physics}\ }\textbf {\bibinfo {volume} {22}},\
  \bibinfo {pages} {083004} (\bibinfo {year} {2020}{\natexlab{a}})}\BibitemShut
  {NoStop}%
\bibitem [{\citenamefont {Xu}\ \emph {et~al.}(2020)\citenamefont {Xu},
  \citenamefont {Flynn}, \citenamefont {Alase}, \citenamefont {Cobanera},
  \citenamefont {Viola},\ and\ \citenamefont {Ortiz}}]{Xu2020Squaring}%
  \BibitemOpen
  \bibfield  {author} {\bibinfo {author} {\bibfnamefont {Q.-R.}\ \bibnamefont
  {Xu}}, \bibinfo {author} {\bibfnamefont {V.~P.}\ \bibnamefont {Flynn}},
  \bibinfo {author} {\bibfnamefont {A.}~\bibnamefont {Alase}}, \bibinfo
  {author} {\bibfnamefont {E.}~\bibnamefont {Cobanera}}, \bibinfo {author}
  {\bibfnamefont {L.}~\bibnamefont {Viola}},\ and\ \bibinfo {author}
  {\bibfnamefont {G.}~\bibnamefont {Ortiz}},\ }\bibfield  {title} {\bibinfo
  {title} {Squaring the fermion: The threefold way and the fate of zero
  modes},\ }\href {https://doi.org/10.1103/PhysRevB.102.125127} {\bibfield
  {journal} {\bibinfo  {journal} {Phys. Rev. B}\ }\textbf {\bibinfo {volume}
  {102}},\ \bibinfo {pages} {125127} (\bibinfo {year} {2020})}\BibitemShut
  {NoStop}%
\bibitem [{\citenamefont {Kumar}\ \emph {et~al.}(2020)\citenamefont {Kumar},
  \citenamefont {Herbut},\ and\ \citenamefont {Ganesh}}]{Kumar2020Dirac}%
  \BibitemOpen
  \bibfield  {author} {\bibinfo {author} {\bibfnamefont {P.~S.}\ \bibnamefont
  {Kumar}}, \bibinfo {author} {\bibfnamefont {I.~F.}\ \bibnamefont {Herbut}},\
  and\ \bibinfo {author} {\bibfnamefont {R.}~\bibnamefont {Ganesh}},\
  }\bibfield  {title} {\bibinfo {title} {{Dirac} {Hamiltonians} for bosonic
  spectra},\ }\href {https://doi.org/10.1103/PhysRevResearch.2.033035}
  {\bibfield  {journal} {\bibinfo  {journal} {Phys. Rev. Research}\ }\textbf
  {\bibinfo {volume} {2}},\ \bibinfo {pages} {033035} (\bibinfo {year}
  {2020})}\BibitemShut {NoStop}%
\bibitem [{\citenamefont {Corticelli}\ \emph {et~al.}(2022)\citenamefont
  {Corticelli}, \citenamefont {Moessner},\ and\ \citenamefont
  {McClarty}}]{McClarty2021SpinSpaceGroups}%
  \BibitemOpen
  \bibfield  {author} {\bibinfo {author} {\bibfnamefont {A.}~\bibnamefont
  {Corticelli}}, \bibinfo {author} {\bibfnamefont {R.}~\bibnamefont
  {Moessner}},\ and\ \bibinfo {author} {\bibfnamefont {P.~A.}\ \bibnamefont
  {McClarty}},\ }\bibfield  {title} {\bibinfo {title} {Spin-space groups and
  magnon band topology},\ }\href {https://doi.org/10.1103/PhysRevB.105.064430}
  {\bibfield  {journal} {\bibinfo  {journal} {Phys. Rev. B}\ }\textbf {\bibinfo
  {volume} {105}},\ \bibinfo {pages} {064430} (\bibinfo {year}
  {2022})}\BibitemShut {NoStop}%
\bibitem [{\citenamefont {Bender}\ \emph {et~al.}(1999)\citenamefont {Bender},
  \citenamefont {Boettcher},\ and\ \citenamefont {Meisinger}}]{bender1999pt}%
  \BibitemOpen
  \bibfield  {author} {\bibinfo {author} {\bibfnamefont {C.~M.}\ \bibnamefont
  {Bender}}, \bibinfo {author} {\bibfnamefont {S.}~\bibnamefont {Boettcher}},\
  and\ \bibinfo {author} {\bibfnamefont {P.~N.}\ \bibnamefont {Meisinger}},\
  }\bibfield  {title} {\bibinfo {title} {{$PT$}-symmetric quantum mechanics},\
  }\href {https://doi.org/10.1063/1.532860} {\bibfield  {journal} {\bibinfo
  {journal} {Journal of Mathematical Physics}\ }\textbf {\bibinfo {volume}
  {40}},\ \bibinfo {pages} {2201} (\bibinfo {year} {1999})}\BibitemShut
  {NoStop}%
\bibitem [{\citenamefont {Bender}(2007)}]{Bender_2007}%
  \BibitemOpen
  \bibfield  {author} {\bibinfo {author} {\bibfnamefont {C.~M.}\ \bibnamefont
  {Bender}},\ }\bibfield  {title} {\bibinfo {title} {Making sense of
  {non-Hermitian} {Hamiltonians}},\ }\href
  {https://doi.org/10.1088/0034-4885/70/6/r03} {\bibfield  {journal} {\bibinfo
  {journal} {Reports on Progress in Physics}\ }\textbf {\bibinfo {volume}
  {70}},\ \bibinfo {pages} {947} (\bibinfo {year} {2007})}\BibitemShut
  {NoStop}%
\bibitem [{\citenamefont
  {Mostafazadeh}(2002{\natexlab{a}})}]{Mostafazadeh2002PseudoI}%
  \BibitemOpen
  \bibfield  {author} {\bibinfo {author} {\bibfnamefont {A.}~\bibnamefont
  {Mostafazadeh}},\ }\bibfield  {title} {\bibinfo {title} {{Pseudo-Hermiticity}
  versus {$PT$} symmetry: The necessary condition for the reality of the
  spectrum of a {non-Hermitian} {Hamiltonian}},\ }\href
  {https://doi.org/10.1063/1.1418246} {\bibfield  {journal} {\bibinfo
  {journal} {Journal of Mathematical Physics}\ }\textbf {\bibinfo {volume}
  {43}},\ \bibinfo {pages} {205} (\bibinfo {year}
  {2002}{\natexlab{a}})}\BibitemShut {NoStop}%
\bibitem [{\citenamefont
  {Mostafazadeh}(2002{\natexlab{b}})}]{Mostafazadeh2002PseudoII}%
  \BibitemOpen
  \bibfield  {author} {\bibinfo {author} {\bibfnamefont {A.}~\bibnamefont
  {Mostafazadeh}},\ }\bibfield  {title} {\bibinfo {title} {{Pseudo-Hermiticity}
  versus {$PT$} symmetry. ii. a complete characterization of {non-Hermitian}
  {Hamiltonians} with a real spectrum},\ }\href
  {https://doi.org/10.1063/1.1461427} {\bibfield  {journal} {\bibinfo
  {journal} {Journal of Mathematical Physics}\ }\textbf {\bibinfo {volume}
  {43}},\ \bibinfo {pages} {2814} (\bibinfo {year}
  {2002}{\natexlab{b}})}\BibitemShut {NoStop}%
\bibitem [{\citenamefont
  {Mostafazadeh}(2002{\natexlab{c}})}]{Mostafazadeh2002PseudoIII}%
  \BibitemOpen
  \bibfield  {author} {\bibinfo {author} {\bibfnamefont {A.}~\bibnamefont
  {Mostafazadeh}},\ }\bibfield  {title} {\bibinfo {title} {{Pseudo-Hermiticity}
  versus {$PT$} symmetry iii: Equivalence of {pseudo-Hermiticity} and the
  presence of antilinear symmetries},\ }\href
  {https://doi.org/10.1063/1.1489072} {\bibfield  {journal} {\bibinfo
  {journal} {Journal of Mathematical Physics}\ }\textbf {\bibinfo {volume}
  {43}},\ \bibinfo {pages} {3944} (\bibinfo {year}
  {2002}{\natexlab{c}})}\BibitemShut {NoStop}%
\bibitem [{\citenamefont {Mostafazadeh}(2006)}]{mostafazadeh2006krein}%
  \BibitemOpen
  \bibfield  {author} {\bibinfo {author} {\bibfnamefont {A.}~\bibnamefont
  {Mostafazadeh}},\ }\bibfield  {title} {\bibinfo {title} {{Krein-space}
  formulation of $\mathcal{PT}$ symmetry, $\mathcal{CPT}$-inner products, and
  {pseudo-Hermiticity}},\ }\href {https://doi.org/10.1007/s10582-006-0388-8}
  {\bibfield  {journal} {\bibinfo  {journal} {Czechoslovak Journal of Physics}\
  }\textbf {\bibinfo {volume} {56}},\ \bibinfo {pages} {919} (\bibinfo {year}
  {2006})}\BibitemShut {NoStop}%
\bibitem [{\citenamefont {Tanaka}(2006{\natexlab{a}})}]{Tanaka2006Krein1}%
  \BibitemOpen
  \bibfield  {author} {\bibinfo {author} {\bibfnamefont {T.}~\bibnamefont
  {Tanaka}},\ }\bibfield  {title} {\bibinfo {title} {{$PT$}-symmetric quantum
  theory defined in a {Krein} space},\ }\href
  {https://doi.org/10.1088/0305-4470/39/22/l04} {\bibfield  {journal} {\bibinfo
   {journal} {Journal of Physics A: Mathematical and General}\ }\textbf
  {\bibinfo {volume} {39}},\ \bibinfo {pages} {L369} (\bibinfo {year}
  {2006}{\natexlab{a}})}\BibitemShut {NoStop}%
\bibitem [{\citenamefont {Tanaka}(2006{\natexlab{b}})}]{Tanaka2006Krein2}%
  \BibitemOpen
  \bibfield  {author} {\bibinfo {author} {\bibfnamefont {T.}~\bibnamefont
  {Tanaka}},\ }\bibfield  {title} {\bibinfo {title} {General aspects of
  {$PT$}-symmetric and {$P$}-self-adjoint quantum theory in a {Krein} space},\
  }\href {https://doi.org/10.1088/0305-4470/39/45/025} {\bibfield  {journal}
  {\bibinfo  {journal} {Journal of Physics A: Mathematical and General}\
  }\textbf {\bibinfo {volume} {39}},\ \bibinfo {pages} {14175} (\bibinfo {year}
  {2006}{\natexlab{b}})}\BibitemShut {NoStop}%
\bibitem [{\citenamefont {Albeverio}\ and\ \citenamefont
  {Kuzhel}(2015)}]{AlbeverioKuzhel2015}%
  \BibitemOpen
  \bibfield  {author} {\bibinfo {author} {\bibfnamefont {S.}~\bibnamefont
  {Albeverio}}\ and\ \bibinfo {author} {\bibfnamefont {S.}~\bibnamefont
  {Kuzhel}},\ }\bibinfo {title} {{$PT$}-symmetric operators in quantum
  mechanics: {Krein} spaces methods},\ in\ \href
  {https://doi.org/10.1002/9781118855300.ch6} {\emph {\bibinfo {booktitle}
  {Non‐Selfadjoint Operators in Quantum Physics}}}\ (\bibinfo  {publisher}
  {John Wiley \& Sons, Ltd},\ \bibinfo {year} {2015})\ Chap.~\bibinfo {chapter}
  {6}, pp.\ \bibinfo {pages} {293--344}\BibitemShut {NoStop}%
\bibitem [{\citenamefont {El-Ganainy}\ \emph {et~al.}(2018)\citenamefont
  {El-Ganainy}, \citenamefont {Makris}, \citenamefont {Khajavikhan},
  \citenamefont {Musslimani}, \citenamefont {Rotter},\ and\ \citenamefont
  {Christodoulides}}]{elganainy2018nonhermitian}%
  \BibitemOpen
  \bibfield  {author} {\bibinfo {author} {\bibfnamefont {R.}~\bibnamefont
  {El-Ganainy}}, \bibinfo {author} {\bibfnamefont {K.~G.}\ \bibnamefont
  {Makris}}, \bibinfo {author} {\bibfnamefont {M.}~\bibnamefont {Khajavikhan}},
  \bibinfo {author} {\bibfnamefont {Z.~H.}\ \bibnamefont {Musslimani}},
  \bibinfo {author} {\bibfnamefont {S.}~\bibnamefont {Rotter}},\ and\ \bibinfo
  {author} {\bibfnamefont {D.~N.}\ \bibnamefont {Christodoulides}},\ }\bibfield
   {title} {\bibinfo {title} {{Non-Hermitian} physics and {$PT$} symmetry},\
  }\href {https://doi.org/10.1038/nphys4323} {\bibfield  {journal} {\bibinfo
  {journal} {Nature Physics}\ }\textbf {\bibinfo {volume} {14}},\ \bibinfo
  {pages} {11} (\bibinfo {year} {2018})}\BibitemShut {NoStop}%
\bibitem [{\citenamefont {Flynn}\ \emph
  {et~al.}(2020{\natexlab{b}})\citenamefont {Flynn}, \citenamefont {Cobanera},\
  and\ \citenamefont {Viola}}]{Flynn2020Restoring}%
  \BibitemOpen
  \bibfield  {author} {\bibinfo {author} {\bibfnamefont {V.~P.}\ \bibnamefont
  {Flynn}}, \bibinfo {author} {\bibfnamefont {E.}~\bibnamefont {Cobanera}},\
  and\ \bibinfo {author} {\bibfnamefont {L.}~\bibnamefont {Viola}},\ }\bibfield
   {title} {\bibinfo {title} {Restoring number conservation in quadratic
  bosonic {Hamiltonians} with dualities},\ }\href
  {https://doi.org/10.1209/0295-5075/131/40006} {\bibfield  {journal} {\bibinfo
   {journal} {{EPL} (Europhysics Letters)}\ }\textbf {\bibinfo {volume}
  {131}},\ \bibinfo {pages} {40006} (\bibinfo {year}
  {2020}{\natexlab{b}})}\BibitemShut {NoStop}%
\bibitem [{\citenamefont {Lein}\ and\ \citenamefont
  {Sato}(2019)}]{Lein2019Krein}%
  \BibitemOpen
  \bibfield  {author} {\bibinfo {author} {\bibfnamefont {M.}~\bibnamefont
  {Lein}}\ and\ \bibinfo {author} {\bibfnamefont {K.}~\bibnamefont {Sato}},\
  }\bibfield  {title} {\bibinfo {title} {{Krein-Schr\"odinger} formalism of
  bosonic {Bogoliubov--de Gennes} and certain classical systems and their
  topological classification},\ }\href
  {https://doi.org/10.1103/PhysRevB.100.075414} {\bibfield  {journal} {\bibinfo
   {journal} {Phys. Rev. B}\ }\textbf {\bibinfo {volume} {100}},\ \bibinfo
  {pages} {075414} (\bibinfo {year} {2019})}\BibitemShut {NoStop}%
\bibitem [{\citenamefont {Bravyi}\ \emph {et~al.}(2011)\citenamefont {Bravyi},
  \citenamefont {DiVincenzo},\ and\ \citenamefont
  {Loss}}]{Bravyi2011Schrieffer}%
  \BibitemOpen
  \bibfield  {author} {\bibinfo {author} {\bibfnamefont {S.}~\bibnamefont
  {Bravyi}}, \bibinfo {author} {\bibfnamefont {D.~P.}\ \bibnamefont
  {DiVincenzo}},\ and\ \bibinfo {author} {\bibfnamefont {D.}~\bibnamefont
  {Loss}},\ }\bibfield  {title} {\bibinfo {title} {{Schrieffer–Wolff}
  transformation for quantum many-body systems},\ }\href
  {https://doi.org/10.1016/j.aop.2011.06.004} {\bibfield  {journal} {\bibinfo
  {journal} {Annals of Physics}\ }\textbf {\bibinfo {volume} {326}},\ \bibinfo
  {pages} {2793} (\bibinfo {year} {2011})}\BibitemShut {NoStop}%
\bibitem [{\citenamefont {Winkler}(2003)}]{Winkler2003Quasi}%
  \BibitemOpen
  \bibfield  {author} {\bibinfo {author} {\bibfnamefont {R.}~\bibnamefont
  {Winkler}},\ }\bibfield  {title} {\bibinfo {title} {Quasi-degenerate
  perturbation theory},\ }in\ \href
  {https://doi.org/10.1007/978-3-540-36616-4_12} {\emph {\bibinfo {booktitle}
  {Spin--Orbit Coupling Effects in Two-Dimensional Electron and Hole
  Systems}}}\ (\bibinfo  {publisher} {Springer Berlin Heidelberg},\ \bibinfo
  {address} {Berlin, Heidelberg},\ \bibinfo {year} {2003})\ pp.\ \bibinfo
  {pages} {201--206}\BibitemShut {NoStop}%
\bibitem [{\citenamefont {Kessler}(2012)}]{Kessler2012Generalized}%
  \BibitemOpen
  \bibfield  {author} {\bibinfo {author} {\bibfnamefont {E.~M.}\ \bibnamefont
  {Kessler}},\ }\bibfield  {title} {\bibinfo {title} {Generalized
  {Schrieffer-Wolff} formalism for dissipative systems},\ }\href
  {https://doi.org/10.1103/PhysRevA.86.012126} {\bibfield  {journal} {\bibinfo
  {journal} {Phys. Rev. A}\ }\textbf {\bibinfo {volume} {86}},\ \bibinfo
  {pages} {012126} (\bibinfo {year} {2012})}\BibitemShut {NoStop}%
\bibitem [{\citenamefont {Kessler}\ \emph {et~al.}(2012)\citenamefont
  {Kessler}, \citenamefont {Giedke}, \citenamefont {Imamoglu}, \citenamefont
  {Yelin}, \citenamefont {Lukin},\ and\ \citenamefont
  {Cirac}}]{Kessler2012Dissipative}%
  \BibitemOpen
  \bibfield  {author} {\bibinfo {author} {\bibfnamefont {E.~M.}\ \bibnamefont
  {Kessler}}, \bibinfo {author} {\bibfnamefont {G.}~\bibnamefont {Giedke}},
  \bibinfo {author} {\bibfnamefont {A.}~\bibnamefont {Imamoglu}}, \bibinfo
  {author} {\bibfnamefont {S.~F.}\ \bibnamefont {Yelin}}, \bibinfo {author}
  {\bibfnamefont {M.~D.}\ \bibnamefont {Lukin}},\ and\ \bibinfo {author}
  {\bibfnamefont {J.~I.}\ \bibnamefont {Cirac}},\ }\bibfield  {title} {\bibinfo
  {title} {Dissipative phase transition in a central spin system},\ }\href
  {https://doi.org/10.1103/PhysRevA.86.012116} {\bibfield  {journal} {\bibinfo
  {journal} {Phys. Rev. A}\ }\textbf {\bibinfo {volume} {86}},\ \bibinfo
  {pages} {012116} (\bibinfo {year} {2012})}\BibitemShut {NoStop}%
\bibitem [{\citenamefont {Shindou}\ \emph
  {et~al.}(2013{\natexlab{a}})\citenamefont {Shindou}, \citenamefont {Ohe},
  \citenamefont {Matsumoto}, \citenamefont {Murakami},\ and\ \citenamefont
  {Saitoh}}]{Shindou2013SpinWave}%
  \BibitemOpen
  \bibfield  {author} {\bibinfo {author} {\bibfnamefont {R.}~\bibnamefont
  {Shindou}}, \bibinfo {author} {\bibfnamefont {J.-i.}\ \bibnamefont {Ohe}},
  \bibinfo {author} {\bibfnamefont {R.}~\bibnamefont {Matsumoto}}, \bibinfo
  {author} {\bibfnamefont {S.}~\bibnamefont {Murakami}},\ and\ \bibinfo
  {author} {\bibfnamefont {E.}~\bibnamefont {Saitoh}},\ }\bibfield  {title}
  {\bibinfo {title} {Chiral spin-wave edge modes in dipolar magnetic thin
  films},\ }\href {https://doi.org/10.1103/PhysRevB.87.174402} {\bibfield
  {journal} {\bibinfo  {journal} {Phys. Rev. B}\ }\textbf {\bibinfo {volume}
  {87}},\ \bibinfo {pages} {174402} (\bibinfo {year}
  {2013}{\natexlab{a}})}\BibitemShut {NoStop}%
\bibitem [{\citenamefont {Shindou}\ and\ \citenamefont
  {Ohe}(2014)}]{Shindou2014Magnetostatic}%
  \BibitemOpen
  \bibfield  {author} {\bibinfo {author} {\bibfnamefont {R.}~\bibnamefont
  {Shindou}}\ and\ \bibinfo {author} {\bibfnamefont {J.-i.}\ \bibnamefont
  {Ohe}},\ }\bibfield  {title} {\bibinfo {title} {Magnetostatic wave analog of
  integer quantum hall state in patterned magnetic films},\ }\href
  {https://doi.org/10.1103/PhysRevB.89.054412} {\bibfield  {journal} {\bibinfo
  {journal} {Phys. Rev. B}\ }\textbf {\bibinfo {volume} {89}},\ \bibinfo
  {pages} {054412} (\bibinfo {year} {2014})}\BibitemShut {NoStop}%
\bibitem [{\citenamefont {Zhou}\ \emph {et~al.}(2020)\citenamefont {Zhou},
  \citenamefont {Wan},\ and\ \citenamefont {Xu}}]{Zhou2020Bosonic}%
  \BibitemOpen
  \bibfield  {author} {\bibinfo {author} {\bibfnamefont {Z.}~\bibnamefont
  {Zhou}}, \bibinfo {author} {\bibfnamefont {L.-L.}\ \bibnamefont {Wan}},\ and\
  \bibinfo {author} {\bibfnamefont {Z.-F.}\ \bibnamefont {Xu}},\ }\bibfield
  {title} {\bibinfo {title} {Topological classification of excitations in
  quadratic bosonic systems},\ }\href
  {https://doi.org/10.1088/1751-8121/abb92b} {\bibfield  {journal} {\bibinfo
  {journal} {J. Phys. A: Math. Theor.}\ }\textbf {\bibinfo {volume} {53}},\
  \bibinfo {pages} {425203} (\bibinfo {year} {2020})}\BibitemShut {NoStop}%
\bibitem [{\citenamefont {Wan}\ \emph {et~al.}(2021)\citenamefont {Wan},
  \citenamefont {Zhou},\ and\ \citenamefont {Xu}}]{Wan2021Squeezing}%
  \BibitemOpen
  \bibfield  {author} {\bibinfo {author} {\bibfnamefont {L.-L.}\ \bibnamefont
  {Wan}}, \bibinfo {author} {\bibfnamefont {Z.}~\bibnamefont {Zhou}},\ and\
  \bibinfo {author} {\bibfnamefont {Z.-F.}\ \bibnamefont {Xu}},\ }\bibfield
  {title} {\bibinfo {title} {Squeezing-induced topological gap opening on
  bosonic bogoliubov excitations},\ }\href
  {https://doi.org/10.1103/PhysRevA.103.013308} {\bibfield  {journal} {\bibinfo
   {journal} {Phys. Rev. A}\ }\textbf {\bibinfo {volume} {103}},\ \bibinfo
  {pages} {013308} (\bibinfo {year} {2021})}\BibitemShut {NoStop}%
\bibitem [{\citenamefont {Tiwari}\ and\ \citenamefont
  {Bzdu\ifmmode~\check{s}\else \v{s}\fi{}ek}(2020)}]{Tiwari2020NodalLine}%
  \BibitemOpen
  \bibfield  {author} {\bibinfo {author} {\bibfnamefont {A.}~\bibnamefont
  {Tiwari}}\ and\ \bibinfo {author} {\bibfnamefont {T.~c.~v.}\ \bibnamefont
  {Bzdu\ifmmode~\check{s}\else \v{s}\fi{}ek}},\ }\bibfield  {title} {\bibinfo
  {title} {Non-abelian topology of nodal-line rings in $\mathcal{PT}$-symmetric
  systems},\ }\href {https://doi.org/10.1103/PhysRevB.101.195130} {\bibfield
  {journal} {\bibinfo  {journal} {Phys. Rev. B}\ }\textbf {\bibinfo {volume}
  {101}},\ \bibinfo {pages} {195130} (\bibinfo {year} {2020})}\BibitemShut
  {NoStop}%
\bibitem [{\citenamefont {Li}\ \emph {et~al.}(2017)\citenamefont {Li},
  \citenamefont {Li}, \citenamefont {Hu}, \citenamefont {Li},\ and\
  \citenamefont {Fang}}]{Li2017Dirac}%
  \BibitemOpen
  \bibfield  {author} {\bibinfo {author} {\bibfnamefont {K.}~\bibnamefont
  {Li}}, \bibinfo {author} {\bibfnamefont {C.}~\bibnamefont {Li}}, \bibinfo
  {author} {\bibfnamefont {J.}~\bibnamefont {Hu}}, \bibinfo {author}
  {\bibfnamefont {Y.}~\bibnamefont {Li}},\ and\ \bibinfo {author}
  {\bibfnamefont {C.}~\bibnamefont {Fang}},\ }\bibfield  {title} {\bibinfo
  {title} {{Dirac} and nodal line magnons in three-dimensional
  antiferromagnets},\ }\href {https://doi.org/10.1103/PhysRevLett.119.247202}
  {\bibfield  {journal} {\bibinfo  {journal} {Phys. Rev. Lett.}\ }\textbf
  {\bibinfo {volume} {119}},\ \bibinfo {pages} {247202} (\bibinfo {year}
  {2017})}\BibitemShut {NoStop}%
\bibitem [{\citenamefont {Yao}\ \emph {et~al.}(2018)\citenamefont {Yao},
  \citenamefont {Li}, \citenamefont {Wang}, \citenamefont {Xue}, \citenamefont
  {Dan}, \citenamefont {Iida}, \citenamefont {Kamazawa}, \citenamefont {Li},
  \citenamefont {Fang},\ and\ \citenamefont {Li}}]{Yao2018Antiferromagnet}%
  \BibitemOpen
  \bibfield  {author} {\bibinfo {author} {\bibfnamefont {W.}~\bibnamefont
  {Yao}}, \bibinfo {author} {\bibfnamefont {C.}~\bibnamefont {Li}}, \bibinfo
  {author} {\bibfnamefont {L.}~\bibnamefont {Wang}}, \bibinfo {author}
  {\bibfnamefont {S.}~\bibnamefont {Xue}}, \bibinfo {author} {\bibfnamefont
  {Y.}~\bibnamefont {Dan}}, \bibinfo {author} {\bibfnamefont {K.}~\bibnamefont
  {Iida}}, \bibinfo {author} {\bibfnamefont {K.}~\bibnamefont {Kamazawa}},
  \bibinfo {author} {\bibfnamefont {K.}~\bibnamefont {Li}}, \bibinfo {author}
  {\bibfnamefont {C.}~\bibnamefont {Fang}},\ and\ \bibinfo {author}
  {\bibfnamefont {Y.}~\bibnamefont {Li}},\ }\bibfield  {title} {\bibinfo
  {title} {Topological spin excitations in a three-dimensional
  antiferromagnet},\ }\href {https://doi.org/10.1038/s41567-018-0213-x}
  {\bibfield  {journal} {\bibinfo  {journal} {Nature Physics}\ }\textbf
  {\bibinfo {volume} {14}},\ \bibinfo {pages} {1011} (\bibinfo {year}
  {2018})}\BibitemShut {NoStop}%
\bibitem [{\citenamefont {Bao}\ \emph {et~al.}(2018)\citenamefont {Bao},
  \citenamefont {Wang}, \citenamefont {Wang}, \citenamefont {Cai},
  \citenamefont {Li}, \citenamefont {Ma}, \citenamefont {Wang}, \citenamefont
  {Ran}, \citenamefont {Dong}, \citenamefont {Abernathy}, \citenamefont {Yu},
  \citenamefont {Wan}, \citenamefont {Li},\ and\ \citenamefont
  {Wen}}]{Bao2018Antiferromagnet}%
  \BibitemOpen
  \bibfield  {author} {\bibinfo {author} {\bibfnamefont {S.}~\bibnamefont
  {Bao}}, \bibinfo {author} {\bibfnamefont {J.}~\bibnamefont {Wang}}, \bibinfo
  {author} {\bibfnamefont {W.}~\bibnamefont {Wang}}, \bibinfo {author}
  {\bibfnamefont {Z.}~\bibnamefont {Cai}}, \bibinfo {author} {\bibfnamefont
  {S.}~\bibnamefont {Li}}, \bibinfo {author} {\bibfnamefont {Z.}~\bibnamefont
  {Ma}}, \bibinfo {author} {\bibfnamefont {D.}~\bibnamefont {Wang}}, \bibinfo
  {author} {\bibfnamefont {K.}~\bibnamefont {Ran}}, \bibinfo {author}
  {\bibfnamefont {Z.-Y.}\ \bibnamefont {Dong}}, \bibinfo {author}
  {\bibfnamefont {D.~L.}\ \bibnamefont {Abernathy}}, \bibinfo {author}
  {\bibfnamefont {S.-L.}\ \bibnamefont {Yu}}, \bibinfo {author} {\bibfnamefont
  {X.}~\bibnamefont {Wan}}, \bibinfo {author} {\bibfnamefont {J.-X.}\
  \bibnamefont {Li}},\ and\ \bibinfo {author} {\bibfnamefont {J.}~\bibnamefont
  {Wen}},\ }\bibfield  {title} {\bibinfo {title} {Discovery of coexisting dirac
  and triply degenerate magnons in a three-dimensional antiferromagnet},\
  }\href {https://doi.org/10.1038/s41467-018-05054-2} {\bibfield  {journal}
  {\bibinfo  {journal} {Nature Communications}\ }\textbf {\bibinfo {volume}
  {9}},\ \bibinfo {pages} {2591} (\bibinfo {year} {2018})}\BibitemShut
  {NoStop}%
\bibitem [{\citenamefont {Yuan}\ \emph {et~al.}(2020)\citenamefont {Yuan},
  \citenamefont {Khait}, \citenamefont {Shu}, \citenamefont {Chou},
  \citenamefont {Stone}, \citenamefont {Clancy}, \citenamefont {Paramekanti},\
  and\ \citenamefont {Kim}}]{Yuan2020}%
  \BibitemOpen
  \bibfield  {author} {\bibinfo {author} {\bibfnamefont {B.}~\bibnamefont
  {Yuan}}, \bibinfo {author} {\bibfnamefont {I.}~\bibnamefont {Khait}},
  \bibinfo {author} {\bibfnamefont {G.-J.}\ \bibnamefont {Shu}}, \bibinfo
  {author} {\bibfnamefont {F.~C.}\ \bibnamefont {Chou}}, \bibinfo {author}
  {\bibfnamefont {M.~B.}\ \bibnamefont {Stone}}, \bibinfo {author}
  {\bibfnamefont {J.~P.}\ \bibnamefont {Clancy}}, \bibinfo {author}
  {\bibfnamefont {A.}~\bibnamefont {Paramekanti}},\ and\ \bibinfo {author}
  {\bibfnamefont {Y.-J.}\ \bibnamefont {Kim}},\ }\bibfield  {title} {\bibinfo
  {title} {{Dirac} magnons in a honeycomb lattice quantum $\mathit{XY}$ magnet
  ${\mathrm{cotio}}_{3}$},\ }\href {https://doi.org/10.1103/PhysRevX.10.011062}
  {\bibfield  {journal} {\bibinfo  {journal} {Phys. Rev. X}\ }\textbf {\bibinfo
  {volume} {10}},\ \bibinfo {pages} {011062} (\bibinfo {year}
  {2020})}\BibitemShut {NoStop}%
\bibitem [{\citenamefont {Elliot}\ \emph {et~al.}(2021)\citenamefont {Elliot},
  \citenamefont {McClarty}, \citenamefont {Prabhakaran}, \citenamefont
  {Johnson}, \citenamefont {Walker}, \citenamefont {Manuel},\ and\
  \citenamefont {Coldea}}]{Elliot2020Visualization}%
  \BibitemOpen
  \bibfield  {author} {\bibinfo {author} {\bibfnamefont {M.}~\bibnamefont
  {Elliot}}, \bibinfo {author} {\bibfnamefont {P.~A.}\ \bibnamefont
  {McClarty}}, \bibinfo {author} {\bibfnamefont {D.}~\bibnamefont
  {Prabhakaran}}, \bibinfo {author} {\bibfnamefont {R.~D.}\ \bibnamefont
  {Johnson}}, \bibinfo {author} {\bibfnamefont {H.~C.}\ \bibnamefont {Walker}},
  \bibinfo {author} {\bibfnamefont {P.}~\bibnamefont {Manuel}},\ and\ \bibinfo
  {author} {\bibfnamefont {R.}~\bibnamefont {Coldea}},\ }\bibfield  {title}
  {\bibinfo {title} {Order-by-disorder from bond-dependent exchange and
  intensity signature of nodal quasiparticles in a honeycomb cobaltate},\
  }\href {https://doi.org/10.1038/s41467-021-23851-0} {\bibfield  {journal}
  {\bibinfo  {journal} {Nature Communications}\ }\textbf {\bibinfo {volume}
  {12}},\ \bibinfo {pages} {3936} (\bibinfo {year} {2021})}\BibitemShut
  {NoStop}%
\bibitem [{\citenamefont {{Xiao}}(2009)}]{Xiao2009Theory}%
  \BibitemOpen
  \bibfield  {author} {\bibinfo {author} {\bibfnamefont {M.-W.}\ \bibnamefont
  {{Xiao}}},\ }\bibfield  {title} {\bibinfo {title} {{Theory of transformation
  for the diagonalization of quadratic {Hamiltonians}}},\ }\href@noop {}
  {\bibfield  {journal} {\bibinfo  {journal} {arXiv e-prints}\ ,\ \bibinfo
  {eid} {arXiv:0908.0787}} (\bibinfo {year} {2009})},\ \Eprint
  {https://arxiv.org/abs/0908.0787} {arXiv:0908.0787 [math-ph]} \BibitemShut
  {NoStop}%
\bibitem [{\citenamefont {Kawaguchi}\ and\ \citenamefont
  {Ueda}(2012)}]{Kawaguchi2012Spinor}%
  \BibitemOpen
  \bibfield  {author} {\bibinfo {author} {\bibfnamefont {Y.}~\bibnamefont
  {Kawaguchi}}\ and\ \bibinfo {author} {\bibfnamefont {M.}~\bibnamefont
  {Ueda}},\ }\bibfield  {title} {\bibinfo {title} {Spinor {Bose–Einstein}
  condensates},\ }\href {https://doi.org/10.1016/j.physrep.2012.07.005}
  {\bibfield  {journal} {\bibinfo  {journal} {Physics Reports}\ }\textbf
  {\bibinfo {volume} {520}},\ \bibinfo {pages} {253 } (\bibinfo {year}
  {2012})}\BibitemShut {NoStop}%
\bibitem [{\citenamefont {Choi}\ \emph {et~al.}(2019)\citenamefont {Choi},
  \citenamefont {Mizoguchi},\ and\ \citenamefont
  {Kim}}]{Choi2019Nonsymmorphic}%
  \BibitemOpen
  \bibfield  {author} {\bibinfo {author} {\bibfnamefont {W.}~\bibnamefont
  {Choi}}, \bibinfo {author} {\bibfnamefont {T.}~\bibnamefont {Mizoguchi}},\
  and\ \bibinfo {author} {\bibfnamefont {Y.~B.}\ \bibnamefont {Kim}},\
  }\bibfield  {title} {\bibinfo {title} {Nonsymmorphic-symmetry-protected
  topological magnons in three-dimensional {Kitaev} materials},\ }\href
  {https://doi.org/10.1103/PhysRevLett.123.227202} {\bibfield  {journal}
  {\bibinfo  {journal} {Phys. Rev. Lett.}\ }\textbf {\bibinfo {volume} {123}},\
  \bibinfo {pages} {227202} (\bibinfo {year} {2019})}\BibitemShut {NoStop}%
\bibitem [{Note1()}]{Note1}%
  \BibitemOpen
  \bibinfo {note} {In general, a matrix transformation $P^\dagger A P$, where
  $P$ is invertible, is known as a congruence transformation. Congruence
  transformations and similarity transformations overlap when $P$ is
  unitary.}\BibitemShut {Stop}%
\bibitem [{Note2()}]{Note2}%
  \BibitemOpen
  \bibinfo {note} {It seems this was initially proven in the physics literature
  using a constructive approach based on the Cholesky decomposition of
  $H_{\protect \mathbf {k}}$~\cite {Colpa1978Diagonalization}, rather than
  using the tools from Krein theory.}\BibitemShut {Stop}%
\bibitem [{Note3()}]{Note3}%
  \BibitemOpen
  \bibinfo {note} {The various expressions of PH symmetry all arise in the same
  way, from the redundancy of the Nambu spinors. On the (Hermitian) Bloch-BdG
  matrix, it reads as $\tau ^1 H_{\protect \mathbf {k}} \tau ^1 = H_{-\protect
  \mathbf {k}}^* = H_{-\protect \mathbf {k}}^\top $. On the transformation
  matrix, it reads as $\tau ^1 T_{\protect \mathbf {k}} \tau ^1 = T_{-\protect
  \mathbf {k}}^*$, making the transformed Bloch-BdG matrix P-H symmetric. On
  the (skew-Hermitian) generator $W_{\protect \mathbf {k}}$, it reads as $\tau
  ^1 W_{\protect \mathbf {k}} \tau ^1 = W_{-\protect \mathbf {k}}^\top = -
  W_{-\protect \mathbf {k}}^*$.}\BibitemShut {Stop}%
\bibitem [{Note4()}]{Note4}%
  \BibitemOpen
  \bibinfo {note} {Note that dynamical stability and thermodynamic stability
  are fully independent concepts: Krein-Hermitian Hamiltonians can be both,
  either, or neither~\cite {Flynn2020Deconstructing}.}\BibitemShut {Stop}%
\bibitem [{\citenamefont {Peano}\ and\ \citenamefont
  {Schulz-Baldes}(2018)}]{Peano2018Topological}%
  \BibitemOpen
  \bibfield  {author} {\bibinfo {author} {\bibfnamefont {V.}~\bibnamefont
  {Peano}}\ and\ \bibinfo {author} {\bibfnamefont {H.}~\bibnamefont
  {Schulz-Baldes}},\ }\bibfield  {title} {\bibinfo {title} {Topological edge
  states for disordered bosonic systems},\ }\href
  {https://doi.org/10.1063/1.5002094} {\bibfield  {journal} {\bibinfo
  {journal} {Journal of Mathematical Physics}\ }\textbf {\bibinfo {volume}
  {59}},\ \bibinfo {pages} {031901} (\bibinfo {year} {2018})}\BibitemShut
  {NoStop}%
\bibitem [{\citenamefont {Schulz-Baldes}\ and\ \citenamefont
  {Villegas-Blas}(2017)}]{SchulzBaldes2017Signatures}%
  \BibitemOpen
  \bibfield  {author} {\bibinfo {author} {\bibfnamefont {H.}~\bibnamefont
  {Schulz-Baldes}}\ and\ \bibinfo {author} {\bibfnamefont {C.}~\bibnamefont
  {Villegas-Blas}},\ }\bibfield  {title} {\bibinfo {title} {Signatures for
  {$J$-Hermitians} and {$J$}-unitaries on {Krein} spaces with real
  structures},\ }\href {https://doi.org/10.1002/mana.201600018} {\bibfield
  {journal} {\bibinfo  {journal} {Mathematische Nachrichten}\ }\textbf
  {\bibinfo {volume} {290}},\ \bibinfo {pages} {1840} (\bibinfo {year}
  {2017})}\BibitemShut {NoStop}%
\bibitem [{\citenamefont {Bracci}\ \emph {et~al.}(1975)\citenamefont {Bracci},
  \citenamefont {Morchio},\ and\ \citenamefont {Strocchi}}]{Bracci1975Wigner}%
  \BibitemOpen
  \bibfield  {author} {\bibinfo {author} {\bibfnamefont {L.}~\bibnamefont
  {Bracci}}, \bibinfo {author} {\bibfnamefont {G.}~\bibnamefont {Morchio}},\
  and\ \bibinfo {author} {\bibfnamefont {F.}~\bibnamefont {Strocchi}},\
  }\bibfield  {title} {\bibinfo {title} {{Wigner}'s theorem on symmetries in
  indefinite metric spaces},\ }\href {https://doi.org/10.1007/BF01608993}
  {\bibfield  {journal} {\bibinfo  {journal} {Communications in Mathematical
  Physics}\ }\textbf {\bibinfo {volume} {41}},\ \bibinfo {pages} {289}
  (\bibinfo {year} {1975})}\BibitemShut {NoStop}%
\bibitem [{Note5()}]{Note5}%
  \BibitemOpen
  \bibinfo {note} {To see this, write $\eta $ in its eigenbasis, $\eta = U \eta
  _\protect \mathrm {d}U^\dagger $, and note that $\eta = (U \protect \sqrt
  {\protect \text {abs} \protect \, \eta _\protect \mathrm {d}}) \protect \text
  {sgn} \protect \, \eta _\protect \mathrm {d} (\protect \sqrt {\protect \text
  {abs} \protect \, \eta _\protect \mathrm {d}} U^\dagger ) = V^\dagger
  \protect \text {sgn} \protect \, \eta _\protect \mathrm {d} V$, where $V =
  \protect \sqrt {\protect \text {abs} \protect \, \eta _\protect \mathrm {d}}
  U^\dagger $ and the $\protect \text {sgn}$ and $\protect \text {abs}$
  functions act elementwise. It is easy to see that $\langle \phi , \psi
  \rangle _\eta = \langle V \phi , V \psi \rangle _{\protect \text {sgn}
  \protect \, \eta _\protect \mathrm {d}}$, $K$ is Krein Hermitian with respect
  to $\eta $ iff $VKV^{-1}$ is Krein Hermitian with respect to $\protect \text
  {sgn} \protect \, \eta _\protect \mathrm {d}$, etc.}\BibitemShut {Stop}%
\bibitem [{Note6()}]{Note6}%
  \BibitemOpen
  \bibinfo {note} {When $\eta $ is diagonal with entries $\pm 1$, the columns
  of a Krein-unitary matrix $T$ are orthonormal with respect to the Krein inner
  product.}\BibitemShut {Stop}%
\bibitem [{Note7()}]{Note7}%
  \BibitemOpen
  \bibinfo {note} {For Krein-Hermitian matrices, dynamical stability is
  equivalent to the concept of ``Krein spectrality''~\cite [Sec.~III\protect
  \,A\protect \,1]{Lein2019Krein}.}\BibitemShut {Stop}%
\bibitem [{\citenamefont {Sakurai}(1994)}]{sakurai}%
  \BibitemOpen
  \bibfield  {author} {\bibinfo {author} {\bibfnamefont {J.~J.}\ \bibnamefont
  {Sakurai}},\ }\href@noop {} {\emph {\bibinfo {title} {Modern Quantum
  Mechanics}}}\ (\bibinfo  {publisher} {Addison-Wesley},\ \bibinfo {address}
  {Reading, Massachusetts},\ \bibinfo {year} {1994})\BibitemShut {NoStop}%
\bibitem [{Note8()}]{Note8}%
  \BibitemOpen
  \bibinfo {note} {In general, a non-orthonogal projection can be written $P =
  \DOTSB \sum@ \slimits@ _{i \in S} e_i^{\protect \vphantom {\dagger }}
  \protect \, \omega _i^\dagger $, where $\{e_i\}$ is a basis of the vector
  space and $S$ is a subset of the basis vectors. Furthermore, $\{\omega _i\}$
  is the basis of the dual space, chosen to satisfy the biorthogonality
  condition $\omega _i^\dagger e_j^{\protect \vphantom {\dagger }} = \delta
  _{ij}$. For a Krein-orthogonal basis $\{t_i\}$, the dual basis vector $\omega
  _i$ depends only on the basis vector $t_i$: $\omega _i = \eta \protect \,
  t_i^{\protect \vphantom {\dagger }} / (t_i^\dagger \eta t_i^{\protect
  \vphantom {\dagger }})$.}\BibitemShut {Stop}%
\bibitem [{Note9()}]{Note9}%
  \BibitemOpen
  \bibinfo {note} {Please bear in mind the difference between the digit ``$0$''
  (zero) and the letter ``o''.}\BibitemShut {Stop}%
\bibitem [{Note10()}]{Note10}%
  \BibitemOpen
  \bibinfo {note} {Clearly, this condition does not uniquely specify $W$ or
  $T$. In Appendix~\ref {app:SW-expansion}, we perturbatively derive the
  ``canonical'' choice~\cite {Kessler2012Generalized} of $W$, with which $W$ is
  block off-diagonal.}\BibitemShut {Stop}%
\bibitem [{\citenamefont {Shindou}\ \emph
  {et~al.}(2013{\natexlab{b}})\citenamefont {Shindou}, \citenamefont
  {Matsumoto}, \citenamefont {Murakami},\ and\ \citenamefont
  {Ohe}}]{Sindou2013Magnonic}%
  \BibitemOpen
  \bibfield  {author} {\bibinfo {author} {\bibfnamefont {R.}~\bibnamefont
  {Shindou}}, \bibinfo {author} {\bibfnamefont {R.}~\bibnamefont {Matsumoto}},
  \bibinfo {author} {\bibfnamefont {S.}~\bibnamefont {Murakami}},\ and\
  \bibinfo {author} {\bibfnamefont {J.-i.}\ \bibnamefont {Ohe}},\ }\bibfield
  {title} {\bibinfo {title} {Topological chiral magnonic edge mode in a
  magnonic crystal},\ }\href {https://doi.org/10.1103/PhysRevB.87.174427}
  {\bibfield  {journal} {\bibinfo  {journal} {Phys. Rev. B}\ }\textbf {\bibinfo
  {volume} {87}},\ \bibinfo {pages} {174427} (\bibinfo {year}
  {2013}{\natexlab{b}})}\BibitemShut {NoStop}%
\bibitem [{\citenamefont {Kondo}\ \emph {et~al.}(2019)\citenamefont {Kondo},
  \citenamefont {Akagi},\ and\ \citenamefont {Katsura}}]{Kondo2019Magnon}%
  \BibitemOpen
  \bibfield  {author} {\bibinfo {author} {\bibfnamefont {H.}~\bibnamefont
  {Kondo}}, \bibinfo {author} {\bibfnamefont {Y.}~\bibnamefont {Akagi}},\ and\
  \bibinfo {author} {\bibfnamefont {H.}~\bibnamefont {Katsura}},\ }\bibfield
  {title} {\bibinfo {title} {Three-dimensional topological magnon systems},\
  }\href {https://doi.org/10.1103/PhysRevB.100.144401} {\bibfield  {journal}
  {\bibinfo  {journal} {Phys. Rev. B}\ }\textbf {\bibinfo {volume} {100}},\
  \bibinfo {pages} {144401} (\bibinfo {year} {2019})}\BibitemShut {NoStop}%
\bibitem [{\citenamefont {Trefethen}\ and\ \citenamefont
  {Embree}(2005)}]{TrefethenEmbreePseudospectra}%
  \BibitemOpen
  \bibfield  {author} {\bibinfo {author} {\bibfnamefont {L.~N.}\ \bibnamefont
  {Trefethen}}\ and\ \bibinfo {author} {\bibfnamefont {M.}~\bibnamefont
  {Embree}},\ }\href@noop {} {\emph {\bibinfo {title} {Spectra and
  Pseudospectra: The Behavior of Nonnormal Matrices and Operators}}}\ (\bibinfo
   {publisher} {Princeton University Press},\ \bibinfo {address} {41 William
  Street, Princeton, New Jersey 08540},\ \bibinfo {year} {2005})\BibitemShut
  {NoStop}%
\bibitem [{\citenamefont {Stewart}\ and\ \citenamefont
  {Sun}(1990)}]{stewart1990matrix}%
  \BibitemOpen
  \bibfield  {author} {\bibinfo {author} {\bibfnamefont {G.~W.}\ \bibnamefont
  {Stewart}}\ and\ \bibinfo {author} {\bibfnamefont {J.}~\bibnamefont {Sun}},\
  }\bibfield  {title} {\bibinfo {title} {Matrix perturbation theory},\ }in\
  \href@noop {} {\emph {\bibinfo {booktitle} {Computer Science and Scientific
  Computing}}},\ \bibinfo {editor} {edited by\ \bibinfo {editor} {\bibfnamefont
  {W.}~\bibnamefont {Reinboldt}}\ and\ \bibinfo {editor} {\bibfnamefont
  {D.}~\bibnamefont {Siewiorek}}}\ (\bibinfo  {publisher} {Academic Press,
  inc.},\ \bibinfo {address} {Boston},\ \bibinfo {year} {1990})\BibitemShut
  {NoStop}%
\bibitem [{Note11()}]{Note11}%
  \BibitemOpen
  \bibinfo {note} {It is easy to convince oneself that $\eta K^0$ is positive
  definite in the ordinary sense if and only if $K^0$ is positive definite with
  respect to the Krein inner product, with the consequences laid out at the end
  of Sec.~\ref {subsec:KreinPrimer}.}\BibitemShut {Stop}%
\bibitem [{\citenamefont {Korm\'anyos}\ \emph {et~al.}(2013)\citenamefont
  {Korm\'anyos}, \citenamefont {Z\'olyomi}, \citenamefont {Drummond},
  \citenamefont {Rakyta}, \citenamefont {Burkard},\ and\ \citenamefont
  {Fal'ko}}]{Kormanyos2013Monolayer}%
  \BibitemOpen
  \bibfield  {author} {\bibinfo {author} {\bibfnamefont {A.}~\bibnamefont
  {Korm\'anyos}}, \bibinfo {author} {\bibfnamefont {V.}~\bibnamefont
  {Z\'olyomi}}, \bibinfo {author} {\bibfnamefont {N.~D.}\ \bibnamefont
  {Drummond}}, \bibinfo {author} {\bibfnamefont {P.}~\bibnamefont {Rakyta}},
  \bibinfo {author} {\bibfnamefont {G.}~\bibnamefont {Burkard}},\ and\ \bibinfo
  {author} {\bibfnamefont {V.~I.}\ \bibnamefont {Fal'ko}},\ }\bibfield  {title}
  {\bibinfo {title} {Monolayer {MoS}${}_{2}$: Trigonal warping, the
  $\ensuremath{\Gamma}$ valley, and spin-orbit coupling effects},\ }\href
  {https://doi.org/10.1103/PhysRevB.88.045416} {\bibfield  {journal} {\bibinfo
  {journal} {Phys. Rev. B}\ }\textbf {\bibinfo {volume} {88}},\ \bibinfo
  {pages} {045416} (\bibinfo {year} {2013})}\BibitemShut {NoStop}%
\bibitem [{\citenamefont {Korm\'anyos}\ \emph {et~al.}(2014)\citenamefont
  {Korm\'anyos}, \citenamefont {Z\'olyomi}, \citenamefont {Drummond},\ and\
  \citenamefont {Burkard}}]{Kormanyos2014SpinOrbit}%
  \BibitemOpen
  \bibfield  {author} {\bibinfo {author} {\bibfnamefont {A.}~\bibnamefont
  {Korm\'anyos}}, \bibinfo {author} {\bibfnamefont {V.}~\bibnamefont
  {Z\'olyomi}}, \bibinfo {author} {\bibfnamefont {N.~D.}\ \bibnamefont
  {Drummond}},\ and\ \bibinfo {author} {\bibfnamefont {G.}~\bibnamefont
  {Burkard}},\ }\bibfield  {title} {\bibinfo {title} {Spin-orbit coupling,
  quantum dots, and qubits in monolayer transition metal dichalcogenides},\
  }\href {https://doi.org/10.1103/PhysRevX.4.011034} {\bibfield  {journal}
  {\bibinfo  {journal} {Phys. Rev. X}\ }\textbf {\bibinfo {volume} {4}},\
  \bibinfo {pages} {011034} (\bibinfo {year} {2014})}\BibitemShut {NoStop}%
\bibitem [{\citenamefont {Korm{\'{a}}nyos}\ \emph
  {et~al.}(2015{\natexlab{a}})\citenamefont {Korm{\'{a}}nyos}, \citenamefont
  {Burkard}, \citenamefont {Gmitra}, \citenamefont {Fabian}, \citenamefont
  {Z{\'{o}}lyomi}, \citenamefont {Drummond},\ and\ \citenamefont
  {Fal'ko}}]{Korm_nyos_2015}%
  \BibitemOpen
  \bibfield  {author} {\bibinfo {author} {\bibfnamefont {A.}~\bibnamefont
  {Korm{\'{a}}nyos}}, \bibinfo {author} {\bibfnamefont {G.}~\bibnamefont
  {Burkard}}, \bibinfo {author} {\bibfnamefont {M.}~\bibnamefont {Gmitra}},
  \bibinfo {author} {\bibfnamefont {J.}~\bibnamefont {Fabian}}, \bibinfo
  {author} {\bibfnamefont {V.}~\bibnamefont {Z{\'{o}}lyomi}}, \bibinfo {author}
  {\bibfnamefont {N.~D.}\ \bibnamefont {Drummond}},\ and\ \bibinfo {author}
  {\bibfnamefont {V.}~\bibnamefont {Fal'ko}},\ }\bibfield  {title} {\bibinfo
  {title} {$\mathbf{k} \cdot \mathbf{p}$ theory for two-dimensional transition
  metal dichalcogenide semiconductors},\ }\href
  {https://doi.org/10.1088/2053-1583/2/2/022001} {\bibfield  {journal}
  {\bibinfo  {journal} {2D Materials}\ }\textbf {\bibinfo {volume} {2}},\
  \bibinfo {pages} {022001} (\bibinfo {year} {2015}{\natexlab{a}})}\BibitemShut
  {NoStop}%
\bibitem [{\citenamefont {Korm{\'{a}}nyos}\ \emph
  {et~al.}(2015{\natexlab{b}})\citenamefont {Korm{\'{a}}nyos}, \citenamefont
  {Burkard}, \citenamefont {Gmitra}, \citenamefont {Fabian}, \citenamefont
  {Z{\'{o}}lyomi}, \citenamefont {Drummond},\ and\ \citenamefont
  {Fal'ko}}]{Korm_nyos_2015_corr}%
  \BibitemOpen
  \bibfield  {author} {\bibinfo {author} {\bibfnamefont {A.}~\bibnamefont
  {Korm{\'{a}}nyos}}, \bibinfo {author} {\bibfnamefont {G.}~\bibnamefont
  {Burkard}}, \bibinfo {author} {\bibfnamefont {M.}~\bibnamefont {Gmitra}},
  \bibinfo {author} {\bibfnamefont {J.}~\bibnamefont {Fabian}}, \bibinfo
  {author} {\bibfnamefont {V.}~\bibnamefont {Z{\'{o}}lyomi}}, \bibinfo {author}
  {\bibfnamefont {N.~D.}\ \bibnamefont {Drummond}},\ and\ \bibinfo {author}
  {\bibfnamefont {V.}~\bibnamefont {Fal'ko}},\ }\bibfield  {title} {\bibinfo
  {title} {Corrigendum: $\mathbf{k} \cdot \mathbf{p}$ theory for
  two-dimensional transition metal dichalcogenide semiconductors (2015 {2D}
  {Mater}. 2 022001)},\ }\href {https://doi.org/10.1088/2053-1583/2/4/049501}
  {\bibfield  {journal} {\bibinfo  {journal} {2D Materials}\ }\textbf {\bibinfo
  {volume} {2}},\ \bibinfo {pages} {049501} (\bibinfo {year}
  {2015}{\natexlab{b}})}\BibitemShut {NoStop}%
\bibitem [{\citenamefont {Beiranvand}\ \emph {et~al.}(2018)\citenamefont
  {Beiranvand}, \citenamefont {Dezfuli},\ and\ \citenamefont
  {Sabaeian}}]{Beiranvand2018TwoBand}%
  \BibitemOpen
  \bibfield  {author} {\bibinfo {author} {\bibfnamefont {K.}~\bibnamefont
  {Beiranvand}}, \bibinfo {author} {\bibfnamefont {A.~G.}\ \bibnamefont
  {Dezfuli}},\ and\ \bibinfo {author} {\bibfnamefont {M.}~\bibnamefont
  {Sabaeian}},\ }\bibfield  {title} {\bibinfo {title} {A two-band spinful k.p
  {Hamiltonian} of monolayer {MoS2} from a nine-band model based on group
  theory},\ }\href {https://doi.org/10.1016/j.spmi.2018.06.033} {\bibfield
  {journal} {\bibinfo  {journal} {Superlattices and Microstructures}\ }\textbf
  {\bibinfo {volume} {120}},\ \bibinfo {pages} {812} (\bibinfo {year}
  {2018})}\BibitemShut {NoStop}%
\bibitem [{\citenamefont {Fang}\ \emph {et~al.}(2015)\citenamefont {Fang},
  \citenamefont {Chen}, \citenamefont {Kee},\ and\ \citenamefont
  {Fu}}]{Fang2015Topological}%
  \BibitemOpen
  \bibfield  {author} {\bibinfo {author} {\bibfnamefont {C.}~\bibnamefont
  {Fang}}, \bibinfo {author} {\bibfnamefont {Y.}~\bibnamefont {Chen}}, \bibinfo
  {author} {\bibfnamefont {H.-Y.}\ \bibnamefont {Kee}},\ and\ \bibinfo {author}
  {\bibfnamefont {L.}~\bibnamefont {Fu}},\ }\bibfield  {title} {\bibinfo
  {title} {Topological nodal line semimetals with and without spin-orbital
  coupling},\ }\href {https://doi.org/10.1103/PhysRevB.92.081201} {\bibfield
  {journal} {\bibinfo  {journal} {Phys. Rev. B}\ }\textbf {\bibinfo {volume}
  {92}},\ \bibinfo {pages} {081201(R)} (\bibinfo {year} {2015})}\BibitemShut
  {NoStop}%
\bibitem [{\citenamefont {Ahn}\ \emph {et~al.}(2018)\citenamefont {Ahn},
  \citenamefont {Kim}, \citenamefont {Kim},\ and\ \citenamefont
  {Yang}}]{Ahn2018Band}%
  \BibitemOpen
  \bibfield  {author} {\bibinfo {author} {\bibfnamefont {J.}~\bibnamefont
  {Ahn}}, \bibinfo {author} {\bibfnamefont {D.}~\bibnamefont {Kim}}, \bibinfo
  {author} {\bibfnamefont {Y.}~\bibnamefont {Kim}},\ and\ \bibinfo {author}
  {\bibfnamefont {B.-J.}\ \bibnamefont {Yang}},\ }\bibfield  {title} {\bibinfo
  {title} {Band topology and linking structure of nodal line semimetals with
  ${Z}_{2}$ monopole charges},\ }\href
  {https://doi.org/10.1103/PhysRevLett.121.106403} {\bibfield  {journal}
  {\bibinfo  {journal} {Phys. Rev. Lett.}\ }\textbf {\bibinfo {volume} {121}},\
  \bibinfo {pages} {106403} (\bibinfo {year} {2018})}\BibitemShut {NoStop}%
\bibitem [{\citenamefont {Rui}\ \emph {et~al.}(2018)\citenamefont {Rui},
  \citenamefont {Zhao},\ and\ \citenamefont {Schnyder}}]{Rui2018NodalLine}%
  \BibitemOpen
  \bibfield  {author} {\bibinfo {author} {\bibfnamefont {W.~B.}\ \bibnamefont
  {Rui}}, \bibinfo {author} {\bibfnamefont {Y.~X.}\ \bibnamefont {Zhao}},\ and\
  \bibinfo {author} {\bibfnamefont {A.~P.}\ \bibnamefont {Schnyder}},\
  }\bibfield  {title} {\bibinfo {title} {Topological transport in dirac
  nodal-line semimetals},\ }\href {https://doi.org/10.1103/PhysRevB.97.161113}
  {\bibfield  {journal} {\bibinfo  {journal} {Phys. Rev. B}\ }\textbf {\bibinfo
  {volume} {97}},\ \bibinfo {pages} {161113(R)} (\bibinfo {year}
  {2018})}\BibitemShut {NoStop}%
\bibitem [{\citenamefont {Li}\ \emph {et~al.}(2018)\citenamefont {Li},
  \citenamefont {Liu}, \citenamefont {Fu}, \citenamefont {Yu}, \citenamefont
  {Yang},\ and\ \citenamefont {Yao}}]{Li2018CuTeO3}%
  \BibitemOpen
  \bibfield  {author} {\bibinfo {author} {\bibfnamefont {S.}~\bibnamefont
  {Li}}, \bibinfo {author} {\bibfnamefont {Y.}~\bibnamefont {Liu}}, \bibinfo
  {author} {\bibfnamefont {B.}~\bibnamefont {Fu}}, \bibinfo {author}
  {\bibfnamefont {Z.-M.}\ \bibnamefont {Yu}}, \bibinfo {author} {\bibfnamefont
  {S.~A.}\ \bibnamefont {Yang}},\ and\ \bibinfo {author} {\bibfnamefont
  {Y.}~\bibnamefont {Yao}},\ }\bibfield  {title} {\bibinfo {title} {Almost
  ideal nodal-loop semimetal in monoclinic {CuTeO}\textsubscript{3} material},\
  }\href {https://doi.org/10.1103/PhysRevB.97.245148} {\bibfield  {journal}
  {\bibinfo  {journal} {Phys. Rev. B}\ }\textbf {\bibinfo {volume} {97}},\
  \bibinfo {pages} {245148} (\bibinfo {year} {2018})}\BibitemShut {NoStop}%
\bibitem [{\citenamefont {Ludwig}(2016)}]{Ludwig_2015}%
  \BibitemOpen
  \bibfield  {author} {\bibinfo {author} {\bibfnamefont {A.~W.~W.}\
  \bibnamefont {Ludwig}},\ }\bibfield  {title} {\bibinfo {title} {Topological
  phases: classification of topological insulators and superconductors of
  non-interacting fermions, and beyond},\ }\href
  {https://doi.org/10.1088/0031-8949/2015/t168/014001} {\bibfield  {journal}
  {\bibinfo  {journal} {Physica Scripta}\ }\textbf {\bibinfo {volume} {T168}},\
  \bibinfo {pages} {014001} (\bibinfo {year} {2016})}\BibitemShut {NoStop}%
\bibitem [{\citenamefont {Zhao}\ and\ \citenamefont {Lu}(2017)}]{Zhao2017PT}%
  \BibitemOpen
  \bibfield  {author} {\bibinfo {author} {\bibfnamefont {Y.~X.}\ \bibnamefont
  {Zhao}}\ and\ \bibinfo {author} {\bibfnamefont {Y.}~\bibnamefont {Lu}},\
  }\bibfield  {title} {\bibinfo {title} {{$PT$}-symmetric real dirac fermions
  and semimetals},\ }\href {https://doi.org/10.1103/PhysRevLett.118.056401}
  {\bibfield  {journal} {\bibinfo  {journal} {Phys. Rev. Lett.}\ }\textbf
  {\bibinfo {volume} {118}},\ \bibinfo {pages} {056401} (\bibinfo {year}
  {2017})}\BibitemShut {NoStop}%
\bibitem [{\citenamefont {Murakami}(2007)}]{Shuichi_Murakami_2007}%
  \BibitemOpen
  \bibfield  {author} {\bibinfo {author} {\bibfnamefont {S.}~\bibnamefont
  {Murakami}},\ }\bibfield  {title} {\bibinfo {title} {Phase transition between
  the quantum spin {Hall} and insulator phases in 3d: emergence of a
  topological gapless phase},\ }\href
  {https://doi.org/10.1088/1367-2630/9/9/356} {\bibfield  {journal} {\bibinfo
  {journal} {New Journal of Physics}\ }\textbf {\bibinfo {volume} {9}},\
  \bibinfo {pages} {356} (\bibinfo {year} {2007})}\BibitemShut {NoStop}%
\bibitem [{\citenamefont {Chiu}\ \emph {et~al.}(2016)\citenamefont {Chiu},
  \citenamefont {Teo}, \citenamefont {Schnyder},\ and\ \citenamefont
  {Ryu}}]{RMP2016Classification}%
  \BibitemOpen
  \bibfield  {author} {\bibinfo {author} {\bibfnamefont {C.-K.}\ \bibnamefont
  {Chiu}}, \bibinfo {author} {\bibfnamefont {J.~C.~Y.}\ \bibnamefont {Teo}},
  \bibinfo {author} {\bibfnamefont {A.~P.}\ \bibnamefont {Schnyder}},\ and\
  \bibinfo {author} {\bibfnamefont {S.}~\bibnamefont {Ryu}},\ }\bibfield
  {title} {\bibinfo {title} {Classification of topological quantum matter with
  symmetries},\ }\href {https://doi.org/10.1103/RevModPhys.88.035005}
  {\bibfield  {journal} {\bibinfo  {journal} {Rev. Mod. Phys.}\ }\textbf
  {\bibinfo {volume} {88}},\ \bibinfo {pages} {035005} (\bibinfo {year}
  {2016})}\BibitemShut {NoStop}%
\bibitem [{\citenamefont {Bradley}\ and\ \citenamefont
  {Cracknell}(2009)}]{BradleyCracknell}%
  \BibitemOpen
  \bibfield  {author} {\bibinfo {author} {\bibfnamefont {C.}~\bibnamefont
  {Bradley}}\ and\ \bibinfo {author} {\bibfnamefont {A.}~\bibnamefont
  {Cracknell}},\ }\href@noop {} {\emph {\bibinfo {title} {The mathematical
  theory of symmetry in solids: representation theory for point groups and
  space groups}}}\ (\bibinfo  {publisher} {Oxford University Press},\ \bibinfo
  {year} {2009})\BibitemShut {NoStop}%
\bibitem [{\citenamefont {Mook}\ \emph {et~al.}(2017)\citenamefont {Mook},
  \citenamefont {Henk},\ and\ \citenamefont {Mertig}}]{Mook2017Ferromagnets}%
  \BibitemOpen
  \bibfield  {author} {\bibinfo {author} {\bibfnamefont {A.}~\bibnamefont
  {Mook}}, \bibinfo {author} {\bibfnamefont {J.}~\bibnamefont {Henk}},\ and\
  \bibinfo {author} {\bibfnamefont {I.}~\bibnamefont {Mertig}},\ }\bibfield
  {title} {\bibinfo {title} {Magnon nodal-line semimetals and drumhead surface
  states in anisotropic pyrochlore ferromagnets},\ }\href
  {https://doi.org/10.1103/PhysRevB.95.014418} {\bibfield  {journal} {\bibinfo
  {journal} {Phys. Rev. B}\ }\textbf {\bibinfo {volume} {95}},\ \bibinfo
  {pages} {014418} (\bibinfo {year} {2017})}\BibitemShut {NoStop}%
\bibitem [{\citenamefont {Newnham}\ \emph {et~al.}(1964)\citenamefont
  {Newnham}, \citenamefont {Fang},\ and\ \citenamefont
  {Santoro}}]{Newnham1964}%
  \BibitemOpen
  \bibfield  {author} {\bibinfo {author} {\bibfnamefont {R.~E.}\ \bibnamefont
  {Newnham}}, \bibinfo {author} {\bibfnamefont {J.~H.}\ \bibnamefont {Fang}},\
  and\ \bibinfo {author} {\bibfnamefont {R.~P.}\ \bibnamefont {Santoro}},\
  }\bibfield  {title} {\bibinfo {title} {{Crystal structure and magnetic
  properties of CoTiO${\sb 3}$}},\ }\href
  {https://doi.org/10.1107/S0365110X64000615} {\bibfield  {journal} {\bibinfo
  {journal} {Acta Crystallographica}\ }\textbf {\bibinfo {volume} {17}},\
  \bibinfo {pages} {240} (\bibinfo {year} {1964})}\BibitemShut {NoStop}%
\bibitem [{\citenamefont {Liu}\ \emph {et~al.}(2020)\citenamefont {Liu},
  \citenamefont {Chaloupka},\ and\ \citenamefont {Khaliullin}}]{Liu2020}%
  \BibitemOpen
  \bibfield  {author} {\bibinfo {author} {\bibfnamefont {H.}~\bibnamefont
  {Liu}}, \bibinfo {author} {\bibfnamefont {J.}~\bibnamefont {Chaloupka}},\
  and\ \bibinfo {author} {\bibfnamefont {G.}~\bibnamefont {Khaliullin}},\
  }\bibfield  {title} {\bibinfo {title} {{Kitaev} spin liquid in $3d$
  transition metal compounds},\ }\href
  {https://doi.org/10.1103/PhysRevLett.125.047201} {\bibfield  {journal}
  {\bibinfo  {journal} {Phys. Rev. Lett.}\ }\textbf {\bibinfo {volume} {125}},\
  \bibinfo {pages} {047201} (\bibinfo {year} {2020})}\BibitemShut {NoStop}%
\bibitem [{\citenamefont {Das}\ \emph {et~al.}(2021)\citenamefont {Das},
  \citenamefont {Voleti}, \citenamefont {Saha-Dasgupta},\ and\ \citenamefont
  {Paramekanti}}]{Das2021Cobaltates}%
  \BibitemOpen
  \bibfield  {author} {\bibinfo {author} {\bibfnamefont {S.}~\bibnamefont
  {Das}}, \bibinfo {author} {\bibfnamefont {S.}~\bibnamefont {Voleti}},
  \bibinfo {author} {\bibfnamefont {T.}~\bibnamefont {Saha-Dasgupta}},\ and\
  \bibinfo {author} {\bibfnamefont {A.}~\bibnamefont {Paramekanti}},\
  }\bibfield  {title} {\bibinfo {title} {{XY} magnetism, {Kitaev} exchange, and
  long-range frustration in the ${J}_{\mathrm{eff}}=\frac{1}{2}$ honeycomb
  cobaltates},\ }\href {https://doi.org/10.1103/PhysRevB.104.134425} {\bibfield
   {journal} {\bibinfo  {journal} {Phys. Rev. B}\ }\textbf {\bibinfo {volume}
  {104}},\ \bibinfo {pages} {134425} (\bibinfo {year} {2021})}\BibitemShut
  {NoStop}%
\bibitem [{Note12()}]{Note12}%
  \BibitemOpen
  \bibinfo {note} {Equivalently, the symmetry $g$ allows a Fourier transform in
  a unit cell smaller than the magnetic unit cell, essentially ``unfolding''
  the band structure~\cite {Elliot2020Visualization, Yuan2020} (see also
  Appendix~\ref {sec:symmetries-implementation}).}\BibitemShut {Stop}%
\bibitem [{\citenamefont {Young}\ and\ \citenamefont
  {Kane}(2015)}]{Kane2015DiracSemimetals}%
  \BibitemOpen
  \bibfield  {author} {\bibinfo {author} {\bibfnamefont {S.~M.}\ \bibnamefont
  {Young}}\ and\ \bibinfo {author} {\bibfnamefont {C.~L.}\ \bibnamefont
  {Kane}},\ }\bibfield  {title} {\bibinfo {title} {{Dirac} semimetals in two
  dimensions},\ }\href {https://doi.org/10.1103/PhysRevLett.115.126803}
  {\bibfield  {journal} {\bibinfo  {journal} {Phys. Rev. Lett.}\ }\textbf
  {\bibinfo {volume} {115}},\ \bibinfo {pages} {126803} (\bibinfo {year}
  {2015})}\BibitemShut {NoStop}%
\bibitem [{\citenamefont {Wigner}(1959)}]{Wigner1959Group}%
  \BibitemOpen
  \bibfield  {author} {\bibinfo {author} {\bibfnamefont {E.}~\bibnamefont
  {Wigner}},\ }\href@noop {} {\emph {\bibinfo {title} {Group Theory and its
  Application to the Quantum Mechanics of Atomic Spectra}}}\ (\bibinfo
  {publisher} {Academic Press Inc.},\ \bibinfo {address} {New York},\ \bibinfo
  {year} {1959})\BibitemShut {NoStop}%
\bibitem [{\citenamefont {Zhitomirsky}\ and\ \citenamefont
  {Chernyshev}(2013)}]{Chernyshev2013Colloquium}%
  \BibitemOpen
  \bibfield  {author} {\bibinfo {author} {\bibfnamefont {M.~E.}\ \bibnamefont
  {Zhitomirsky}}\ and\ \bibinfo {author} {\bibfnamefont {A.~L.}\ \bibnamefont
  {Chernyshev}},\ }\bibfield  {title} {\bibinfo {title} {Colloquium:
  Spontaneous magnon decays},\ }\href
  {https://doi.org/10.1103/RevModPhys.85.219} {\bibfield  {journal} {\bibinfo
  {journal} {Rev. Mod. Phys.}\ }\textbf {\bibinfo {volume} {85}},\ \bibinfo
  {pages} {219} (\bibinfo {year} {2013})}\BibitemShut {NoStop}%
\bibitem [{\citenamefont {Mook}\ \emph {et~al.}(2021)\citenamefont {Mook},
  \citenamefont {Plekhanov}, \citenamefont {Klinovaja},\ and\ \citenamefont
  {Loss}}]{Mook2021Interaction}%
  \BibitemOpen
  \bibfield  {author} {\bibinfo {author} {\bibfnamefont {A.}~\bibnamefont
  {Mook}}, \bibinfo {author} {\bibfnamefont {K.}~\bibnamefont {Plekhanov}},
  \bibinfo {author} {\bibfnamefont {J.}~\bibnamefont {Klinovaja}},\ and\
  \bibinfo {author} {\bibfnamefont {D.}~\bibnamefont {Loss}},\ }\bibfield
  {title} {\bibinfo {title} {Interaction-stabilized topological magnon
  insulator in ferromagnets},\ }\href
  {https://doi.org/10.1103/PhysRevX.11.021061} {\bibfield  {journal} {\bibinfo
  {journal} {Phys. Rev. X}\ }\textbf {\bibinfo {volume} {11}},\ \bibinfo
  {pages} {021061} (\bibinfo {year} {2021})}\BibitemShut {NoStop}%
\bibitem [{\citenamefont {Zhang}\ \emph {et~al.}(2019)\citenamefont {Zhang},
  \citenamefont {Miao}, \citenamefont {Wang}, \citenamefont {Lin},
  \citenamefont {Cao}, \citenamefont {Fabbris}, \citenamefont {Said},
  \citenamefont {Liu}, \citenamefont {Lei}, \citenamefont {Fang}, \citenamefont
  {Weng},\ and\ \citenamefont {Dean}}]{Zhang2019Phononic}%
  \BibitemOpen
  \bibfield  {author} {\bibinfo {author} {\bibfnamefont {T.~T.}\ \bibnamefont
  {Zhang}}, \bibinfo {author} {\bibfnamefont {H.}~\bibnamefont {Miao}},
  \bibinfo {author} {\bibfnamefont {Q.}~\bibnamefont {Wang}}, \bibinfo {author}
  {\bibfnamefont {J.~Q.}\ \bibnamefont {Lin}}, \bibinfo {author} {\bibfnamefont
  {Y.}~\bibnamefont {Cao}}, \bibinfo {author} {\bibfnamefont {G.}~\bibnamefont
  {Fabbris}}, \bibinfo {author} {\bibfnamefont {A.~H.}\ \bibnamefont {Said}},
  \bibinfo {author} {\bibfnamefont {X.}~\bibnamefont {Liu}}, \bibinfo {author}
  {\bibfnamefont {H.~C.}\ \bibnamefont {Lei}}, \bibinfo {author} {\bibfnamefont
  {Z.}~\bibnamefont {Fang}}, \bibinfo {author} {\bibfnamefont {H.~M.}\
  \bibnamefont {Weng}},\ and\ \bibinfo {author} {\bibfnamefont {M.~P.~M.}\
  \bibnamefont {Dean}},\ }\bibfield  {title} {\bibinfo {title} {Phononic
  helical nodal lines with $\mathcal{PT}$ protection in {$\mathrm{MoB}_2$}},\
  }\href {https://doi.org/10.1103/PhysRevLett.123.245302} {\bibfield  {journal}
  {\bibinfo  {journal} {Phys. Rev. Lett.}\ }\textbf {\bibinfo {volume} {123}},\
  \bibinfo {pages} {245302} (\bibinfo {year} {2019})}\BibitemShut {NoStop}%
\bibitem [{\citenamefont {Ponce}\ \emph {et~al.}(2019)\citenamefont {Ponce},
  \citenamefont {van Zon}, \citenamefont {Northrup}, \citenamefont {Gruner},
  \citenamefont {Chen}, \citenamefont {Ertinaz}, \citenamefont {Fedoseev},
  \citenamefont {Groer}, \citenamefont {Mao}, \citenamefont {Mundim},
  \citenamefont {Nolta}, \citenamefont {Pinto}, \citenamefont {Saldarriaga},
  \citenamefont {Slavnic}, \citenamefont {Spence}, \citenamefont {Yu},\ and\
  \citenamefont {Peltier}}]{Ponce2019Deploying}%
  \BibitemOpen
  \bibfield  {author} {\bibinfo {author} {\bibfnamefont {M.}~\bibnamefont
  {Ponce}}, \bibinfo {author} {\bibfnamefont {R.}~\bibnamefont {van Zon}},
  \bibinfo {author} {\bibfnamefont {S.}~\bibnamefont {Northrup}}, \bibinfo
  {author} {\bibfnamefont {D.}~\bibnamefont {Gruner}}, \bibinfo {author}
  {\bibfnamefont {J.}~\bibnamefont {Chen}}, \bibinfo {author} {\bibfnamefont
  {F.}~\bibnamefont {Ertinaz}}, \bibinfo {author} {\bibfnamefont
  {A.}~\bibnamefont {Fedoseev}}, \bibinfo {author} {\bibfnamefont
  {L.}~\bibnamefont {Groer}}, \bibinfo {author} {\bibfnamefont
  {F.}~\bibnamefont {Mao}}, \bibinfo {author} {\bibfnamefont {B.~C.}\
  \bibnamefont {Mundim}}, \bibinfo {author} {\bibfnamefont {M.}~\bibnamefont
  {Nolta}}, \bibinfo {author} {\bibfnamefont {J.}~\bibnamefont {Pinto}},
  \bibinfo {author} {\bibfnamefont {M.}~\bibnamefont {Saldarriaga}}, \bibinfo
  {author} {\bibfnamefont {V.}~\bibnamefont {Slavnic}}, \bibinfo {author}
  {\bibfnamefont {E.}~\bibnamefont {Spence}}, \bibinfo {author} {\bibfnamefont
  {C.-H.}\ \bibnamefont {Yu}},\ and\ \bibinfo {author} {\bibfnamefont {W.~R.}\
  \bibnamefont {Peltier}},\ }\bibfield  {title} {\bibinfo {title} {Deploying a
  top-100 supercomputer for large parallel workloads: The {Niagara}
  supercomputer},\ }in\ \href {https://doi.org/10.1145/3332186.3332195} {\emph
  {\bibinfo {booktitle} {Proceedings of the Practice and Experience in Advanced
  Research Computing on Rise of the Machines (Learning)}}},\ \bibinfo {series
  and number} {PEARC '19}\ (\bibinfo  {publisher} {Association for Computing
  Machinery},\ \bibinfo {address} {New York, NY, USA},\ \bibinfo {year}
  {2019})\BibitemShut {NoStop}%
\bibitem [{\citenamefont {Loken}\ \emph {et~al.}(2010)\citenamefont {Loken},
  \citenamefont {Gruner}, \citenamefont {Groer}, \citenamefont {Peltier},
  \citenamefont {Bunn}, \citenamefont {Craig}, \citenamefont {Henriques},
  \citenamefont {Dempsey}, \citenamefont {Yu}, \citenamefont {Chen},
  \citenamefont {Dursi}, \citenamefont {Chong}, \citenamefont {Northrup},
  \citenamefont {Pinto}, \citenamefont {Knecht},\ and\ \citenamefont
  {Zon}}]{Loken2010SciNet}%
  \BibitemOpen
  \bibfield  {author} {\bibinfo {author} {\bibfnamefont {C.}~\bibnamefont
  {Loken}}, \bibinfo {author} {\bibfnamefont {D.}~\bibnamefont {Gruner}},
  \bibinfo {author} {\bibfnamefont {L.}~\bibnamefont {Groer}}, \bibinfo
  {author} {\bibfnamefont {R.}~\bibnamefont {Peltier}}, \bibinfo {author}
  {\bibfnamefont {N.}~\bibnamefont {Bunn}}, \bibinfo {author} {\bibfnamefont
  {M.}~\bibnamefont {Craig}}, \bibinfo {author} {\bibfnamefont
  {T.}~\bibnamefont {Henriques}}, \bibinfo {author} {\bibfnamefont
  {J.}~\bibnamefont {Dempsey}}, \bibinfo {author} {\bibfnamefont {C.-H.}\
  \bibnamefont {Yu}}, \bibinfo {author} {\bibfnamefont {J.}~\bibnamefont
  {Chen}}, \bibinfo {author} {\bibfnamefont {L.~J.}\ \bibnamefont {Dursi}},
  \bibinfo {author} {\bibfnamefont {J.}~\bibnamefont {Chong}}, \bibinfo
  {author} {\bibfnamefont {S.}~\bibnamefont {Northrup}}, \bibinfo {author}
  {\bibfnamefont {J.}~\bibnamefont {Pinto}}, \bibinfo {author} {\bibfnamefont
  {N.}~\bibnamefont {Knecht}},\ and\ \bibinfo {author} {\bibfnamefont {R.~V.}\
  \bibnamefont {Zon}},\ }\bibfield  {title} {\bibinfo {title} {{SciNet}:
  Lessons learned from building a power-efficient top-20 system and data
  centre},\ }\href {https://doi.org/10.1088/1742-6596/256/1/012026} {\bibfield
  {journal} {\bibinfo  {journal} {Journal of Physics: Conference Series}\
  }\textbf {\bibinfo {volume} {256}},\ \bibinfo {pages} {012026} (\bibinfo
  {year} {2010})}\BibitemShut {NoStop}%
\bibitem [{Note13()}]{Note13}%
  \BibitemOpen
  \bibinfo {note} {See Supplementary Figure~4 of Ref.~\protect \rev@citealp
  {Elliot2020Visualization} for a depiction of these lines, which, as explained
  in the reference, project onto the corner K points of the 2D hexagonal
  Brillouin zone.}\BibitemShut {Stop}%
\bibitem [{\citenamefont {Brinkman}\ and\ \citenamefont
  {Elliott}(1966)}]{brinkman1966space}%
  \BibitemOpen
  \bibfield  {author} {\bibinfo {author} {\bibfnamefont {W.}~\bibnamefont
  {Brinkman}}\ and\ \bibinfo {author} {\bibfnamefont {R.~J.}\ \bibnamefont
  {Elliott}},\ }\bibfield  {title} {\bibinfo {title} {Space group theory for
  spin waves},\ }\href {https://doi.org/10.1063/1.1708514} {\bibfield
  {journal} {\bibinfo  {journal} {Journal of Applied Physics}\ }\textbf
  {\bibinfo {volume} {37}},\ \bibinfo {pages} {1457} (\bibinfo {year}
  {1966})}\BibitemShut {NoStop}%
\bibitem [{\citenamefont {Brinkman}\ \emph {et~al.}(1966)\citenamefont
  {Brinkman}, \citenamefont {Elliott},\ and\ \citenamefont
  {Peierls}}]{brinkman1966theory}%
  \BibitemOpen
  \bibfield  {author} {\bibinfo {author} {\bibfnamefont {W.~F.}\ \bibnamefont
  {Brinkman}}, \bibinfo {author} {\bibfnamefont {R.~J.}\ \bibnamefont
  {Elliott}},\ and\ \bibinfo {author} {\bibfnamefont {R.~E.}\ \bibnamefont
  {Peierls}},\ }\bibfield  {title} {\bibinfo {title} {Theory of spin-space
  groups},\ }\href {https://doi.org/10.1098/rspa.1966.0211} {\bibfield
  {journal} {\bibinfo  {journal} {Proceedings of the Royal Society of London.
  Series A. Mathematical and Physical Sciences}\ }\textbf {\bibinfo {volume}
  {294}},\ \bibinfo {pages} {343} (\bibinfo {year} {1966})}\BibitemShut
  {NoStop}%
\bibitem [{\citenamefont {Brinkman}(1967)}]{brinkman1967magnetic}%
  \BibitemOpen
  \bibfield  {author} {\bibinfo {author} {\bibfnamefont {W.}~\bibnamefont
  {Brinkman}},\ }\bibfield  {title} {\bibinfo {title} {Magnetic symmetry and
  spin waves},\ }\href {https://doi.org/10.1063/1.1709692} {\bibfield
  {journal} {\bibinfo  {journal} {Journal of Applied Physics}\ }\textbf
  {\bibinfo {volume} {38}},\ \bibinfo {pages} {939} (\bibinfo {year}
  {1967})}\BibitemShut {NoStop}%
\end{thebibliography}%

\end{document}